\documentclass[twocolumn]{aastex61}
\pdfoutput=1 
\usepackage{amsmath,amstext}
\usepackage[T1]{fontenc}
\usepackage{apjfonts} 
\usepackage[figure,figure*]{hypcap}

\newcommand{\kms}{{\hbox {km\thinspace s$^{-1}$}}}
\newcommand{\Lsun}{{\hbox {L$_\odot$}}}
\newcommand{\Msun}{{\hbox {M$_\odot$}}}

\newcommand{\cmd}{{\hbox {cm$^{-2}$}}}

\newcommand{\tdust}{{\hbox {$T_{\mathrm{dust}}$}}}

\shorttitle{The Greenhouse Effect in Buried Galactic Nuclei}
\shortauthors{Gonz\'alez-Alfonso \& Sakamoto}

\begin{document}

\title{The Greenhouse Effect in Buried Galactic Nuclei and the
  Resonant HCN Vibrational Emission} 

\author{Eduardo Gonz\'alez-Alfonso}
\affiliation{Universidad de Alcal\'a, Departamento de F\'{\i}sica
     y Matem\'aticas, Campus Universitario, E-28871 Alcal\'a de Henares,
     Madrid, Spain}
\author{Kazushi Sakamoto}
\affiliation{Academia Sinica, Institute of Astronomy and Astrophysics,
  Taipei 10617, Taiwan}

\begin{abstract}
Recent interferometric observations have shown bright HCN emission from the
$\nu_2=1$ vibrational state arising in buried nuclear regions of galaxies,
indicating an efficient pumping of the $\nu_2=1$ state
through absorption of $14$ $\mu$m continuum photons.  
We have modeled the continuum and HCN vibrational line emission in these
regions, characterized by high column densities of dust and high luminosities,
with a spherically symmetric approach, simulating both a central heating
source (AGN) and a compact nuclear starburst (SB). We find that
when the H$_2$ columns become very high, $N_{\mathrm{H2}}\gtrsim10^{25}$ \cmd,
trapping of continuum photons within the nuclear region dramatically
enhances the dust temperature (\tdust) in the inner regions,
even though the predicted spectral energy distribution
as seen from outside becomes relatively cold.
The models thus predict bright continuum at millimeter
wavelengths for luminosity surface brightness
(averaged over the model source) of $\sim10^{8}$ \Lsun\,pc$^{-2}$.
This {\it greenhouse} effect significantly enhances the mean mid-infrared
intensity within the dusty volume,
populating the $\nu_2=1$ state to the extent that
the HCN vibrational lines become optically thick.
AGN models yield higher \tdust\ in the inner regions
  and higher peak (sub)millimeter continuum brightness than SB models,
  but similar HCN vibrational $J=3-2$ and $4-3$ emission owing to both
    optical depth effects and a moderate impact of high \tdust\ on these low-$J$
    lines. The observed HCN vibrational
emission in several galaxies can be accounted for with a HCN abundance of
$\sim10^{-6}$ (relative to H$_2$) and luminosity surface brightness in the
range $(0.5-2)\times10^{8}$ \Lsun\,pc$^{-2}$, predicting a far-infrared
photosphere with $T_{\mathrm{dust}}\sim80-150$\,K --in agreement with the
values inferred from far-infrared molecular absorption.
\end{abstract}


\section{Introduction}

Buried galactic nuclei (hereafter BGN)
are compact ($\lesssim100$\,pc) nuclear regions in galaxies with both
high gas column densities ($\gtrsim10^{24}$\,H$_2$\,cm$^{-2}$)
and high luminosity surface brightnesses
($\gtrsim10^{7}$\,L$_{\odot}$\,pc$^{-2}$).
Both starbursts and buried, accreting supermassive black holes
  (i.e., active galactic nuclei, AGNs) are suspected to power BGNs,
  although the dominant luminosity sources in individual nuclei and in the
  entire BGN population are unclear and under debate.
In the local Universe, they are mostly observed in (Ultra)-Luminous
Infrared Galaxies, (U)LIRGs, and when present they contribute significantly
to the high luminosities of these galaxies \citep{soi00}.
  Obscured nuclei may be formed through the loss
of angular momentum of the gas associated with a merger event or a strong bar
\citep[e.g.][]{hop08},
or via dissipative contraction of gas-rich disks that become unstable
due to cold inflows \citep{dek14}. All these mechanisms involve neutral inflows
that are indeed observed in some (U)LIRGs via inverse P-Cygni profiles or
redshifted absorption in [O {\sc i}]\,63\,$\mu$m and/or ground-state OH
\citep{gon12,gon17,fal15,fal17}.
At higher redshifts, mergers and cold inflows are probably more common
than locally, and so are the formation of obscured nuclei;
they have been identified from the high infrared to X-ray luminosity ratio and
their negligible emission at optical wavelengths \citep{cha17}, though their
statistical significance is still not well addressed.

Given the extreme extinction that characterize BGNs, the most promising way
to identify and give insight into their physics is through continuum
studies with high angular resolution at millimeter wavelengths and
with molecular spectroscopy from the infrared to centimeter wavelengths.
Interferometric observations have measured $0.8-3$ mm continuum
  brightness temperatures of hundreds of Kelvin at the scale of a few tens
  of parsecs or less in the nuclei of some (U)LIRGs
  \citep[e.g.,][]{sak13,sak17,aal19}. Attributing the continuum to thermal
  emission from dust (as supported by radio spectral energy distribution),
  the strongly peaked emission suggests high luminosity surface densities in
  the nuclei. As a guide, a 100 K blackbody sphere has a luminosity surface
  density of $L_{\mathrm{IR}}/(\pi R^2)\approx6\times 10^7$\,L$_{\odot}$\,pc$^{-2}$.
  (We will see later the effect of gas and dust covering such a source.)
  These nuclei also need to
  have the high column densities of BGNs in order for dust to have a
  significant optical depth that can explain these high brightness
  temperatures. 
At far-infrared (far-IR) wavelengths, BGNs are characterized by high-lying
(i.e., lower-level energy $E_{\mathrm{low}}\gtrsim300$\,K) absorption lines of
  various molecular species against the continuum --usually associated with a
strong [C {\sc ii}] deficit
\citep[e.g.][and references therein]{gon15}. The OH\,65\,$\mu$m doublet
(with $E_{\mathrm{low}}\approx300$\,K), observed in a number of
(U)LIRGs with {\it Herschel}/PACS,
has been used as a far-IR signature of these regions;
however, not all sources
that show deep OH\,65\,$\mu$m absorption are similar as they
span an important range in column densities and molecular excitation, with a
threshold of $N_{\mathrm{H}}\sim10^{24}$\,cm$^{-2}$ above which the doublet is
saturated.

To overcome the curtain of obscuring dust, the observation of high-lying
molecular lines in emission at (sub)millimeter and centimeter wavelengths
also provide high angular resolution and are thus highly useful.
Of special interest are the vibrationally excited HCN $\nu_2$ and
HC$_3$N $\nu_7$ and $\nu_6$ lines.
Probably due to the combination of high columns, which protect
these species from photoionization, and high temperatures \citep{har10},
cyanopolynes attain high abundances and their vibrational lines are prominent
in these regions in spite of their involved high-energy levels ($>1000$\,K
for HCN). Following the first extragalactic detection of the
HCN\,$\nu_2=1^f\,J=3-2$ and $J=4-3$ lines in the prototypical BGN
NGC~4418 \citep{sak10}, the HCN vibrational lines have been detected and imaged
in a number of BGNs \citep{aal15a,aal15b,mar16,ima16a,ima16b,fal18},
indicating a very efficient pumping of the vibrational states
through absorption of $14$\,$\mu$m continuum photons -the wavelength
of the vibrational band detected in strong absorption towards the
same objects by \cite{lah07}.

The HCN vibrational lines are not exclusive of BGNs, but have been also
detected (including vibrational lines of the isotopologues H$^{13}$CN and
HC$^{15}$N) in galactic hot cores \citep{rol11a,rol11b,rol11c} although
with significantly lower $L_{\mathrm{HCN\,\nu2\,3-2}}/L_{\mathrm{IR}}$ ratio in the
case of Sgr\,B2(M) \citep{aal15b}. One interesting point that
\cite{rol11b,rol11c} emphasized is the role of the continuum optical depth
in explaining the HCN vibrational emission. Once the dust becomes optically
thick to its own radiation, the dust emission is trapped and its diffusion
enhances the inner dust temperature ($T_{\mathrm{dust}}$). Dust is heated through
absorption of infrared photons coming from the full $4\pi$\,sr even if the
heating source is located at the center, i.e. backwarming
\cite[first discussed by][]{row82} is key to increase $T_{\mathrm{dust}}$.
This ``greenhouse'' effect is very efficient in raising $T_{\mathrm{dust}}$ as it
works in the continuum, i.e. at all relevant wavelengths
--rather than through bands of molecules at specific wavelength
ranges, as in the atmosphere of the Earth\footnote{An additional obvious
    difference
    is that the heating source in the case of the Earth is external, with
    photons
    penetrating the atmosphere owing to its transparency at their wavelengths,
    while in the present case the heating source is internal; nevertheless,
    the ultimate reason for the extra heating is in both cases the trapping of
    radiation in the infrared and implied backwarming.}.
In addition, the greenhouse effect
is evidently taking place in BGNs, as the far-IR molecular line absorption
observed with {\it Herschel}/PACS in these galaxies demonstrates that the
continuum behind is optically thick in the far-IR \citep{gon15} and, in
some extreme cases, even at millimeter wavelengths \citep{sak13,sak17,sco17}.

In this paper, we develop on the greenhouse effect in BGNs to quantitatively
explore to which extent the observed fluxes of the HCN vibrational lines can
be understood upon values of physical parameters that are inferred from
independent methods; specifically, the column densities, luminosity surface
densities, absolute luminosities, and HCN abundances. An oversimplified 
spherical symmetry is used in the present study that, nevertheless, gives
a solid basis on the problem because of the reduced number of involved free
parameters. We model the $T_{\mathrm{dust}}$ profile in \S\ref{sec:cont} for
pure AGN and pure SB models; these $T_{\mathrm{dust}}$ profiles are used in
\S\ref{modelshcn} to model the HCN vibrational emission and to compare the
modeling results with observations. Our main results, including the use
of spherical symmetry, are discussed in
\S\ref{sec:discussion}, and the conclusions are summarized in 
\S\ref{sec:conclusions}. Predictions for HC$_3$N and HNC vibrational
emission will be studied in a future work.

\section{Models for the continuum} \label{modelscont}
\label{sec:cont}
  
\subsection{Description of the models}
\label{sec:descont}

The models for the continuum aim to compute the dust temperature
($T_{\mathrm{dust}}$) profile in the source and the emergent spectral enery
distribution (SED), assuming spherical symmetry. These
models were used but only briefly described in \cite{gon99}, and we
describe them in more detail in Appendix~\ref{app}.

We performed two types of models according to the (dominant) nature of the
heating source(s): ``AGN'' models assume a central source of heating, while
starburst (``SB'') models simulate a deposition of energy distributed
across the source.  
In both models, it is assumed that the radiation
from the heating source(s) is locally absorbed by dust and re-emitted
in the infrared. This approach, which is a good approximation because
of the high column densities that characterize these obscured
regions\footnote{X-rays will leak out in our models below with column
    densities $<10^{24}$\,cm$^{-2}$, but they represent $\lesssim10$\%
  of the bolometric luminosity \citep{ris04,lus12}.},
imply that the bulk of the dust is heated by mid- and far-infrared
radiation, and hence scattering of radiation can be neglected. In
AGN models, the central heating source is a blackbody with temperature
$T_{\mathrm{cen}}=1300$\,K and a radius $R_{\mathrm{cen}}$ that is
set to match the required luminosity
$L_{\mathrm{IR}}^{\mathrm{AGN}}$.\footnote{We use $L_{\mathrm{IR}}$
    as an equivalent of the bolometric luminosity because the bulk
    of the luminosity in our models is emitted in the infrared.}
In SB models, there is no central source and the heating of
shell $m$ ($\Gamma_m$) due to stars is assumed to be proportional
to both the density of dust and the total dust mass of the shell,
$\Gamma_m\propto \rho_m M_m$, normalized to give the required
luminosity ($L_{\mathrm{IR}}^{\mathrm{SB}}=\sum_m\Gamma_m$).
Evidently, the pure AGN models are highly idealized in sources with
  high column densities, where star formation is unavoidable, and represent
  an extreme limit still useful to potentially address, from comparison
  with SB models, the possible presence of an
  extremely buried and energetically dominant AGN.
  In addition, the SB models obviously smooth out the variation of dust
    temperatures within any shell as a result of star formation therein,
    so that the dust temperatures should be considered as mass-averaged 
    within the shell.
    On the other hand, while both types of models aim to simulate the
    BGN as a single source, they can also be applied to
    a collection of independent (not radiatively interacting) sources,
    provided that the
    parameters listed below are applicable to each source of the ensemble.

\begin{figure}
\begin{center}
\includegraphics[angle=0,scale=.42]{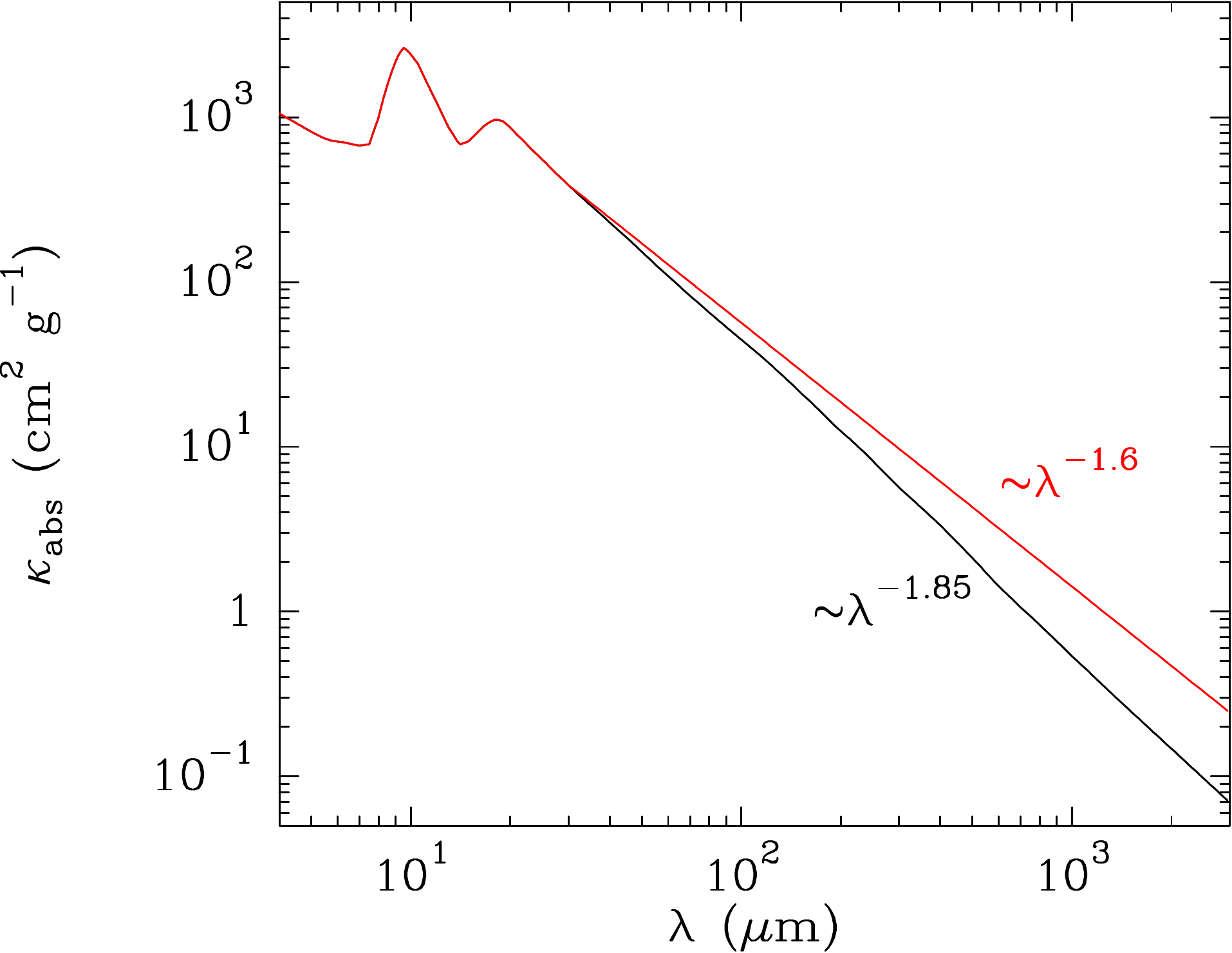}
\end{center}
\caption{The two curves of mass absorption coefficient of dust as a function of
  wavelength considered in this work. We use as fiducial the red curve,
  with an emissivity index of $\beta=1.6$ and
  $\kappa_{\mathrm{abs}}=1.2$\,cm$^2$\,g$^{-1}$ of dust at $\lambda=1.1$\,mm.
 }    
\label{kabs}
\end{figure}

The density profile across the source is described with a power-law
$\rho\propto r^{-q}$, with $q=1,1.5$. In all our models, a small cavity
with radius $R_{\mathrm{int}}=R_{\mathrm{out}}/17$ is adopted, where
$R_{\mathrm{out}}$ is the outer radius of the source
($R_{\mathrm{int}}>R_{\mathrm{cen}}$ in all AGN models).
Figure~\ref{kabs} shows the two curves for the mass absorption coefficient
of dust, $\kappa_{\mathrm{abs}}$, considered in this work. The black curve is the
same as used in \cite{gon14}, with an emissivity index from the far-IR to
millimeter wavelengths of $\beta=1.85$ and
$\kappa_{\mathrm{250\,\mu m}}=8.2$\,cm$^2$\,g$^{-1}$ of dust
at 250\,$\mu$m. According to \cite{pla11}, however, and for a gas-to-dust ratio
of 100, $\kappa_{\mathrm{250\,\mu m}}$ is significantly higher,
  $\approx14$\,cm$^2$\,g$^{-1}$; in addition, $\kappa_{\mathrm{abs}}$ 
  increases with increasing dust temperature \citep{men98}. Therefore,
we have also used the red curve with the same profile as the black curve
at $\lambda<30$\,$\mu$m but with $\beta=1.6$ at longer wavelengths. At
$1.1$mm, which is the wavelength of the HCN\,$\nu_2=1^f\,J=3-2$ transition,
the black and red curves give $0.43$ and $1.2$\,cm$^2$\,g$^{-1}$ of dust,
which we expect to bracket the actual value.
We have used as fiducial the red curve, which
gives significant optical depth at $1.1$mm for the maximum column density
considered in this work ($N_{\mathrm{H2}}=10^{25}$\,cm$^{-2}$), thus
allowing us to study the effect of absorption of dust emission by
the HCN vibrational line (\S\ref{sec:spat}). Nevertheless, the
$T_{\mathrm{dust}}$ profiles obtained with both $\kappa_{\nu}$-curves are
basically indistinguishable, as $T_{\mathrm{dust}}$ depends on the optical depths
at $\lambda<100$\,$\mu$m. Only results involving the continuum emission at
(sub)millimeter wavelengths show differences between the two
$\kappa_{\nu}$-curves. Our values of $\kappa_{\mathrm{abs}}$ between
  7.5 and 50\,$\mu$m agree within 30\% with the values tabulated by
  \cite{dra03}.

   \begin{table*}
      \caption{Model parameters}
         \label{tab:par}
\begin{center}
          \begin{tabular}{lccl}   
            \hline
            \noalign{\smallskip}
            Parameter & Fiducial & Explored & Meaning \\
                      & value    & range    &         \\
            \noalign{\smallskip}
            \hline
            \noalign{\smallskip}
AGN-SB  &   & Both & AGN or starburst (SB) generation of the luminosity \\
$\Sigma_{\mathrm{IR}}$ (\Lsun\,pc$^{-2}$) & $(5.5-11)\times10^{7}$ &
$(1.4-22) \times10^{7}$
& Surface brightness $=L_{\mathrm{IR}}/(\pi R_{\mathrm{out}}^2)$ \\
$N_{\mathrm{H2}}$ (cm$^{-2}$)   &   $10^{25}$ & $10^{23}-10^{25}$
& Column density of H$_2$ along a radial path (from $R_{\mathrm{int}}$ to $R_{\mathrm{out}}$)  \\
$q$      & 1.0 & $1.0-1.5$ & Gas and dust density profile ($n\propto r^{-q}$) \\
$R_{\mathrm{out}}/R_{\mathrm{int}}$  & $17$ & $-$ & Outer-to-inner radius of the source \\
$X_{\mathrm{HCN}}/\Delta V$ ((km\,s$^{-1}$)$^{-1}$)   &
$1.5\times10^{-8}$ & $-$ & HCN abundance (relative to H$_2$) per unit
velocity interval \\ 
$\Delta V$ (km s$^{-1}$) &    $67$  &  $-$ & Gas velocity dispersion
(one-dimensional FWHM) \\
$\Delta\Omega$ (arc\,sec$^2$)   &   $1.1\times10^{-2}$   &  $-$ & Solid angle
$=\pi R_{\mathrm{out}}^2/D^2$, 
relevant for absolute values\\
            \noalign{\smallskip}
            \hline
         \end{tabular} 
\end{center}
   \end{table*}

As shown by \cite{ive97},
the solution of radiative transfer (i.e. the $T_{\mathrm{dust}}$ profile and
normalized SED) basically depends on dimensionless parameters, but we opt
here to use astrophysical parameters applied to the sources of interest:
the $T_{\mathrm{dust}}$ profile as a function of the normalized radius
($r_n=r/R_{\mathrm{out}}$) depends on 
the spatial distribution of the heating source(s)
(AGN or SB), the luminosity surface density (characterized as
$\Sigma_{\mathrm{IR}}=L_{\mathrm{IR}}/(\pi R_{\mathrm{out}}^2)$),
the density profile ($q$), and the H$_2$ column density $N_{\mathrm{H2}}$.
The latter is determined by assuming a gas-to-dust ratio
of 100 by mass; for reference, a radial optical depth at 100\,$\mu$m of
$\tau_{100}=1$ corresponds to $N_{\mathrm{H2}}\approx0.6\times10^{24}$\,cm$^{-2}$.

\subsection{Fiducial values}

Our models are applied to obscured galaxy nuclei where vibrationally
excited HCN has been detected, and the fiducial values listed
in Table~\ref{tab:par} for
$N_{\mathrm{H2}}$, $\Sigma_{\mathrm{IR}}$ and $q$ can account for
most of these observations, as shown below.
We use as fiducial $N_{\mathrm{H2}}$ the maximum value considered in
this work, $10^{25}$\,cm$^{-2}$, characteristic of buried sources
with bright HCN vibrational emission
\citep[e.g. NGC~4418 and Zw~049;][]{sak10,sak13,cos13,gon12,fal15}.
More extreme values of $N_{\mathrm{H2}}\gtrsim10^{26}$\,cm$^{-2}$ have been
inferred toward the western nucleus of Arp~220 \citep{sco17,sak17} but,
due to the inclination of the disk, these may not be representative
of the column averaged over solid angles. We thus conservatively treat
the extreme values of $N_{\mathrm{H2}}>10^{25}$\,cm$^{-2}$ by assuming that
the $T_{\mathrm{dust}}$ profile for $N_{\mathrm{H2}}=10^{25}$\,cm$^{-2}$
is approximately valid (\S\ref{sec:spat}).
On the other side, our simulations also cover relatively low values of
  $N_{\mathrm{H2}}<10^{24}$\,cm$^{-2}$ (Table~\ref{tab:par}); hence
  non-BGN sources are also considered.
We also adopt as fiducial high values of
$\Sigma_{\mathrm{IR}}$, $(0.55-1.1)\times10^8$\,\Lsun\,pc$^{-2}$,
as well as $q=1$, yielding
$L_{\mathrm{IR}}/M_{\mathrm{gas}}=(480-960)$\,\Lsun/M$_{\odot}$
for fiducial values. Higher values of
$L_{\mathrm{IR}}/M_{\mathrm{gas}}$ may represent sources where negative feedback
has cleared the nuclear region from molecular gas.

For a given set of parameters that determine the $T_{\mathrm{dust}}$
  profile (AGN/SB, $\Sigma_{\mathrm{IR}}$, $q$, and $N_{\mathrm{H2}}$),
  the absolute flux densities shown below are proportional to the solid angle
$\Delta\Omega\equiv \pi R_{\mathrm{out}}^2/D^2$, where $D$ is
the distance to the source\footnote{In case of high redshift sources,
flux densities scale proportional to
$\pi R_{\mathrm{out}}^2 (1+z) /D_L^2$, where $D_L$ is the
luminosity distance.}.
As shown in \S\ref{datacomp}, $\Delta\Omega$ is 
in the range $(1-3)\times10^{-2}$\,arc\,sec$^2$ for nearby LIRGs
with bright HCN vibrational emission, and we adopt as fiducial
$\Delta\Omega=1.1\times10^{-2}$\,arc\,sec$^2$.

For a given set of values for the parameters in Table~\ref{tab:par},
  $D$ is the only free parameter
  required to obtain the values of $R_{\mathrm{out}}$, $L_{\mathrm{IR}}$ and
  $M_{\mathrm{dust}}$:
\begin{eqnarray}
  R_{\mathrm{out}} & = & 
 28.6 \, \frac{D}{100\,\mathrm{Mpc}} \,
 \left(\frac{\Delta\Omega}{1.1\times10^{-2}\, \mathrm{arc\,sec^2}}\right)^{1/2}
 \, \mathrm{pc} \\
 L_{\mathrm{IR}} & = & 1.4\times10^{11} \,
 \frac{\Sigma_{\mathrm{IR}}}{5.5\times10^7\,\mathrm{L_{\odot}\,pc^{-2}}}
 \, \left(\frac{D}{100\,\mathrm{Mpc}}\right)^2 \nonumber \\
 & \times &  \frac{\Delta\Omega}{1.1\times10^{-2}\, \mathrm{arc\,sec^2}}
 \, \mathrm{L_{\odot}} \\
 M_{\mathrm{dust}} & = & (1.7-2.9)\times10^6 \,
 \frac{N_{\mathrm{H2}}}{10^{25}\,\mathrm{cm^{-2}}} \,
 \left(\frac{D}{100\,\mathrm{Mpc}}\right)^2 \, \nonumber \\
 & \times & \frac{\Delta\Omega}{1.1\times10^{-2}\, \mathrm{arc\,sec^2}} \,
 \, \mathrm{M_{\odot}},
\end{eqnarray}
    where the two values of $M_{\mathrm{dust}}$ correspond to $q=1.5-1.0$,
      respectively. For fixed fiducial parameters
        (AGN/SB, $\Sigma_{\mathrm{IR}}$,
  $N_{\mathrm{H2}}$, $q$, and $R_{\mathrm{out}}/R_{\mathrm{int}}$),   
  absolute luminosities and masses are 
  $\propto R_{\mathrm{out}}^2$.



\subsection{Results}

\begin{figure*}
\begin{center}
\includegraphics[angle=0,scale=.60]{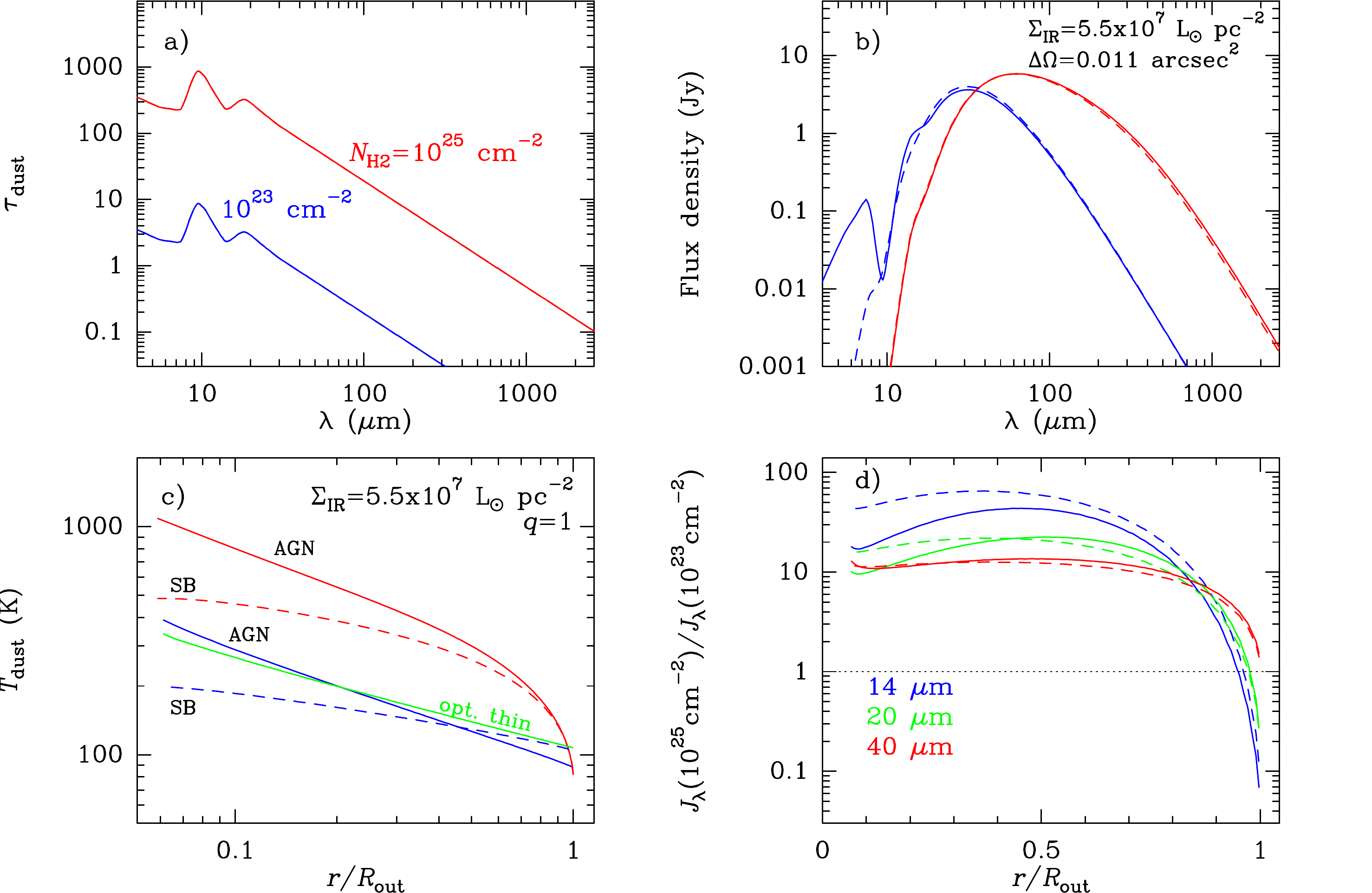}
\end{center}
\caption{Results of two continuum models for an AGN-dominated source (solid
  lines) and two models for a starburst-dominated source (dashed lines), 
  illustrating the greenhouse effect. The four models, shown with blue and
  red lines in panels a-c, have the same luminosity surface density
  ($\Sigma_{\mathrm{IR}}=5.5\times10^{7}$\,\Lsun\,pc$^{-2}$), solid angle
  ($\Delta\Omega=1.1\times10^{-2}$\,arc\,sec$^2$),
  and the density varies as $r^{-1}$ ($q=1$). The models differ only in the
  column density (panel a): $N_{\mathrm{H2}}=10^{25}$ (red) and
  $N_{\mathrm{H2}}=10^{23}$ (blue) \cmd. The predicted SED of the blue
  (thinner) model is warmer than the red (thicker) model (panel
  b), because in the former the observer penetrates much inside the cocoon of
  dust with warmer material. Nevertheless, \tdust\ is on average much higher 
  for the red (thicker) models (panel c), due to trapping of continuum
  photons. In panel c, the green curve indicates the optically thin
    limit for the AGN model, with $T_{\mathrm{dust}}\propto r^{-0.4}$ (see text).
  Within the cocoon of dust, the radiation field at 14, 20, and 40 $\mu$m,
  responsible for the excitation of HCN $\nu_2=1$, HC$_3$N $\nu_6=1$, and
  HC$_3$N $\nu_7=1$ states, is dramatically enhanced for the thick models.
  The calculated mean intensities
  $<J>$ at the wavelengths of the HCN and HC$_3$N ro-vibrational lines
  have large enhancement
  factors ($\gtrsim10$, panel d). Therefore, the 
  HCN and HC$_3$N ro-vibrational lines enter in  
  ``resonance'' with the greenhouse effect under conditions of high columns,
  generating strong cyanopolyne emission from the excited vibrational states.
 }    
\label{greenhouse}
\end{figure*}

Figure~\ref{greenhouse} compares results for two AGN models (solid lines)
and two SB models (dashed lines) that only differ in the column density,
$N_{\mathrm{H2}}=10^{23}$\,cm$^{-2}$ (blue curves) and
$10^{25}$\,cm$^{-2}$ (red curves, see panel a); all other parameters correspond
to the fiducial values. In both models (AGN and SB) with
$N_{\mathrm{H2}}=10^{25}$\,cm$^{-2}$, trapping of continuum photons is extremely
efficient, thereby strongly increasing $T_{\mathrm{dust}}$ in the innermost
regions relative to models with the same luminosity but lower $N_{\mathrm{H2}}$
(see Fig.~\ref{greenhouse}c). Nevertheless, the mid-IR continuum in these
models can only probe the externalmost shells, and the apparently paradoxical
effect of a colder SED but a warmer $T_{\mathrm{dust}}$ over the bulk of the
source (relative to models with the same luminosity but lower columns,
see panels b and c) is obtained for extreme $N_{\mathrm{H2}}$.
This greenhouse effect is also illustrated
in panel d, which shows an enhancement of the mean intensity at mid-IR
wavelengths within the cocoon of dust by a factor of $\gtrsim10$.
This is the radiation
field that pumps the vibrationally excited states of the cyanopolynes, which
will then enter in resonance with the greenhouse effect to produce strong
vibrational emission.

The radiative diffusion timescale,
  $t_{\mathrm{diff}}\sim \tau_{\mathrm{dust}}\,R_{\mathrm{out}}/c$ is evaluated for
  $25$\,$\mu$m photons and $N_{\mathrm{H2}}=10^{25}$\,cm$^{-2}$ as
  $t_{\mathrm{diff}}\sim 10^4\,(\tau_{\mathrm{dust}}/200)\,
  (R_{\mathrm{out}}/20\,\mathrm{pc})$\,yr. This is much shorter than the dynamical
time scale ($>10^7$\,yr) and BGNs will attain radiative equilibrium.

While the predicted SEDs for AGN and SB models are basically indistinguishable
for $N_{\mathrm{H2}}=10^{25}$\,cm$^{-2}$ (even at $\lambda<10$\,$\mu$m, 
  outside of Fig.~\ref{greenhouse}b),
the AGN models have significantly higher
$T_{\mathrm{dust}}$ in the innermost regions (Fig.~\ref{greenhouse}c).
In SB models, the infrared luminosity generated per unit radial interval is
$d\,L_{\mathrm{IR}}/dr\propto r^{2(1-q)}$, which is flat for $q=1$. An important
fraction of the luminosity is thus generated close to the surface with more
chance to escape, thus decreasing $T_{\mathrm{dust}}$ in the
innermost regions relative to AGN models.

The green curve in Fig.~\ref{greenhouse}c indicates the $T_{\mathrm{dust}}$
  profile in the optically thin limit for parameters other than the column
  density equal to fiducial values
  ($\Sigma_{\mathrm{IR}}=5.5\times10^{7}$\,\Lsun\,pc$^{-2}$, $q=1$) and the
  AGN approach.
  The slope of this curve is $s=-d\log T_{\mathrm{dust}}/d\log r\approx0.4$,
  lower than the value of $0.5$
  expected for {\it grey} dust grains (i.e. $\kappa_{\mathrm{abs}}$ independent
  of $\lambda$) due to the decreasing thermal cooling efficiency with
  decreasing $T_{\mathrm{dust}}$. The AGN model with $N_{\mathrm{H2}}=10^{23}$ \cmd\
  shows a similar $T_{\mathrm{dust}}$ profile, though already showing some
  effects of trapping, but the $N_{\mathrm{H2}}=10^{25}$ \cmd\ AGN model shows
  a higher slope of $s\approx0.6$ for $r/R_{\mathrm{out}}<0.5$ and even higher
  in the external regions. By contrast, the SB model with
  $N_{\mathrm{H2}}=10^{25}$ \cmd\ shows
  a slope of only $s\lesssim0.2$ in the inner $r/R_{\mathrm{out}}<0.3$ region.
  
  The steep slope $s$ of $T_{\mathrm{dust}}(r)$ in AGN models with high
    $N_{\mathrm{H2}}$ implies that,
    for thermal equilibrium between dust and gas, the gas in the inner regions
    will be unstable under
    adiabatic radial displacements. The criterion for convective
    instability translates into the condition $s>q(\gamma-1)$, where
    $\gamma$ is the adiabatic index of the gas.
    For $T_{\mathrm{dust}}\gtrsim200$\,K, the excited rotational levels of H$_2$
    are populated and $\gamma\sim1.4$, so that the instability criterion
    $s>0.4q$ is met in AGN models. For SB models, however, the innermost
    $r/R_{\mathrm{out}}<0.3$ regions are stable. In the outermost layers of
    both AGN and SB models $s>1$ and these regions, where the far-IR molecular
    absorption is produced, will be convective. Convection
    in BGNs is expected to generate turbulence, and would also modify the
    temperature (and density) structures of our models,
    but its quantitative assessment is beyond the scope of this paper.
  

\begin{figure}
\begin{center}
\includegraphics[angle=0,scale=.33]{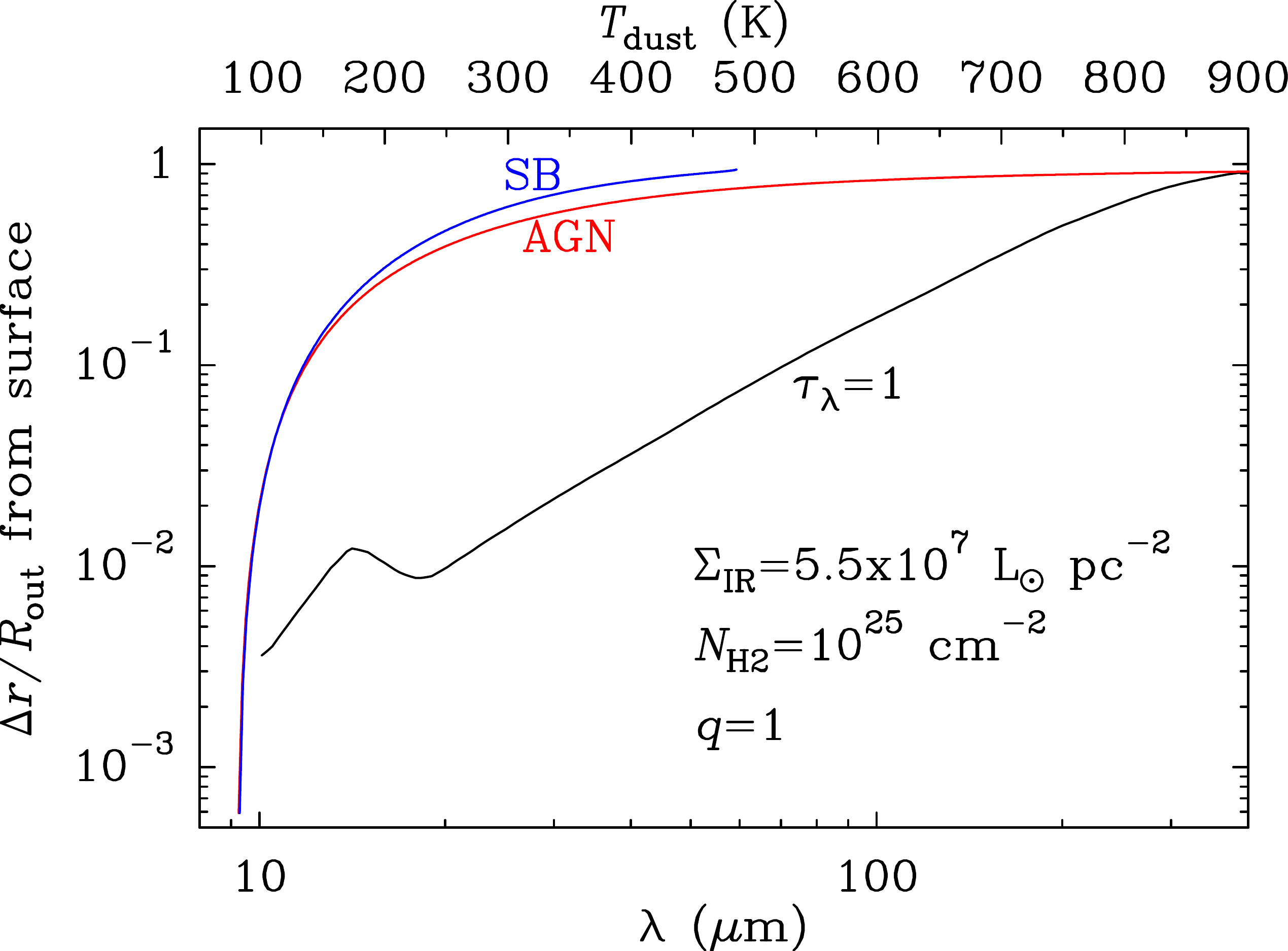}
\end{center}
\caption{The black line indicates the thickness of the photosphere
  ($\Delta r/R_{\mathrm{out}}$ from
  the surface for $\tau_{\lambda}=1$, along the sightline that passes
    through the center of the source)
  as a function of wavelength (lower horizontal axis)
  for $N_{\mathrm{H2}}=10^{25}$\,cm$^{-2}$ and $q=1$.
  The colored lines show the calculated \tdust\ profile 
  (upper horizontal axis)  as a function of $\Delta r/R_{\mathrm{out}}$
  for the two models with
  $N_{\mathrm{H2}}=10^{25}$\,cm$^{-2}$ of Fig.~\ref{greenhouse} (and
  fiducial values in Table~\ref{tab:par}).
} 
\label{photos}
\end{figure}

\begin{figure}
\begin{center}
\includegraphics[angle=0,scale=.70]{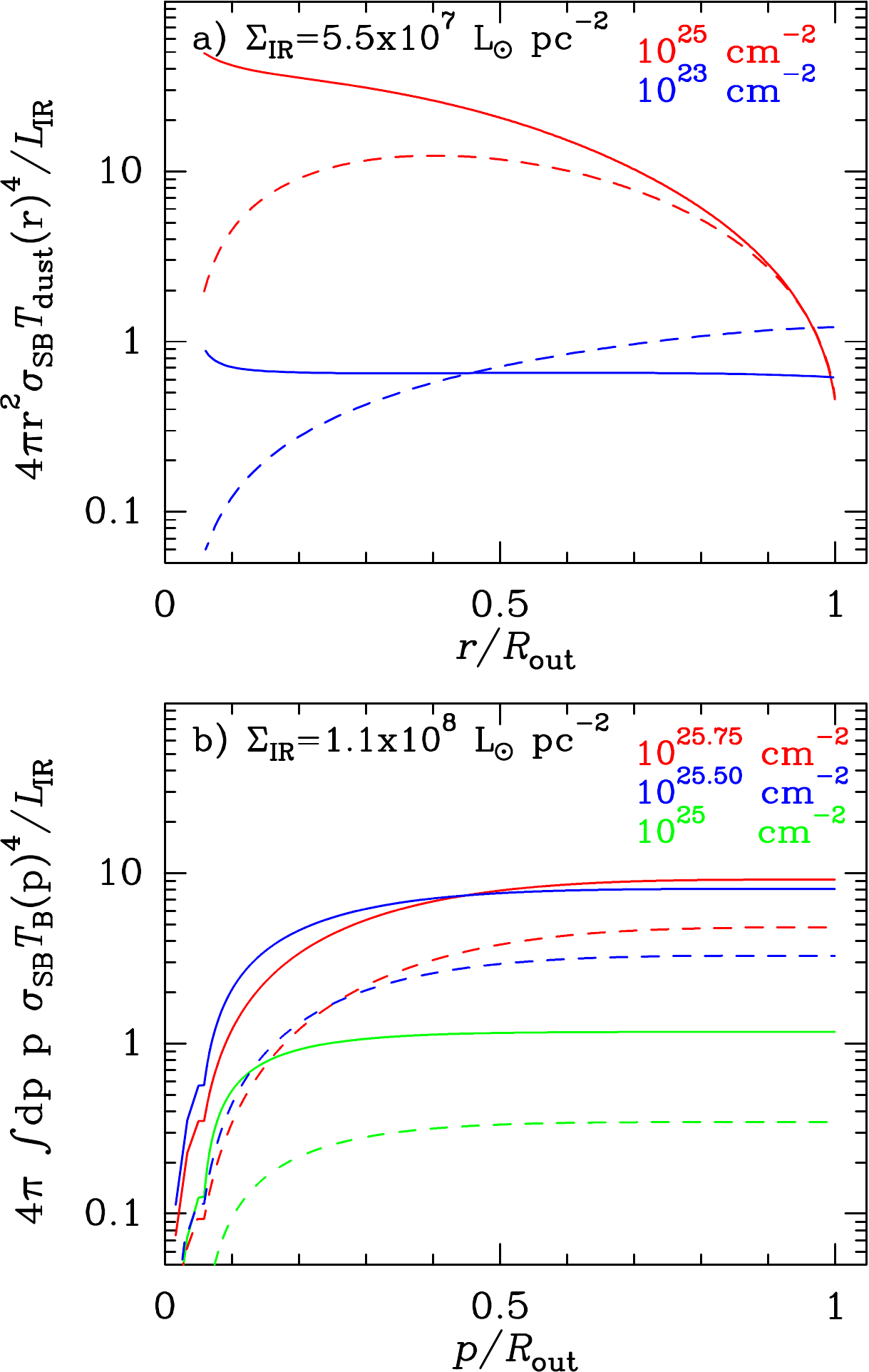}
\end{center}
\caption{a) The ratio of
  $\Upsilon_{\mathrm{IR}}\equiv4\pi r^2\sigma_{\mathrm{SB}}T_{\mathrm{dust}}(r)^4$ to
  the luminosity of the source for the same models as in Fig.~\ref{greenhouse}.
  b) The ratio of
$\Psi_{\mathrm{IR}}\equiv2\int_0^p dp'\,2\pi\,p'\,\sigma_{\mathrm{SB}}T_{\mathrm{B}}(p')^4$
    to the luminosity of the source, where $T_{\mathrm{B}}(p')$ is the brightness
    temperature at $1.1$\,mm for impact parameter $p'$. Results are shown for
    $\Sigma_{\mathrm{IR}}=1.1\times10^8$\,L$_{\odot}$\,pc$^{-2}$ and three values
    of $N_{\mathrm{H2}}$; solid and dashed lines correspond to AGN and SB models,
  respectively.
 }    
\label{overlir}
\end{figure}

We show in Fig.~\ref{photos} the ``photosphere'' effect that we may expect in
nuclei with high $N_{\mathrm{H2}}=10^{25}$\,cm$^{-2}$ \citep{gon12}. With these
extreme columns and $q=1$, the observer can only penetrate $\lesssim20$\% of
the source radius for $\lambda\lesssim100$\,$\mu$m. In this external region,
both the AGN and SB models of Fig.~\ref{greenhouse}
($\Sigma_{\mathrm{IR}}=5.5\times10^7$\,L$_{\odot}$\,pc$^{-2}$) yield
$T_{\mathrm{dust}}$ between 85 and 160 K, in general agreement with the
values inferred from far-IR molecular absorption lines in the most
buried sources (NGC~4418, Arp~220, Zw~049.057).
At (sub)millimeter wavelengths, the penetration
is constrained by the optical depth of the observed line.

One important implication of the greenhouse effect is that
$\Upsilon_{\mathrm{IR}}\equiv4\pi r^2\sigma_{\mathrm{SB}}T_{\mathrm{dust}}(r)^4$
is not conserved through
the source, but is much higher than $L_{\mathrm{IR}}$ over most volume
for high $N_{\mathrm{H2}}$. This is illustrated in Fig.~\ref{overlir}a for
the same models as in Fig.~\ref{greenhouse}.
The infrared luminosity inferred from $T_{\mathrm{dust}}$ and the
  apparent size may be overestimated by a large factor when selectively
  probing the innermost regions of the BGN.
  An alternative way to estimate the
  source luminosity is integrating the inferred
  $\sigma_{\mathrm{SB}}\,T_{\mathrm{dust}}^4$ over the sky plane, using the
  observed $T_{\mathrm{dust}}$ distribution, and multiplying
  by 2 to account for the two faces of the disk \citep{wil14,sak17}.
  We have performed a similar
  calculation in Fig.~\ref{overlir}b ($\Psi_{\mathrm{IR}}$)
  by using the brightness temperature
  $T_{\mathrm{B}}(1.1\,\mathrm{mm})$, rather than $T_{\mathrm{dust}}$, as a
  function of the impact parameter $p$. Results strongly depend on
  the optical depth at the observed wavelength ($1.1$\,mm in our case).
  For $N_{\mathrm{H2}}=10^{25}$\,cm$^{-2}$,
  $\tau_{\mathrm{dust}}(1.1\,\mathrm{mm})\approx0.4$ (Fig.~\ref{greenhouse}a) and 
  $\Psi_{\mathrm{IR}}$(AGN) gives a good estimate to
  $L_{\mathrm{IR}}$ while $\Psi_{\mathrm{IR}}$(SB) underestimates it
  by a factor of $\approx3$.
  However, once the continuum at $1.1$\,mm becomes optically thick,
  $\Psi_{\mathrm{IR}}$ may overestimate $L_{\mathrm{IR}}$ by an important factor.

\begin{figure*}
\begin{center}
\includegraphics[angle=0,scale=.6]{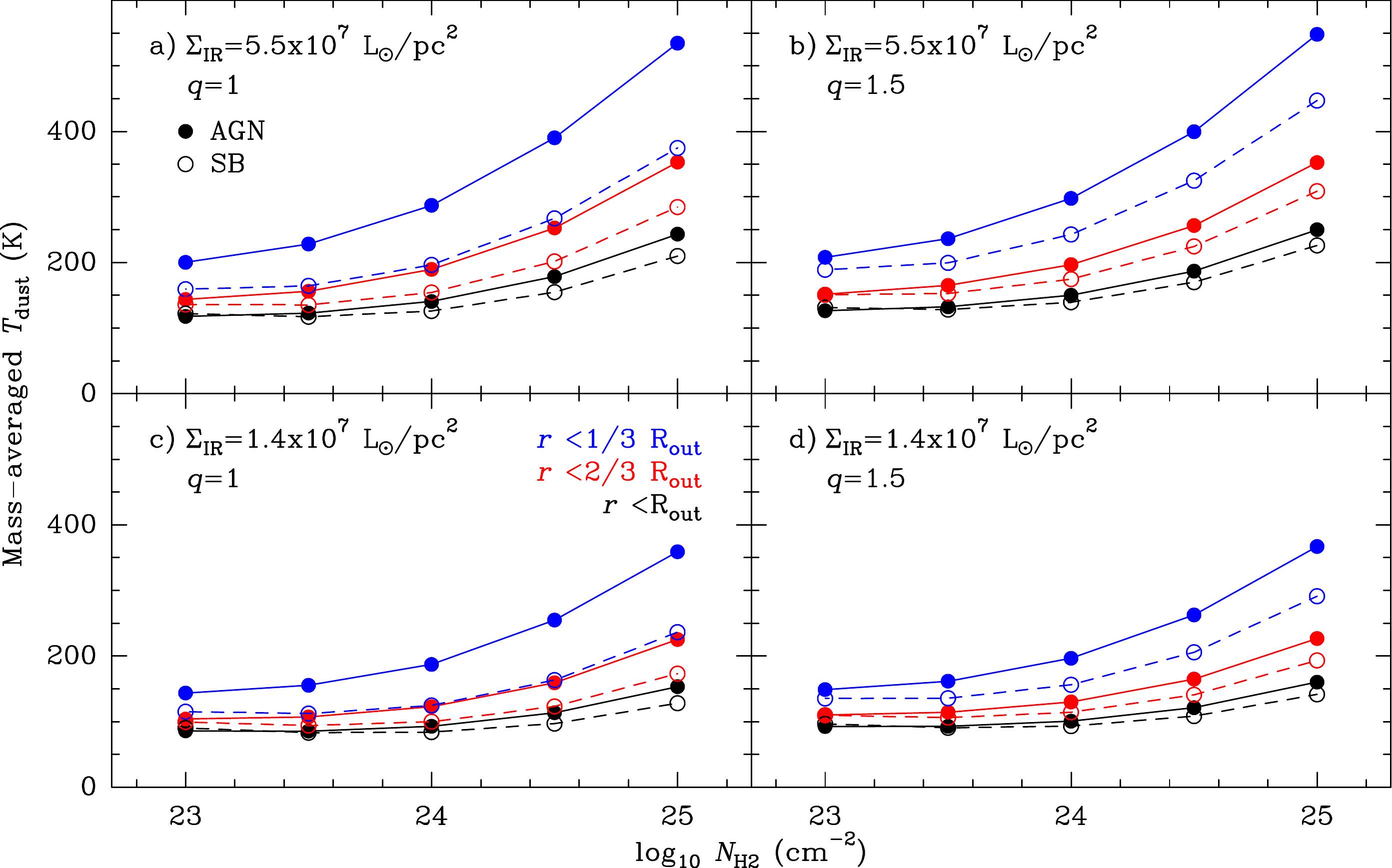}
\end{center}
\caption{Dependence of the mass-averaged $T_{\mathrm{dust}}$ on the surface
  brightness ($\Sigma_{\mathrm{IR}}=L_{\mathrm{IR}}/\pi R_{\mathrm{out}}^2$),
  density profile ($q$), and fractional volume over which the average is
  performed. Full circles (solid lines)
  and open circles (dashed lines) indicate AGN and SB models, respectively.
  As labeled in panel c, blue symbols show $<T_{\mathrm{dust}}>$ for the
  innermost $r<R_{\mathrm{out}}/3$ region, red symbols for the
  $r<2R_{\mathrm{out}}/3$ region, and black symbols show $<T_{\mathrm{dust}}>$ for
  the entire source. Upper panels show results for
  $\Sigma_{\mathrm{IR}}=5.5\times10^7$\,L$_{\odot}$\,pc$^{-2}$, with both
  (a) $q=1$ and (b) $q=1.5$, and lower panels show results for
  $\Sigma_{\mathrm{IR}}=1.4\times10^7$\,L$_{\odot}$\,pc$^{-2}$.
} 
\label{tdaver}
\end{figure*}

Figure~\ref{tdaver} shows the dependence of $<T_{\mathrm{dust}}>$, the
mass-averaged $T_{\mathrm{dust}}$
($\int \rho(r)\,T_{\mathrm{dust}}(r)\, r^2 dr/\int \rho(r)\,r^2 dr$),
on model parameters. In order to describe spatial variation of
$T_{\mathrm{dust}}$, the average is performed over 3 radial intervals:
the innermost $r<R_{\mathrm{out}}/3$ region (blue symbols),
the $r<2R_{\mathrm{out}}/3$ region (red symbols), and the entire
source (black symbols), and is shown as a function of $N_{\mathrm{H2}}$
and for both
AGN (filled symbols) and SB (open symbols) models. The different panels
show results for $\Sigma_{\mathrm{IR}}=5.5\times10^7$ (upper panels)
and $\Sigma_{\mathrm{IR}}=1.4\times10^7$\,L$_{\odot}$\,pc$^{-2}$ (lower panels),
and for $q=1.0$ and $1.5$ (left-hand and right-hand panels, respectively).
While there is in all models a contrast between the temperature in the
innermost regions and the value averaged over the full source, the strongest
contrast corresponds to models with $N_{\mathrm{H2}}\gtrsim10^{24}$\,cm$^{-2}$,
giving in AGN models $<T_{\mathrm{dust}}>\gtrsim200-300$\,K
for $r<R_{\mathrm{out}}/3$ and
$\Sigma_{\mathrm{IR}}=(1.4-5.5)\times10^{7}$\,\Lsun\,pc$^{-2}$, respectively.
While the AGN models yield similar $<T_{\mathrm{dust}}>$ for $q=1$ and
$q=1.5$, the SB models give significatively higher temperatures for $q=1.5$,
because the radiation is in the latter case more centrally generated.

Appendix~\ref{app} shows the $T_{\mathrm{dust}}$ profiles calculated
for most performed
models, and the results of fitting these profiles to analytic curves that give
accurate results for $T_{\mathrm{dust}}$ within 10\% at all radii.



\subsection{Radiation pressure}

\begin{figure*}
\begin{center}
\includegraphics[angle=0,scale=.58]{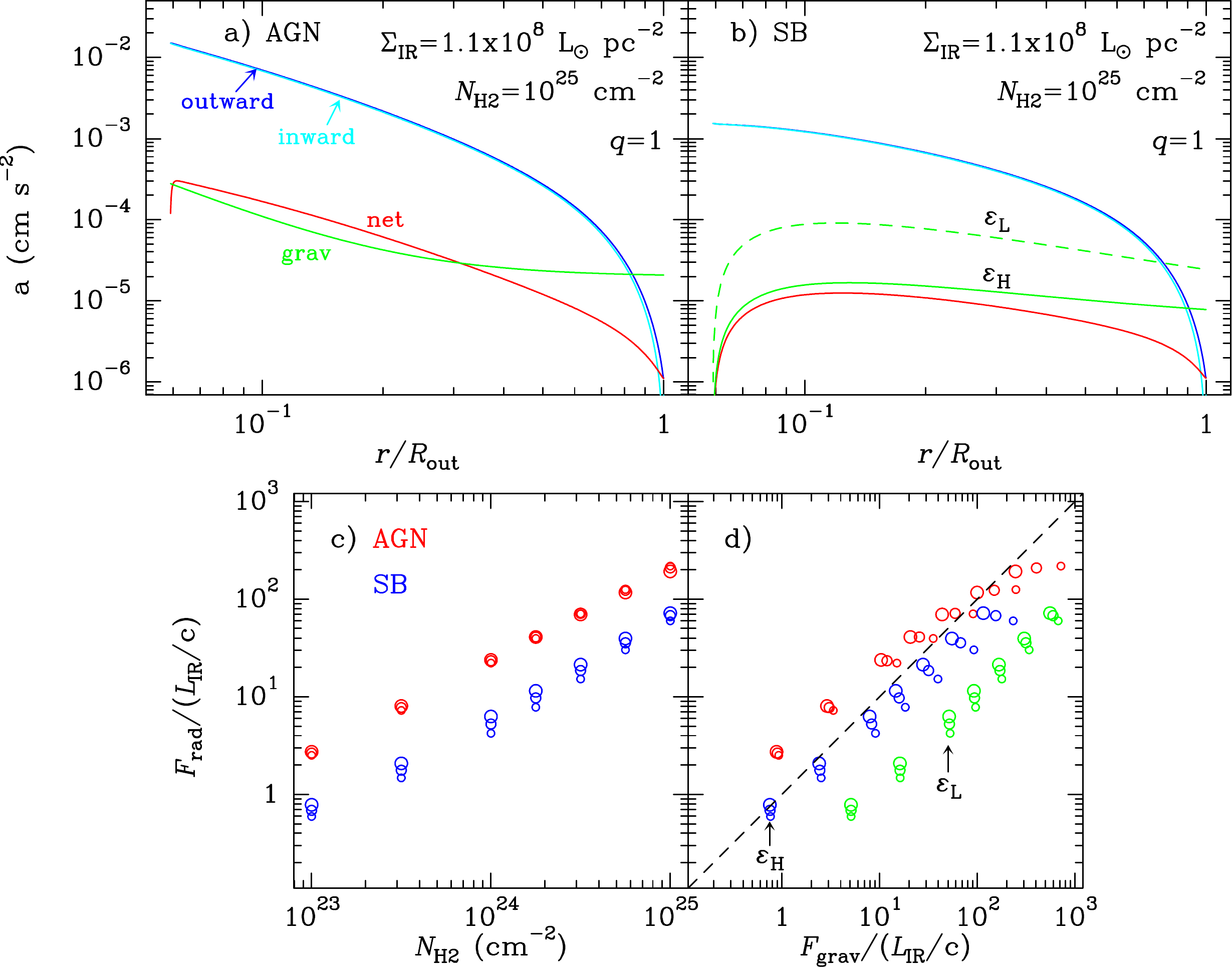}
\end{center}
\caption{{\it Upper}: Radial profiles of the acceleration (force per unit
  gas mass) due to radiation pressure on dust, for (a) AGN  and (b) SB
  models, with parameters specified. 
  The dark and light blue curves show
  the outward and inward accelerations, respectively, and the net (outward)
  acceleration is shown in red. The green curves show the inward acceleration
  due to gravity that, for the SB model,
    has been calculated for two values of $\epsilon$, the light-to-mass ratio
    of the stellar population (denoted as $\epsilon_{\mathrm{H}}$ and
    $\epsilon_{\mathrm{L}}$, see text).
  {\it Lower}: The net force due to radiation pressure is plotted as a function
  of (c) $N_{\mathrm{H2}}$ and (d) the net force due to gravity,
  for both AGN (red) and SB (blue and green) models.
  Small, medium, and large
  circles indicate $\Sigma_{\mathrm{IR}}=5.5\times10^{7}$, $1.1\times10^{8}$, and
  $2.2\times10^{8}$\,L$_{\odot}$\,pc$^{-2}$. All values are scale invariant.
 }    
\label{hcngreenhouseradpress}
\end{figure*}

We evaluate in Fig.~\ref{hcngreenhouseradpress}
the interplay between radiation pressure and gravity
  in our modeled sources.
As backwarming is key to evaluate $T_{\mathrm{dust}}$ in models with high column
densities, {\it backpressure} is equally important to compute the net outward
force due to radiation pressure (see Appendix~\ref{app}).
In Fig.~\ref{hcngreenhouseradpress}a-b, the radial profiles of the outward,
inward, and net (outward$-$inward) forces per unit gas mass (i.e. the
acceleration) due to radiation are shown for the AGN and SB models with
$\Sigma_{\mathrm{IR}}=1.1\times10^{8}$\,L$_{\odot}$\,pc$^{-2}$,
$N_{\mathrm{H2}}=10^{25}$\,cm$^{-2}$, and $q=1$.

We have estimated the inward force due to gravity
as follows: for SB models,
we compute the stellar mass
in every shell $m$ as $M_*(m)=L_{\mathrm{IR}}(m)/\epsilon$,
where $L_{\mathrm{IR}}(m)$ is the luminosity generated in shell $m$
and $\epsilon$ is the light-to-mass ratio of the current stellar
population. We first note that there is a threshold value of
$\epsilon$, $\epsilon_{\mathrm{th}}$, below which the radiation pressure
  support against gravity is not possible. Using
$F_{\mathrm{edd}}=4\pi G\,c\,\Sigma_g\,\kappa_F\,f_g^{-1}$
\citep[e.g.][]{and11}, where $F_{\mathrm{edd}}$ is the Eddington flux,
$\Sigma_g$ is the gas surface density, $\kappa_F$ is the Rosseland mean
opacity (assumed independent of $T_{\mathrm{dust}}$), and $f_g$ is the gas
fraction, combined with
$f_g^{-1}=1+\epsilon^{-1}\Sigma_g^{-1}F_{\mathrm{edd}}$, we get
$F_{\mathrm{edd}}=\epsilon_{\mathrm{th}}\Sigma_g/(1-\epsilon_{\mathrm{th}}/\epsilon)$
where
\begin{equation}
  \epsilon_{\mathrm{th}}=\frac{4\pi G\,c}{\kappa_F} =
  1.3\times10^3 \left(\frac{\kappa_F}{10\,\mathrm{cm^2\,g^{-1}}}\right)^{-1}
    \, \,\mathrm{L_{\odot}/M_{\odot}}.
\end{equation}  
If $\epsilon<\epsilon_{\mathrm{th}}$, radiation pressure support is unattainable
regardless of the gas column density and gas fraction, and the source is
intrinsically sub-Eddington. A top-heavy stellar intial mass function,
combined with a young age, appear to be strong constraints for global
radiation pressure support in starburst galaxies.
Following \cite{for03} we choose two values for $\epsilon$ that lie
above and below $\epsilon_{\mathrm{th}}$: a high value of
  $\epsilon_{\mathrm{H}}=1700$\,L$_{\odot}$/M$_{\odot}$,
  which corresponds to the modeled value for
a young starburst with a \cite{sal55} initial mass function and a lower
cutoff mass of 1\,\Msun, and a lower value of
$\epsilon_{\mathrm{L}}=250$\,L$_{\odot}$/M$_{\odot}$, representing a more aged
burst. 
For $N_{\mathrm{H2}}=10^{25}$\,cm$^{-2}$ and
$\Sigma_{\mathrm{IR}}=1.1\times10^{8}$\,L$_{\odot}$\,pc$^{-2}$, 
the gas fractions are $f_{\mathrm{gas}}=0.63$ and $0.33$ for $\epsilon_{\mathrm{H}}$
and $\epsilon_{\mathrm{L}}$, respectively, decreasing
for lower values of $N_{\mathrm{H2}}$.


For AGN models, we simply assume that $M_*(m)=3\times M_{\mathrm{gas}}(m)$ in
every shell, which can be attributed to an old stellar population.
The central mass is assumed to be
$M_{\mathrm{central}}=2\times10^{-4}\,L_{\mathrm{IR}}$ (in solar units).
This is a factor 6 above
the Eddington limit ($M_{\mathrm{BH}}=3\times10^{-5}\,L_{\mathrm{Edd}}$),
but $M_{\mathrm{central}}$ should include both the black hole mass and the
surrounding gas feeding it. The values of the acceleration in
Fig.~\ref{hcngreenhouseradpress}a-b are scale invariant.

Figure~\ref{hcngreenhouseradpress}a-b shows that, in the AGN and
  SB-$\epsilon_{\mathrm{H}}$ models, the force due to
  radiation pressure is close to the gravity force 
  in the innermost regions, but gravity overcomes radiation pressure in the
  external regions where $T_{\mathrm{dust}}$ drops.
  As expected, radiation pressure cannot support
  the structure in the SB model with $\epsilon_{\mathrm{L}}$.

Considering each modeled source as a whole,
Fig.~\ref{hcngreenhouseradpress}c-d plots the net force on the whole gas
due to radiation pressure, normalized to $L_{\mathrm{IR}}/c$, as a function of
$N_{\mathrm{H2}}$ and of the corresponding net force due to gravity.
Both AGN and SB models with
$\Sigma_{\mathrm{IR}}=(5.5-22)\times10^{7}$\,L$_{\odot}$\,pc$^{-2}$ are plotted.
For $N_{\mathrm{H2}}=10^{25}$\,cm$^{-2}$, the AGN models yield
$F_{\mathrm{rad}}/(L_{\mathrm{IR}}/c)\approx200$, the continuum optical depth at
$\approx25$\,$\mu$m (Fig.~\ref{greenhouse}a). Again, all plotted values are
scale invariant. In SB models with $\epsilon_{\mathrm{L}}$,
$F_{\mathrm{grav}}$ is much higher than $F_{\mathrm{rad}}$ for all columns
  and $\Sigma_{\mathrm{IR}}$ values. The AGN and SB ($\epsilon_{\mathrm{H}}$)
  models are closer to the Eddington limit. In addition, we find that
$F_{\mathrm{rad}}$ can overcome gravity for
  AGN models with moderate $N_{\mathrm{H2}}<10^{24}$\,cm$^{-2}$, though this becomes
  hard in sources with very high column densities. 
  Since real systems are expected to be a combination of our pure
  AGN and SB models with several ages,
  this result could shed some light on the lack of
wide angle outflows in OH, in BGNs with extreme column densities
\citep{fal19}, provided that the AGN is not luminous enough to
  generate a hot bubble that would drive an
  energy-conserving outflow  \citep[e.g.][]{fau12,ric17}.
We conclude that our models may represent BGNs close to
radiation pressure support \citep{sco03}, though feedback through
a hot bubble or winds is probably required
to launch a wide-angle outflow in real systems
with typical column densities of
$\mathrm{a\,\,few}\,\times10^{23}$\,cm$^{-2}$ \citep{gon17}.

\section{Models for HCN} \label{modelshcn}

\subsection{Description of the models}

The models for HCN include 25 rotational levels -ignoring hyperfine
structure- in the ground vibrational
state ($v=0$) and, because of the $l-$doubling in the $\nu_2=1$ bending state,
up to 48 levels in $\nu_2=1$, with a maximum energy above the ground level
of 2300\,K ($\nu_2=1,\,J=24$) and giving a total of 165 transitions
(including the direct $l$-type transitions within $\nu_2=1$ at centimeter
  wavelengths).
The models use the $T_{\mathrm{dust}}$ profiles obtained in the
previous section, and assume thermal equilibrium between dust and gas
($T_{\mathrm{gas}}=T_{\mathrm{dust}}$). For HCN, however, there is no need of
the large number of shells required for $T_{\mathrm{dust}}$ calculations
(Appendix~\ref{app}), so that the $T_{\mathrm{dust}}$ profiles were smoothed
and 30 shells were used for molecular calculations. The approach described in
\cite{gon97,gon99} was used to calculate the equilibrium populations and
emergent spectra, and the molecular excitation by dust-emitted
photons was treated assuming that gas and dust are mixed.
Absorption of line emitted photons by dust (extinction) is 
  taken into account for all transitions.
Calculations for HCN include overlaps between the Q-branch ro-vibrational
lines, as well as between the blended $\nu=0$ and $\nu_2=1^e$
  rotational lines. For simplicity, no velocity gradients are
included.

\begin{figure}
\begin{center}
\includegraphics[angle=0,scale=.48]{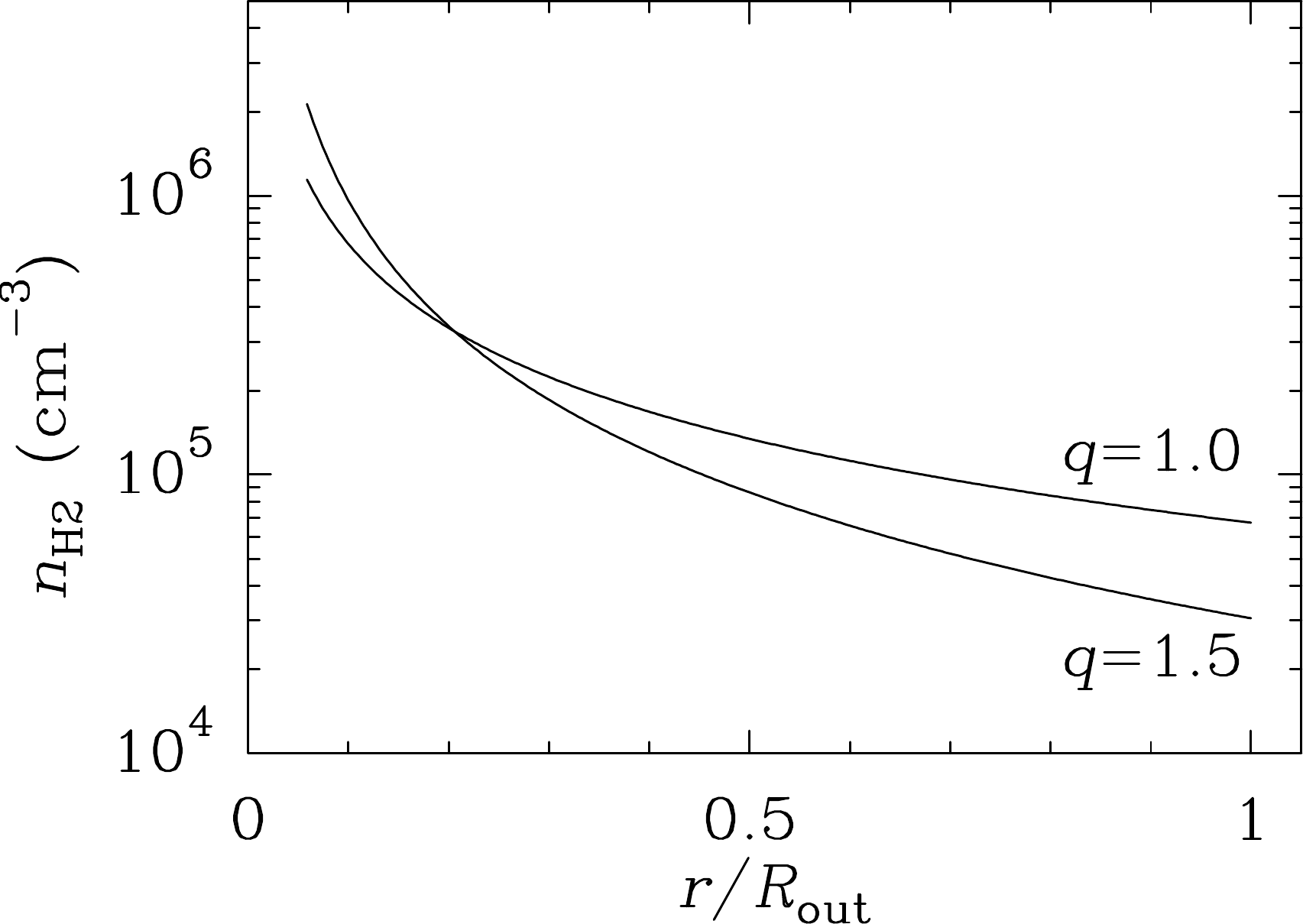}
\end{center}
\caption{The density profile for $N_{\mathrm{H2}}=10^{25}$ \cmd,
  $R_{\mathrm{out}}=17$\,pc, and $q=1.0-1.5$. In spherical symmetry,
  densities scale as
  $\propto N_{\mathrm{H2}}\,R_{\mathrm{out}}^{-1}$, but 
  we neglect the $R_{\mathrm{out}}^{-1}$ dependence to account for
    more general geometries (see text).
} 
\label{dens}
\end{figure}

Collisional excitation from the ground to the $\nu_2=1$
vibrational state is ignored, as well as among levels within
  the $\nu_2=1$ state. Collisional rates among rotational levels of
the ground $\nu=0$ state are taken from \cite{dum10}.
Unlike the continuum models developed in \S\ref{sec:cont},
  line excitation models
  have the source size ($R_{\mathrm{out}}$) as an independent parameter in
  addition to $N_{\mathrm{H2}}$ and others in Table~\ref{tab:par}.
This is because collisional excitation depends on the local
gas density $n_{\mathrm{H2}}$, which in spherical symmetry is proportional to
$N_{\mathrm{H2}}\,R_{\mathrm{out}}^{-1}$. Despite this dependence on $R_{\mathrm{out}}$,
we use a simplified approach in our modeling to adopt the $n_{\mathrm{H2}}$
profiles (as a function of $r/R_{\mathrm{out}}$) in Fig.~\ref{dens} for any
$R_{\mathrm{out}}$, and scale $n_{\mathrm{H2}}$ with $N_{\mathrm{H2}}$.
The density profiles in Fig.~\ref{dens} are exact for $R_{\mathrm{out}}=17$\,pc.
This simplification shrinks the
model-parameter space and is partly justified by noting that $N_{\mathrm{H2}}$,
$n_{\mathrm{H2}}$, and $R_{\mathrm{out}}$ do not necessarily have a direct link in
realistic situations where the gas distribution is neither smooth nor
spherically symmetric:
in a flat structure like an inclined disk, the solid angle subtended by the
source yields a scale length in the plane of sky
$R_{\mathrm{out}}=D\sqrt{\Delta\Omega/\pi}$, but
the scale length along the line of sight $N_{\mathrm{H2}}/n_{\mathrm{H2}}$ is an
independent parameter. We thus unlink the values of $n_{\mathrm{H2}}$ from
$R_{\mathrm{out}}$ to approximately account for more
general geometries, and discuss below
the impact of the adopted density profiles on results.



Assuming that the HCN abundance relative to H$_2$, $X_{\mathrm{HCN}}$, is
uniform across the source, the excitation of HCN for a given continuum model
and density profile depends on $N_{\mathrm{HCN}}/\Delta V$, the HCN column
density (along a radial path) per unit of velocity interval. Line broadening
is simulated
with a microturbulent approach. For given $N_{\mathrm{HCN}}/\Delta V$,
the emergent line fluxes are proportional to the velocity dispersion $\Delta V$.
In NGC~4418, pure rotational HCN lines have been detected in absorption
with {\it Herschel}/PACS at far-IR wavelengths
($135-190$\,$\mu$m, up to at least $J=25-24$), suggesting 
high column densities of HCN \citep{gon12}.
For $\mathrm{HCN/H_2O=0.1-0.3}$ and $\mathrm{H_2O/H\sim10^{-5}}$, we
expect $X_{\mathrm{HCN}}\sim10^{-6}$ in these nuclear regions, which we
adopt as fiducial value. A similar HCN abundance was derived by
  \cite{lah07} in BGNs from the analysis of the 14\,$\mu$m band, and
chemical calculations also favor high $X_{\mathrm{HCN}}$ in very warm regions
\citep{har10}.
On the other hand, $\Delta V$ should approximately characterize the
velocity dispersion measured in the nuclear regions
of (U)LIRGs after correcting for the rotation velocity; i.e. $\Delta V$
is the velocity dispersion along a typical line of sight through the
nucleus. In the nuclear regions of ULIRGs, $\Delta V$ is high
($\gtrsim100$\,km\,s$^{-1}$) as measured from the
CO lines \citep[e.g.][]{dow98}, but CO probably probes more extended gas than
that associated with the HCN vibrational emission. In
the LIRG IC~860, where the HCN\,$\nu_2=1^f\,J=3-2$ line is unblended from
the neighboring HCO$^+$\,$3-2$ line, the HCN vibrational line
has $\mathrm{FWHM}=130$\,\kms\ \citep{aal15b},
with some broadening attributed to the rotation.
We adopt a fiducial $\Delta V=67$\,\kms\ with the simplified assumption
  that $\Delta V$ is uniform and hence independent of the impact parameter
  $p$, and then
$N_{\mathrm{HCN}}/\Delta V=1.5\times10^{17}$\,cm$^{-2}$/(km s$^{-1}$) for
$N_{\mathrm{H2}}=10^{25}$\,cm$^{-2}$.
Since $N_{\mathrm{HCN}}$ is uniquely determined by $N_{\mathrm{H2}}$ and
$X_{\mathrm{HCN}}$, we list in Table~\ref{tab:par} the fiducial values for
$X_{\mathrm{HCN}}/\Delta V$ and $\Delta V$.

As it is the case for the continuum, line fluxes are proportional to
  $\Delta\Omega=\pi R_{\mathrm{out}}^2/D^2$ for fixed values of the fiducial
  parameters in Table~\ref{tab:par}.


\subsection{Results}

\subsubsection{HCN excitation and line optical depths}
\label{sec:excit}

\begin{figure*}
\begin{center}
\includegraphics[angle=0,scale=.58]{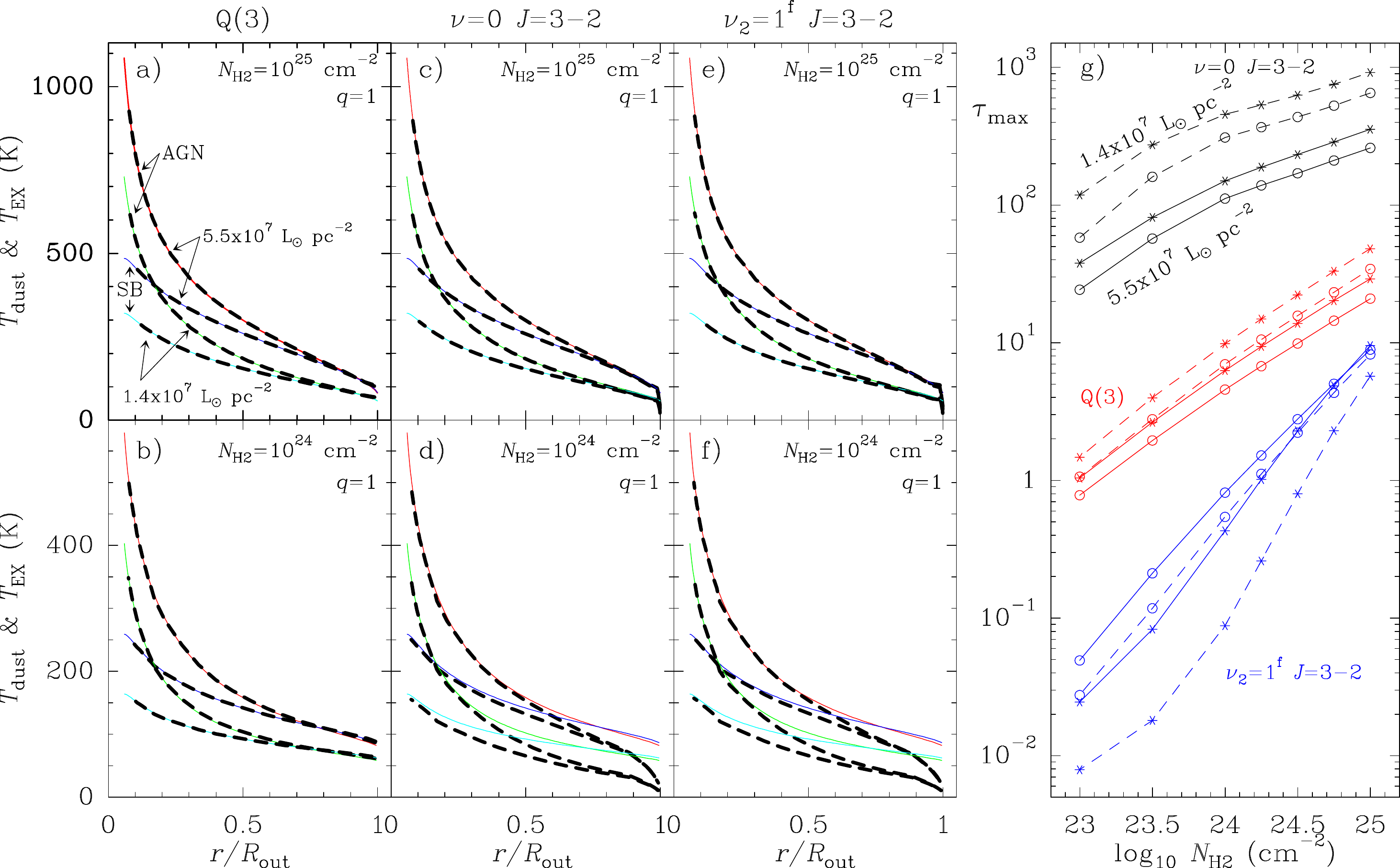}
\end{center}
\caption{
    a-f) Comparison between the $T_{\mathrm{dust}}$ profiles (colored curves)
  and the excitation temperature of the HCN Q(3) (a-b), $\nu=0\,J=3-2$ (c-d),
  and $\nu_2=1^f\,J=3-2$ (e-f) lines (overplotted dashed black curves).
  In our models we assume $T_{\mathrm{gas}}=T_{\mathrm{dust}}$. As
  indicated in panel a, the red-green curves indicate AGN models with
  $\Sigma_{\mathrm{IR}}=(5.5-1.4)\times10^7$\,L$_{\odot}$\,pc$^{-2}$, while the
  light-blue and blue curves show the analogous SB models. Upper panels show
  results for $N_{\mathrm{H2}}=10^{25}$\,cm$^{-2}$, and lower panels for
  $N_{\mathrm{H2}}=10^{24}$\,cm$^{-2}$; other parameters have fiducial values.
  g) The maximum line optical depth (at line center and along the sightline
  that crosses the source tangent to the inner cavity) of the same lines
  as before ($\nu=0\,J=3-2$ in black, Q(3) in red, and $\nu_2=1^f\,J=3-2$
  in blue) as a function of $N_{\mathrm{H2}}$. Solid and dashed lines correspond
  to $\Sigma_{\mathrm{IR}}=5.5\times10^7$ and
  $1.4\times10^7$\,L$_{\odot}$\,pc$^{-2}$, respectively, and circles and
  starred symbols indicate AGN and SB models, respectively.
 }    
\label{excit}
\end{figure*}
\begin{figure}
\begin{center}
\includegraphics[angle=0,scale=.65]{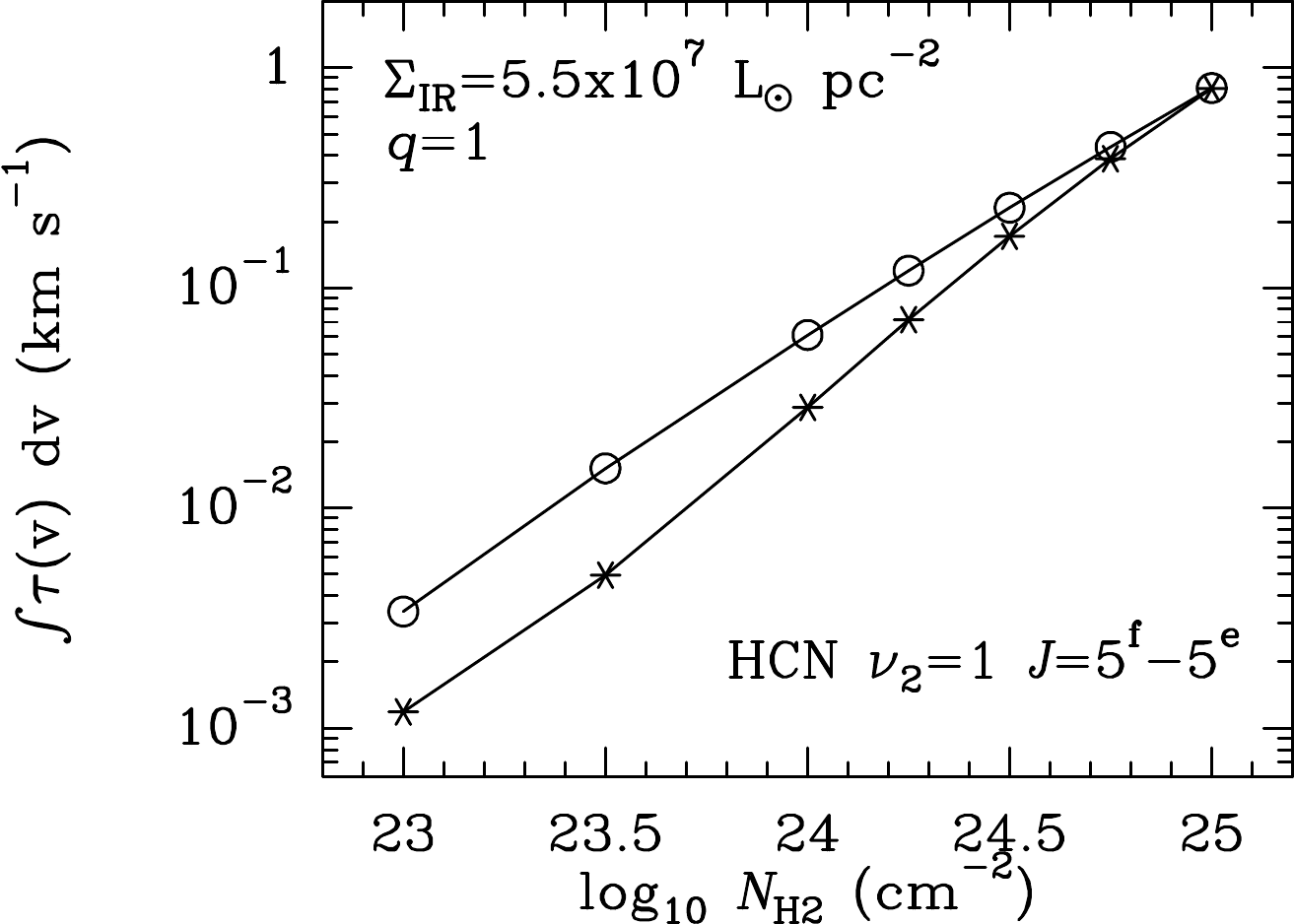}
\end{center}
\caption{
  The velocity-integrated line optical depth along a radial path of the
  direct $l$-type HCN $\nu_2=1\,J=5$ line at $6.73$\,GHz, for AGN
  (circles) and SB (stars) models with
  $\Sigma_{\mathrm{IR}}=5.5\times10^7$\,L$_{\odot}$\,pc$^{-2}$ and $q=1$.
 }    
\label{ltype}
\end{figure}

The excitation of HCN is illustrated in Fig.~\ref{excit}a-f through the
comparison between the $T_{\mathrm{dust}}$ profiles and the excitation
temperatures ($T_{\mathrm{EX}}$) of the HCN Q(3), $\nu=0\,J=3-2$, and
$\nu_2=1^f\,J=3-2$ lines. As shown in the upper panels (a, c, and e),
the excitation is extremely simple for $N_{\mathrm{H2}}=10^{25}$\,cm$^{-2}$,
as the three lines are in LTE at $T_{\mathrm{gas}}=T_{\mathrm{dust}}$ even for
moderate $\Sigma_{\mathrm{IR}}=1.4\times10^7$\,L$_{\odot}$\,pc$^{-2}$.
Collisional excitation is able to
thermalize the HCN $\nu=0$ low-$J$ levels at $T_{\mathrm{gas}}$, and the Q(3)
line, connecting the $\nu=0\,J=3$ and $\nu_2=1^f\,J=3$ levels, is also
thermalized at the local $T_{\mathrm{dust}}$. As a consequence, the
$\nu_2=1^f\,J=3-2$ line is also thermalized.

The HCN ground-state levels can be excited either via
collisional events and/or by radiative pumping to the $\nu_2$ state and
subsequent decay. If the column density is decreased by a
factor 10 (Fig.~\ref{excit}b-d-f, with $N_{\mathrm{H2}}=10^{24}$\,cm$^{-2}$),
$n_{\mathrm{H2}}$ drops by the same factor (Fig.~\ref{dens})
and collisions are unable to thermalize the $\nu=0$ low-$J$ lines
in the external regions (Fig.~\ref{excit}d). The departure from LTE occurs
at $T_{\mathrm{dust}}\lesssim200$\,K, because radiative excitation is still able
to thermalize the $\nu=0$ low-$J$ levels at higher $T_{\mathrm{dust}}$.
Nevertheless, the Q(3) line is still in LTE at all radii (Fig.~\ref{excit}b)
because the continuum at 14\,$\mu$m remains very optically thick
(Fig.~\ref{greenhouse}a). Therefore, the excitation of the
$\nu_2=1^f\,J=3-2$ line mimics that of the ground $\nu=0\,J=3-2$ line,
hence showing the same departure from LTE (Fig.~\ref{excit}d-f).

The maximum optical depths through the source ($\tau_{\mathrm{max}}$)
of the quoted lines are also shown in Fig.~\ref{excit}g.
The HCN $\nu=0\,J=3-2$ line is very optically thick in all models,
and as a surface tracer it is unuseful to probe the inner regions
of BGNs where radiative excitation is important. 
The Q(3) line is also saturated.
For fixed $N_{\mathrm{H2}}$, these transitions show higher
$\tau_{\mathrm{max}}$ in models with lower
overall excitation, because the population is accumulated in the
low-$J$ levels. By contrast, the HCN $\nu_2=1^f\,J=3-2$ line obviously
shows higher $\tau_{\mathrm{max}}$ in higher excitation models,
and the mere detection of the line provides evidence for environments
with extreme physical conditions.
For our fiducial $X_{\mathrm{HCN}}/\Delta V$,
this line becomes optically thick
for $N_{\mathrm{H2}}>10^{24}$\,cm$^{-2}$ and
$\Sigma_{\mathrm{IR}}\gtrsim5\times10^7$\,L$_{\odot}$\,pc$^{-2}$.

We have also checked the excitation and optical depth of the
  direct $l-$type transitions ($\Delta J=0$) in the $\nu_2=1$ state
  at centimeter wavelengths. Three of these lines ($J=4,5,6$) were
  detected in absorption towards Arp\,220  with the Arecibo telescope
  \citep{sal08}. In our models, however, these lines are slightly
  inverted, although with weak amplification ($|\tau|<0.3$ for the
  highest $N_{\mathrm{H2}}=10^{25}$\,cm$^{-2}$). The inversion is due to
  overlap effects among the ro-vibrational lines that pump the $\nu_2=1$
  state, and involves a tiny perturbation in the relative populations
  of the $e$ and $f$ levels because of the low splitting due to
  $l-$type doubling \citep[see discussion by][]{tho03}. Since collisional
  coupling among the involved levels, which is not included in our models,
  is expected to quench the maser in high density regions, we have 
  estimated the velocity-integrated optical depth of the $J=5$ line
  at $6.73$\,GHz by assuming that the
  sum of the upper and lower level populations remains unchanged, and that
  the levels are thermalized at the local $T_{\mathrm{gas}}$. Results for
  $\int\tau(v)\,dv$ along a radial path, as a function of
  $N_{\mathrm{H2}}$ for $\Sigma_{\mathrm{IR}}=5.5\times10^7$\,L$_{\odot}$\,pc$^{-2}$
  and $q=1$, are shown in Fig.~\ref{ltype}.
  For $N_{\mathrm{H2}}=10^{25}$\,cm$^{-2}$,
  $\int\tau(v)\,dv\sim0.8$\,km\,s$^{-1}$ with very similar values for the
  AGN and SB models. Most of the absorption
  is produced in the inner layers ($r/R_{\mathrm{out}}\lesssim0.3$) where
  $T_{\mathrm{dust}}\gtrsim300$\,K.
  In Arp~220, the value measured by \cite{sal08} for this line is
  $\approx5$\,km\,s$^{-1}$, which independently indicates the enormous
  columns of warm HCN gas in the nuclear region of this galaxy.

\subsubsection{The impact of the greenhouse effect on the HCN vibrational
  emission in buried galactic nuclei}

\begin{figure*}
\begin{center}
\includegraphics[angle=0,scale=.58]{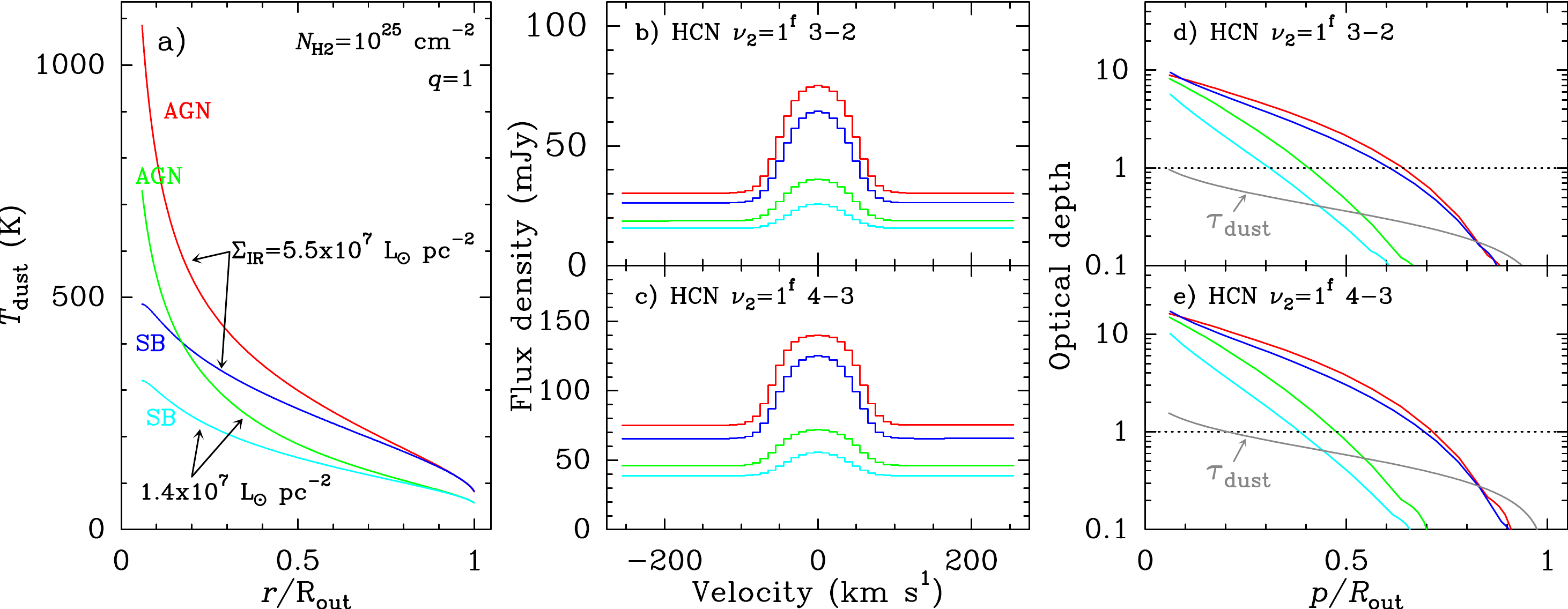}
\end{center}
\caption{Detailed results for the same models as in Fig.~\ref{excit}a-c-e.
    a) The $T_{\mathrm{dust}}$ profiles.
  (b-c) The line profiles and (d-e) optical depths
  at line center of the HCN $\nu_2=1^f\,J=3-2$ and $4-3$ transitions, with
  fiducial parameters
  ($N_{\mathrm{HCN}}/\Delta V=1.5\times10^{17}$\,cm$^{-2}$/(km\,s$^{-1}$) and
  $\Delta V=67$\,km\,s$^{-1}$). In panels d-e, the grey curves indicate
    the optical depth of the continuum at the wavelengths of the lines
    ($1.1$ and $0.84$\,mm).
  In panels b-c, a solid angle of
  $\Delta\Omega=1.1\times10^{-2}$ arc\,sec$^2$ 
  is adopted for the flux density scale.
  After subtracting the continuum, the HCN $\nu_2=1^f\,J=3-2$ line fluxes  
  are $4.6$ (red), $3.7$ (blue), $1.7$ (green), and
    $0.87$\,Jy\,km\,s$^{-1}$ (light blue),
    while the $\nu_2=1^f\,J=4-3$ line fluxes are $7.4$ (red), $6.3$ (blue),
    $2.7$ (green), and $1.6$\,Jy\,km\,s$^{-1}$ (light blue).
 }    
\label{hcngreenhouse}
\end{figure*}

Figure~\ref{hcngreenhouse} illustrates the impact of the greenhouse effect
on the HCN $\nu_2=1^f\,J=3-2$ and $J=4-3$ emission, by comparing
in detail results of the same models as in Fig.~\ref{excit}a,c,e
  (with $N_{\mathrm{H2}}=10^{25}$\,cm$^{-2}$ and
  $\Sigma_{\mathrm{IR}}=(1.4-5.5)\times10^7$\,\Lsun\,pc$^{-2}$).
As shown in Fig.~\ref{hcngreenhouse}d-e,
the HCN $\nu_2=1^f\,J=3-2$ and $J=4-3$ lines are
optically thick along lines of sight that cross regions with
$T_{\mathrm{dust}}\gtrsim200$\,K.  These are the regions that mostly contribute
to the line fluxes, so that the fluxes are nearly proportional to the solid
angle subtended by the region with temperatures above $\sim200$\,K.
For $\Sigma_{\mathrm{IR}}=5.5\times10^7$\,\Lsun\,pc$^{-2}$, this region
is about twice the size of the model with $1.4\times10^7$\,\Lsun\,pc$^{-2}$,
and hence the modeled fluxes are a factor $\sim4$ higher.

As discussed above (Fig.~\ref{excit}a,c,e),
  the low-$J$ $\nu=0$, $\nu_2=1-0$, and $\nu_2=1$ lines are in LTE
  at the local $T_{\mathrm{gas}}=T_{\mathrm{dust}}$ for the high
  $N_{\mathrm{H2}}=10^{25}$\,cm$^{-2}$ value considered
in Fig.~\ref{hcngreenhouse}, and hence the excitation of the
  $\nu_2=1$ rotational lines is higher in AGN than in SB models.
However, the higher excitation of AGN models mainly affect the high-$J$
  levels of the $\nu_2=1$ state, while the
  relatively low-lying HCN\,$\nu_2=1^f\,J=3,4$ levels only show a moderate
  increase of populations ($\lesssim50$\% for
  $\Sigma_{\mathrm{IR}}\gtrsim5.5\times10^7$\,\Lsun\,pc$^{-2}$) relative to SB
  models\footnote{The high $T_{\mathrm{dust}}$ regions of AGN models
    efficiently populate the $\nu=0$ high-$J$ levels at the expense of the
    low-$J$ levels, so that the $\nu=0\,J=2-5$ levels (which pump the
    $\nu_2=1\,J=3-4$) are less populated in AGN than in SB models (with
    otherwise the same parameters). This effect partially compensates for
    the higher vibrational excitation of AGN models.}. 
In addition, and owing to  
the quoted line opacity effects in the $\nu_2=1$ lines, the
innermost regions of the AGN model where $T_{\mathrm{dust}}$ is very high
are not probed, so that the AGN and SB models with the same
$\Sigma_{\mathrm{IR}}$ yield similar line fluxes.
The differences in HCN $\nu_2=1$ fluxes between AGN and SB models are 
  mostly due to the different spatial scales over which the lines are optically
  thick, and are larger for moderate $\Sigma_{\mathrm{IR}}$
  (Fig.~\ref{hcngreenhouse}d-e).
One way to observationally
check that the $\nu_2=1$ lines saturate is observing both the $3-2$ and $4-3$
lines. In the optically thick regime, the flux ratio $4-3/3-2$
(both in Jy\,km\,s$^{-1}$) is
$(\nu_{4-3}/\nu_{3-2})^2\approx1.8$. A ratio slightly higher
($1.9-2$) may also be expected because the $4-3$ line is optically thick
over a slightly larger spatial extent
(this effect is again more pronounced for moderate $\Sigma_{\mathrm{IR}}$,
Fig.~\ref{hcngreenhouse}). However, the opposite effect is obtained
($\nu_2=1^f\,4-3/3-2<1.8$) if the (sub)millimeter
continuum is bright, because the absorption of the continuum by the lines
is stronger for the $\nu_2=1^f\,J=4-3$ line (\S\ref{overall}); this effect
dominates over the different emitting areas in the AGN models,
and in the SB model with high
$\Sigma_{\mathrm{IR}}$, of Fig.~\ref{hcngreenhouse}.

For models with $N_{\mathrm{H2}}=10^{25}$\,cm$^{-2}$, and since the $\nu_2=1$
lines become optically thick at
$T_{\mathrm{EX}}\approx T_{\mathrm{dust}}\gtrsim200$\,K, line fluxes can be
estimated as
\begin{equation}
  F_{\mathrm{HCN\,\nu2}}\,(\mathrm{Jy\,km\,s^{-1}}) \sim
  10^{23} \frac{2k}{\lambda^2}\,T_{\mathrm{EX}}\,
  \Delta\Omega_{\tau=1}\,\Delta V,
  \label{eq:fhcnv2}
\end{equation}
where $k$ is the Boltzmann constant and $\Delta\Omega_{\tau=1}$ is the solid
angle subtended by the region where the line saturates (i.e. where
$T_{\mathrm{dust}}\approx200$\,K). For the $\nu_2=1^f\,J=3-2$ line, taking
$\Delta\Omega_{\tau=1}=0.6^2\Delta\Omega$ for the SB model with
$\Sigma_{\mathrm{IR}}=5.5\times10^7$\,\Lsun\,pc$^{-2}$
(Fig.~\ref{hcngreenhouse}; $\Delta\Omega$ in Table~\ref{tab:par}) 
and $\Delta V\sim100$\,\kms\
(higher than 67\,\kms\ due to broadening by line opacity effects),
eq.~(\ref{eq:fhcnv2}) gives $4.1$\,Jy\,km\,s$^{-1}$,
similar to the actual value.
Using $\Delta\Omega_{\tau=1}=0.3^2\Delta\Omega$ for the
$\Sigma_{\mathrm{IR}}=1.4\times10^7$\,\Lsun\,pc$^{-2}$ model,
eq.~(\ref{eq:fhcnv2}) yields $1$\,Jy\,km\,s$^{-1}$, also in rough agreement
with the modeled value.
Equation~(\ref{eq:fhcnv2}), and specifically the value of
  $T_{\mathrm{EX}}=200$\,K for the photosphere of the HCN vibrational emission,
can be observationally checked if the ro-vibrational
line is spatially resolved and $\Delta\Omega_{\tau=1}$ is estimated, but
we remark that it is only valid for $N_{\mathrm{HCN}}=10^{19}$\,cm$^{-2}$.

A few models for H$^{13}$CN, assuming
  an abundance ratio relative to HCN of $1/60$, have also been computed
  with fiducial parameters 
  ($\Sigma_{\mathrm{IR}}=1.1\times10^8$\,\Lsun\,pc$^{-2}$). The fluxes predicted
  for the H$^{13}$CN\,$\nu_2=1^f\,J=3-2$ and $J=4-3$ lines are factors
  $20-25$ and $15-18$ weaker than the HCN fluxes of the same lines,
  respectively. Despite the above H$^{13}$CN\,$\nu_2=1^f$ lines are optically
  thin, the AGN and SB models yield similar fluxes for them because the higher
  $T_{\mathrm{dust}}$ of the AGN models mainly affect the high-$J$ levels 
  of the $\nu_2=1$ state and the increase of the $\nu_2=1^f\,J=3-4$
  level populations is moderate.

\subsubsection{Spatial profiles}
\label{sec:spat}
              
\begin{figure*}
\begin{center}
\includegraphics[angle=0,scale=.5]{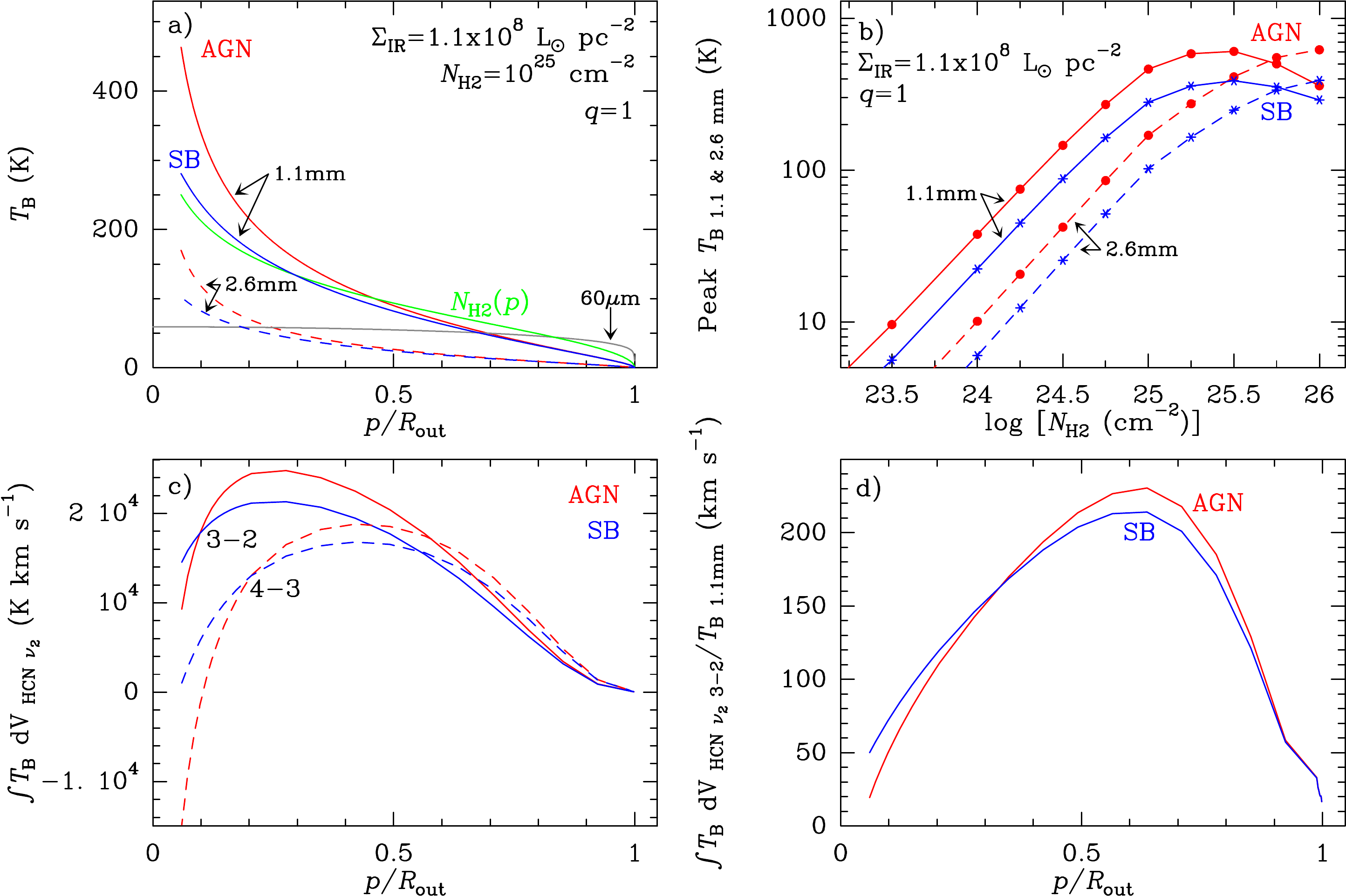}
\end{center}
\caption{a) The brightness $T_{\mathrm{B}}$ of the continuum at 1.1\,mm
  (solid colored curves) and $2.6$\,mm (dashed) as a function of the
  impact parameter, for the fiducial models (AGN in red and SB in blue)
  with $\Sigma_{\mathrm{IR}}=1.1\times10^8$\,L$_{\odot}$\,pc$^{-2}$.
  For comparison, the green curve indicates the $N_{\mathrm{H2}}(p)$ profile
    (arbitrarily scaled), and the grey curve shows the $T_{\mathrm{B}}$
    profile at 60\,$\mu$m (basically the same for AGN and SB).
  b) The peak
  brightness temperature at 1.1 and $2.6$\,mm as a function of the
  H$_2$ column density, for
  $\Sigma_{\mathrm{IR}}=1.1\times10^8$\,L$_{\odot}$\,pc$^{-2}$ and $q=1$.
  c) The brightness of the velocity-integrated (above the continuum)
  HCN $\nu_2=1^f\,J=3-2$ (solid) and $J=4-3$ (dashed) lines
  as a function of the impact parameter, for the same
  models as in panel a. d) The spatial profile of the
  HCN $\nu_2=1^f\,J=3-2$ line flux-to-continuum ratio.
}    
\label{hcngreenhousepar}
\end{figure*}

Since the continuum at millimeter wavelengths is less optically thick than
the HCN $\nu_2=1^f\,J=3-2$ line (Fig.~\ref{hcngreenhouse}d-e),
and due to the greenhouse effect that
is responsible for the high $T_{\mathrm{dust}}$ in the innermost regions,
the millimeter-wave continuum and HCN vibrational
emission are expected to show quite different spatial profiles
(Fig.~\ref{hcngreenhousepar}). The continuum strongly peaks 
toward the center, with a profile steeper than the 
$N_{\mathrm{H2}}(p)$ profile of our spherically symmetric models
(indicated with the green curve in Fig.~\ref{hcngreenhousepar}a;
$N_{\mathrm{H2}}(p)$ attains $10^{25}$\,cm$^{-2}$ at $p/R_{\mathrm{out}}\approx0.45$).
By contrast, the brightness of the continuum at 60\,$\mu$m, for which
  a well defined photosphere exists (Fig.~\ref{photos}),
  is nearly flat (grey curve in Fig.~\ref{hcngreenhousepar}a).
  The bulk of the source luminosity is emitted in the far-IR
  (Fig.~\ref{greenhouse}b), so that the continuum at millimeter wavelengths,
  while nicely probing the warm inner regions, 
may underestimate the area over which the luminosity is re-emitted.
On the other hand, the velocity-integrated line brightness above the
  continuum of the HCN $\nu_2=1^f\,J=3-2$ and $J=4-3$ lines show 
a drop of emission toward the center (Fig.~\ref{hcngreenhousepar}c)
because the line is formed in front of the bright
continuum, thus absorbing it
\footnote{Due to absorption of the
    continuum by the line, there is little continuum emission at line center,
    so that the apparent drop of line emission is
    due to the subtraction of the continuum
    adjacent to the line (i.e. free from line absorption).}.
The continuum brightness at $1.1$\,mm behind the HCN $\nu_2=1^f\,J=3-2$
  photosphere is still insufficient to produce an absorption line (i.e. negative
  $\int T_{\mathrm{B}}\,dv$) toward the center, but the $4-3$ line shows a net
  absorption in the AGN model owing to the increasing optical depth of the
  continuum and the line.
For sufficiently high
$N_{\mathrm{H2}}$, we thus expect a ring-like emission for
both the HCN $\nu_2=1^f\,J=3-2$ and $J=4-3$ lines,
as recently observed in IC~860 \citep{aal19},
and a strongly varying HCN-to-continuum brightness ratio
(Fig.~\ref{hcngreenhousepar}d). Note that significant absorption of the
continuum by the HCN $\nu_2=1^f\,J=3-2$ line is expected when the
HCN line flux-to-continuum ratio, with units of velocity
  (Fig.~\ref{hcngreenhousepar}d), 
  becomes comparable to or lower than the 
  linewidth of the HCN vibrational line, meaning that the
    total (velocity-integrated) continuum 
  absorbed by the line is comparable to the line flux.
Absorption of thermal dust continuum by molecular lines at (sub)millimeter
wavelengths has been inferred in several galactic nuclei
  \citep[e.g.][]{sak09,mar16,ala18}.

Our models with $N_{\mathrm{H2}}=10^{25}$\,cm$^{-2}$ still have too low
continuum optical depth at $2.6$\,mm ($\tau_{\mathrm{2.6\,mm}}\approx0.1$
along a radial path, Fig.~\ref{greenhouse}a) to 
account for the high brightness observed at this wavelength toward Arp~220W
\citep[after subtracting the plasma (free-free and synchrotron) emission;][]{sco17,sak17}.
To estimate the peak values of $T_{\mathrm{B}}$ due to thermal dust emission
in the millimeter at higher column densities, we have simply
assumed that the $T_{\mathrm{dust}}$ profile remains the same as 
for $N_{\mathrm{H2}}=10^{25}$\,cm$^{-2}$.
Since $T_{\mathrm{dust}}$ increases as
$N_{\mathrm{H2}}$ increases due to enhanced trapping of radiation, the 
inferred values of $T_{\mathrm{B}}$, shown in 
Fig.~\ref{hcngreenhousepar}b, can be considered lower limits for
$N_{\mathrm{H2}}>10^{25}$\,cm$^{-2}$. Even so, the 2.6\,mm continuum is expected
to attain brightnesses of $\sim600$ (AGN) and $\sim400$\,K (SB),
the former similar to the strong maximum at 3\,mm inferred in Arp~220W
from high angular resolution observations \citep{sak17}.

Figure~\ref{hcngreenhousepar} is based on the red $\kappa_{\nu}$-curve of
Fig.~\ref{kabs} that, with a value of $1.2$\,cm$^2$\,g$^{-1}$ of dust at
1.1\,mm, gives $\tau_{\mathrm{1.1mm}}\approx0.4$ for
$N_{\mathrm{H2}}=10^{25}$\,cm$^{-2}$ (Fig.~\ref{greenhouse}).
If, however, the black $\kappa_{\nu}$-curve is used,
the brightness of the 1.1\,mm continuum toward the center would be
significantly lower and the HCN $\nu_2=1^f\,J=3-2$ emission would also peak
toward the center. In
this case, $N_{\mathrm{H2}}\gtrsim10^{25.5}$\,cm$^{-2}$ would be required to
obtain the drop of HCN vibrational emission towards the peak of continuum
emission. In addition, high brightnesses of the 2.6\,mm continuum
would only be obtained for $N_{\mathrm{H2}}\sim10^{26.5}$\,cm$^{-2}$.

\subsubsection{Overall modeling results}
\label{overall}

\begin{figure*}
\begin{center}
\includegraphics[angle=0,scale=.58]{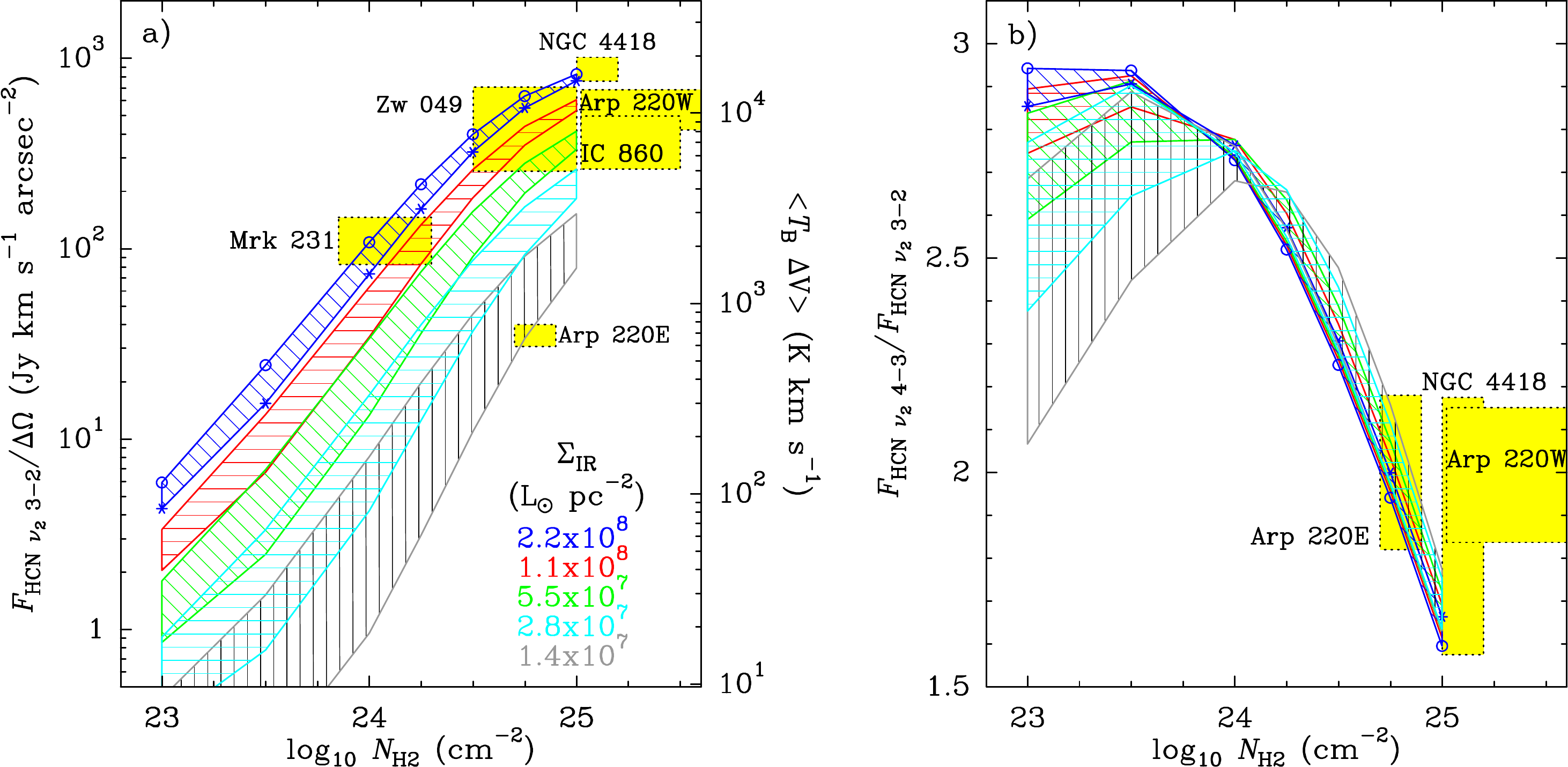}
\end{center}
\caption{a) The flux of the HCN $\nu_2=1^f\,J=3-2$ line per unit of
  solid angle of the source, as a function of the H$_2$
  column density. The right-hand axis gives the source-averaged
    velocity-integrated brightness of the line. Each hatched region
  corresponds to a value of the continuum surface brightness
  $\Sigma_{\mathrm{IR}}$ as indicated, and is delimited by AGN (higher values)
  and SB (lower values) models. Other model parameters have fiducial values
  ($q=1$, $R_{\mathrm{out}}/R_{\mathrm{int}}=17$,
  $X_{\mathrm{HCN}}/\Delta V=1.5\times10^{-8}$\,(km\,s$^{-1}$)$^{-1}$,
  $\Delta V=67$\,km\,s$^{-1}$).
  b) The HCN $\nu_2=1^f\,J=4-3$ to $\nu_2=1^f\,J=3-2$ flux ratio, both
  in Jy\,km\,s$^{-1}$, for the same models as in the left-hand panel. The yellow
  rectangles indicate the most plausible ranges for both axis in several
    galaxies (see \S\ref{datacomp}).
 }    
\label{hcngreenhousecolH2}
\end{figure*}
\begin{figure}
\begin{center}
\includegraphics[angle=0,scale=.53]{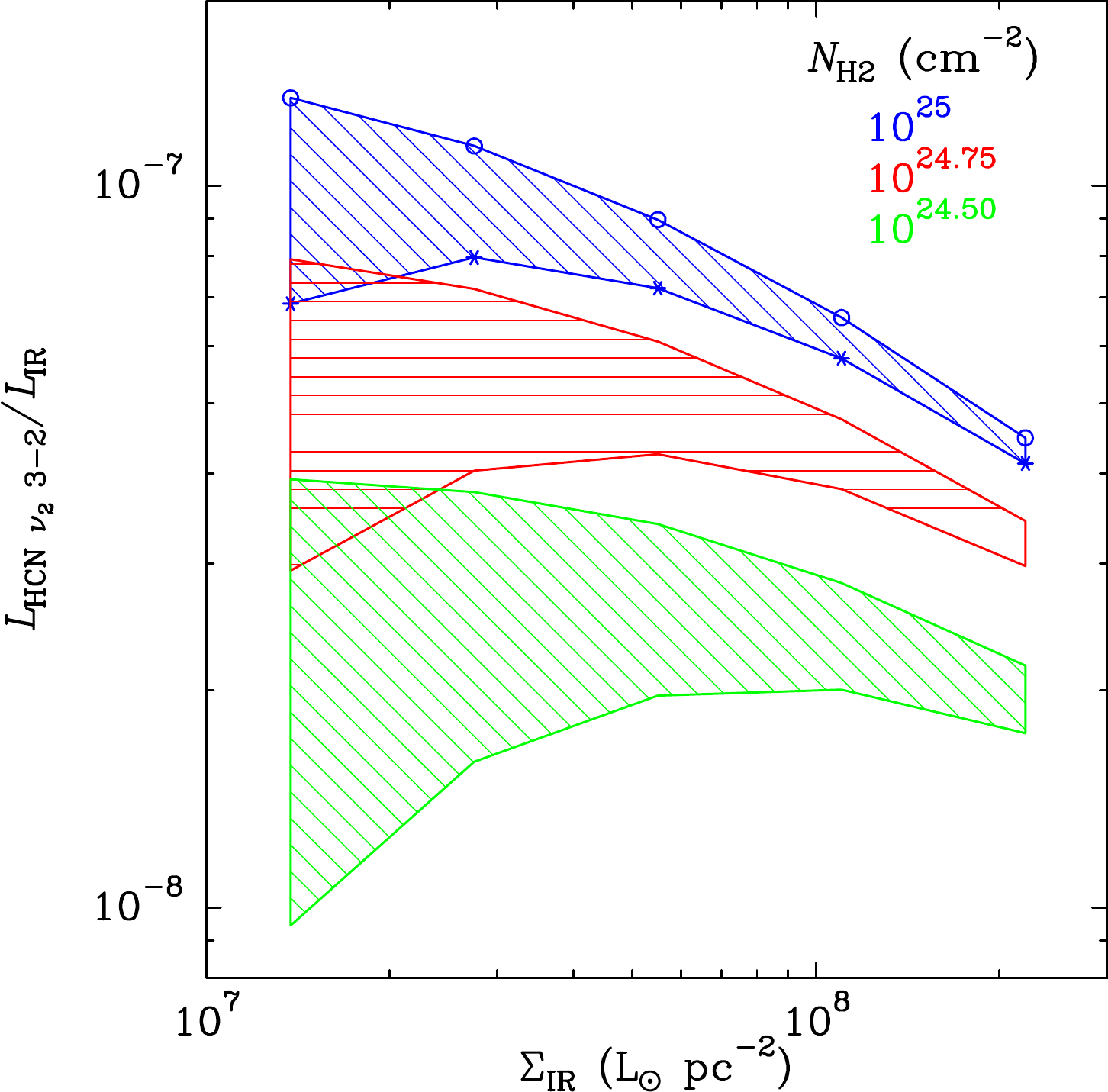}
\end{center}
\caption{The ratio of the HCN $\nu_2=1^f\,J=3-2$ luminosity to the infrared
  luminosity of the source, as a function of $\Sigma_{\mathrm{IR}}$.
  Each hatched region
  corresponds to a value of the H$_2$
  column density as indicated, and is delimited by AGN (higher values)
  and SB (lower values) models. Other model parameters have fiducial values
  ($q=1$, $R_{\mathrm{out}}/R_{\mathrm{int}}=17$,
  $X_{\mathrm{HCN}}/\Delta V=1.5\times10^{-8}$\,(km\,s$^{-1}$)$^{-1}$,
  $\Delta V=67$\,km\,s$^{-1}$).
 }    
\label{hcngreenhousesigmair}
\end{figure}

Overall results for the HCN $\nu_2=1$ fluxes in both AGN and SB models,
as a function of $N_{\mathrm{H2}}$
and for all explored values of $\Sigma_{\mathrm{IR}}$, are displayed in
Fig.~\ref{hcngreenhousecolH2}. Other parameters take fiducial values
(e.g., $q=1$, Table~\ref{tab:par}). Since line fluxes are proportional
  to the solid angle $\Delta\Omega$ of the source if all other parameters
  in Table~\ref{tab:par} are fixed, the average brightness
  $F_{\mathrm{HCN\,\nu2\,3-2}}/\Delta\Omega$ in panel a does not depend on
  $\Delta\Omega$. Its dependence on other parameters ($N_{\mathrm{H2}}$,
  $\Sigma_{\mathrm{IR}}$, and the choice of AGN/SB) is plotted in
  Fig.~\ref{hcngreenhousecolH2}a, such that the modeled values can be
  compared with observations to constrain these parameters. For our fiducial
  value $X_{\mathrm{HCN}}/\Delta V=1.5\times10^{-8}$\,(km\,s$^{-1}$)$^{-1}$,
  $F_{\mathrm{HCN\,\nu2\,3-2}}/\Delta\Omega$ is also proportional
  to $\Delta V$ and the modeled curves
  in Fig.~\ref{hcngreenhousecolH2}a would have to be vertically shifted by a
  factor $\Delta V/(67\,\mathrm{km\,s^{-1}})$, where $\Delta V$
  is the actual intrinsic
  velocity dispersion (i.e. corrected for systemic motions as rotation and
  for broadening by optically thick effects) of the considered
  source. For the purposes of this paper, we assume that
  $\Delta V=67\,\mathrm{km\,s^{-1}}$ is a sufficiently good approximation
  for the sources displayed in Fig.~\ref{hcngreenhousecolH2}, given that
  higher $\Delta V$ (as in Arp\,220W) would be partially compensated by a lower 
  $X_{\mathrm{HCN}}/\Delta V$ if $X_{\mathrm{HCN}}$ is nearly independent of
  $\Delta V$.
The range of solid angles that we estimate for the sources is discussed
below in detail (\S\ref{datacomp}) and listed in Table~\ref{tab:solid}.

Each hatched colored region in Fig.~\ref{hcngreenhousecolH2} indicates
model results for a given
$\Sigma_{\mathrm{IR}}$, delimited by AGN (open circles) and SB (starred symbols)
models. In panel b, the flux ratio of the HCN $\nu_2=1^f\,J=4-3$ to the
$\nu_2=1^f\,J=3-2$ lines is plotted for the same models as in the left-hand
panel, and compared with available data as well (see \S\ref{datacomp}).

For moderate columns, the dependence of $F_{\mathrm{HCN\,\nu2\,3-2}}/\Delta\Omega$
on $N_{\mathrm{H2}}$ is supralinear, because of the increase of both
$T_{\mathrm{dust}}$ and $N_{\mathrm{HCN}}$ as $N_{\mathrm{H2}}$ rises. However,
the dependence becomes sublinear for
  $N_{\mathrm{H2}}>5\times10^{24}$\,cm$^{-2}$
  and $\Sigma_{\mathrm{IR}}>10^8$\,L$_{\odot}$\,pc$^{-2}$, and 
the curves flatten at the highest $N_{\mathrm{H2}}$. At these extreme values,
the HCN vibrational emission is optically thick over most of the continuum
source. At the highest $N_{\mathrm{HCN}}=10^{19}$\,cm$^{-2}$
(i.e. $N_{\mathrm{H2}}=10^{25}$\,cm$^{-2}$) we consider in this work,
the flux ratio
$\nu_2=1^f\,4-3/3-2$ is lower than the optically thick limit
because of the enhanced absorption of the continuum by the $\nu_2=1^f\,J=4-3$
line (see Fig.~\ref{hcngreenhouse}d-e for a comparison of the
$\nu_2=1^f\,J=3-2$ and $J=4-3$ optical depths).
The use of the black $\kappa_{\nu}$-curve of Fig.~\ref{kabs} would
increase the calculated ratios to values closer to the optically thick limit
of $1.8$, and the values of $F_{\mathrm{HCN\,\nu2\,3-2}}/\Delta\Omega$ would also
increase by $\sim25$\% at $N_{\mathrm{H2}}=10^{25}$\,cm$^{-2}$ (results for
  lower $N_{\mathrm{H2}}$ are more similar).

The dependence of $F_{\mathrm{HCN\,\nu2\,3-2}}/\Delta\Omega$ on $N_{\mathrm{H2}}$
in Fig.~\ref{hcngreenhousecolH2}
becomes steeper as $N_{\mathrm{H2}}$ and $\Sigma_{\mathrm{IR}}$ decrease, and
more so for SB models. This effect is partially due to the adopted density
profiles. As shown in \S\ref{sec:excit} for the $Q(3)$ transition,
the ro-vibrational HCN lines at 14\,$\mu$m are thermalized to 
$T_{\mathrm{dust}}$, so that the excitation of the 
$\nu_2=1^f\,J=3-2$ and $J=4-3$ lines mimics the excitation of the corresponding
$\nu=0$ transitions (Fig.~\ref{excit}).
In regions with $T_{\mathrm{dust}}\gtrsim200$\,K,
  radiative excitation is enough to keep
the $\nu=0$ $J=3-2$ and $4-3$ nearly thermalized at
$T_{\mathrm{EX}}=T_{\mathrm{dust}}=T_{\mathrm{gas}}$ 
even for low $N_{\mathrm{H2}}$, but the line excitation in both
  vibrational states becomes subthermal for lower $T_{\mathrm{dust}}$
(Fig.~\ref{excit}d,f). 
Therefore, if the densities were higher than in
our models (Fig.~\ref{dens}), the dependence of
$F_{\mathrm{HCN\,\nu2\,3-2}}/\Delta\Omega$ on $N_{\mathrm{H2}}$ would not be as steep
as in Fig.~\ref{hcngreenhousecolH2} for low $\Sigma_{\mathrm{IR}}$.
This also explains the turnaround seen
in Fig.~\ref{hcngreenhousecolH2}b in the $\nu_2=1^f\,4-3/3-2$ flux ratio,
as the subthermal excitation affects more deeply the $4-3$ line than the
  $3-2$ transition.
If the densities were higher, the
turnaround of the $\nu_2=1^f\,4-3/3-2$ flux ratio with decreasing
$N_{\mathrm{H2}}$ would not be so pronounced. For the highest $N_{\mathrm{H2}}$
considered, however, our HCN fluxes are maximum and no higher values
would be obtained with higher adopted densities (for our fiducial values).

Therefore, the values of $L_{\mathrm{HCN\,\nu2\,3-2}}/L_{\mathrm{IR}}$ shown in
Fig.~\ref{hcngreenhousesigmair} are also maximum for our adopted
fiducial parameters and
continuum opacity at $1.1$\,mm\footnote{Somewhat higher values would be
  obtained with the black $\kappa_{\nu}$-curve of Fig.~\ref{kabs} as the
  optical depth and continuum brightness at $1.1$\,mm would be lower
  and thus the effect of absorption
  of this continuum by the HCN line (\S\ref{sec:spat}) would be less
  important. The increase of $\Delta V$ would also boost
  $L_{\mathrm{HCN\,\nu2\,3-2}}/L_{\mathrm{IR}}$.}.
For fixed $N_{\mathrm{H2}}$ and high $\Sigma_{\mathrm{IR}}$,
  $L_{\mathrm{HCN\,\nu2\,3-2}}/L_{\mathrm{IR}}$ drops with increasing
  $\Sigma_{\mathrm{IR}}$. This is because a
  further increase of $\Sigma_{\mathrm{IR}}$ is not followed
  by a proportional increase of $\Delta\Omega_{\tau=1}$ (eq.~\ref{eq:fhcnv2}),
  and produces in addition brighter millimeter
continuum emission and thus more absorption by the HCN line.
Both $L_{\mathrm{HCN\,\nu2\,3-2}}$ and
$L_{\mathrm{IR}}$ are calculated in spherical symmetry but both are
  optically thick, so that the ratio of the apparent luminosities is
probably valid for more general geometries.
In general, the BGN infrared luminosity is a fraction of the 
  $L_{\mathrm{IR}}$ of the host galaxy,
  but the vast majority of the HCN vibrational
  emission arises from the nucleus --unless the galaxy has other
  off-nuclear, buried sources. Hence, the
  $L_{\mathrm{HCN\,\nu2\,3-2}}/L_{\mathrm{IR}}$ values for the nucleus
  plotted in Fig.~\ref{hcngreenhousesigmair} are upper limits to the
  corresponding ratios for the entire galaxy.

\subsubsection{Comparison with data}
\label{datacomp}

As stated above, $\Delta\Omega=\pi R_{\mathrm{out}}^2/D^2$
in Fig.~\ref{hcngreenhousecolH2}a is
the solid angle of the cocoon covering the luminosity source(s),
rather than of the less extended
HCN vibrational emission, so that comparison with data can be performed
even if the HCN $\nu_2=1$ lines are spatially unresolved. However, 
the radius $R_{\mathrm{out}}$ of the cocoon over which the luminosity 
of the embedded sources is reemitted, must be estimated.
  In buried sources, the luminosity is mainly emitted at far-IR
  wavelengths (Fig.~\ref{greenhouse}b), for which a nearly flat profile is
  found (Fig.~\ref{hcngreenhousepar}a); therefore, $R_{\mathrm{out}}$ can be
  empirically defined as the radius of the far-IR source.
The caveat is that the far-IR extent may be underestimated by measurements
  of millimeter-wave continuum, which is strongly peaked toward the center
  (Fig.~\ref{hcngreenhousepar}a).
In principle, interferometric measurements at shorter (submillimeter)
wavelengths are better suited to probe the source extent.
Alternatively, $\Delta\Omega$ can be estimated from the
analysis of high-lying far-IR molecular
absorption\footnote{The extremely buried nuclei are best traced by
    far-IR molecular
  lines with $E_{\mathrm{low}}>500$\,K, such as the H$_2$O\,$7_{07}-6_{16}$ at
  $72$\,$\mu$m; low-lying lines usually probe in addition more extended,
  colder components with lower column densities.}, as these lines are
directly probing the far-IR photosphere and thus the full extent of
  the source. On the other hand,
spatially extended continuum unassociated directly with the source of
HCN vibrational emission should be ignored.
A suitable criterion would be to exclude surrounding regions where the
inferred H$_2$ column density translates into optically thin far-IR
emission (i.e. $<5\times10^{23}$\,cm$^{-2}$),
  as the greenhouse effect no longer takes place.
We only consider in
the following sources with HCN vibrational emission where $\Delta\Omega$
can be reasonably estimated.

Analysis of interferometric measurements usually involve Gaussian curves
to fit sizes; to compare with our spherical models, we use the equivalent
size of a uniform disk,
i.e. $R_{\mathrm{out}}=0.8\times\mathrm{FWHM}$ \citep[Appendix~A in][]{sak08}. 
The values of $\Delta\Omega$ for all individual sources used in
Fig.~\ref{hcngreenhousecolH2}, listed in Table~\ref{tab:solid}
  together with the inferred $\Sigma_{\mathrm{IR}}$, $L_{\mathrm{IR}}$, and
  $T_{\mathrm{dust}}$ throughout the far-IR photosphere,
are justified in continuation.

   \begin{table*}
     \caption{Estimated solid angles $\Delta\Omega$ for the BGNs considered
       in \S\ref{datacomp}, line fluxes of the HCN vibrational lines, and
 inferred values of $\Sigma_{\mathrm{IR}}$, $L_{\mathrm{IR}}$, and $T_{\mathrm{dust}}$
     at the far-IR photosphere}
         \label{tab:solid}
\begin{center}
          \begin{tabular}{lcccccccccc}   
            \hline
            \noalign{\smallskip}
            Source & $D_L$ & $R_{\mathrm{out}}$ &  $\Delta\Omega$ &
            Ref$^{\mathrm{a}}$ & $F_{\mathrm{HCN\,\nu2\,3-2}}$ & 
            $F_{\mathrm{HCN\,\nu2\,4-3}}$ & Ref$^{\mathrm{b}}$ &
            $\Sigma_{\mathrm{IR}}$ &
            $L_{\mathrm{IR}}$ & $T_{\mathrm{dust}}$-photo\\
            & (Mpc)  & (pc) & ($10^{-2}$ arc\,sec$^2$) & &
            (Jy\,km\,s$^{-1}$) &  (Jy\,km\,s$^{-1}$) & &
            ($10^7$\,L$_{\odot}$\,pc$^{-2}$) & ($10^{11}$\,L$_{\odot}$) & (K) \\
            (1) & (2) & (3) & (4) & (5) & (6) & (7) & (8) & (9) & (10)& (11) \\
            \noalign{\smallskip}
            \hline
            \noalign{\smallskip}
NGC 4418 & 34 & $11.7-13.5$ & $1.6-2.1$ & 1,2 & 16 & 30 & 3 & 22 & $0.9-1.3$ & $115-160$ \\ 
Arp 220W & 85 & $47-60$ & $4.1-6.7$ & 1,4,5 & 28 & 55.0 & 6 & 11 & $7.6-12$ & $100-130$ \\ 
Arp 220E & 85 & $87-90$ & $14-15$ & 4,5 & 5.1 & 10.2 & 6 & $0.5-1$ & $1.3-2.4$ & $50-65$ \\ 
Zw 049.057 & 56 & $15-25$ & $1.0-2.7$ & 7 & 6.8 &  & 8 & $5-20$ & $1.0-1.4$ & $80-160$ \\ 
IC 860   & 59 & $14.5-20$ & $0.8-1.5$ & 11 & 4 &  & 8 & $5.5$ & $0.4-0.7$ & $80-105$ \\
Mrk 231 & 192 & $55-73$ & $1.1-1.9$ & 9 & 1.6 &  & 10 & $11-22$ & $18-21$ & $95-180$ \\ 
            \noalign{\smallskip}
            \hline
         \end{tabular} 
\end{center}
\tablecomments{(1) Galaxy name;
  (2) Luminosity distance;
  (3) Estimate for the outer radius of the cocoon reemitting the luminosity
  generated by the embedded source(s), based on both the measured sizes in
  the (sub)millimeter and on the calculated sizes from far-IR molecular
  absorption (see text); (4) Solid angle, calculated as in Table~\ref{tab:par};
  (5) References for $R_{\mathrm{out}}$; (6) Flux of the HCN\,$\nu_2=1^f\,J=3-2$
  line; (7) Flux of the HCN\,$\nu_2=1^f\,J=4-3$ line; (8) References for the
  HCN\,$\nu_2=1^f$ fluxes; (9) Estimated value of $\Sigma_{\mathrm{IR}}$ based
  on the location of sources in Fig.~\ref{hcngreenhousecolH2}a;
  (10) Luminosity of the BGN, calculated as  
  $L_{\mathrm{IR}}=\Sigma_{\mathrm{IR}}\,\Delta\Omega\,D_L^2$;
  (11) Range of $T_{\mathrm{dust}}$ values throughout the far-IR photosphere,
  defined as $\tau_{60\,\mu m}\leq0.5$ from the surface, as predicted by
  the best-fit continuum models.}
\begin{list}{}{}
\item[References:] (1) \cite{gon12} (2) \cite{sak13} (3) \cite{sak10}
  (4) \cite{sak17} (5) \cite{wil14} (6) \cite{mar16} (7) \cite{fal15}
  (8) \cite{aal15b} (9) \cite{gon14b} (10) \cite{aal15a} (11) \cite{aal19}
\end{list}
   \end{table*}

\underline{NGC 4418}: an extremely compact and warm source of far-IR
emission was identified from the strong and very excited molecular lines
of H$_2$O, OH, and other species including HCN \citep{gon12}. The radius
of this source was estimated to be $\approx10$\,pc at an adopted $D=29$\,Mpc,
i.e. $\Delta\Omega=1.6\times10^{-2}$\,arc\,sec$^2$. From high-angular
resolution observations at $\approx350$\,GHz, \cite{sak13} estimated
a radius of $\approx13.5$\,pc at an adopted $D=34$\,Mpc, i.e.
$\Delta\Omega=2.1\times10^{-2}$\,arc\,sec$^2$. We have used these two values
of $\Delta\Omega$ and a flux of the HCN line of $16$\,Jy\,km\,s$^{-1}$
\citep{sak10} to give
$F_{\mathrm{HCN\,\nu2\,3-2}}/\Delta\Omega=760-1000$\,Jy\,km\,s$^{-1}$\,arc\,sec$^{-2}$.
Only the maximum value of
$\Sigma_{\mathrm{IR}}=2.2\times10^8$\,L$_{\odot}$\,pc$^{-2}$ can account for this
brightness (Fig.~\ref{hcngreenhousecolH2}), yielding a source luminosity of
$D^2\,\Delta\Omega\,\Sigma_{\mathrm{IR}}=(0.9-1.3)\times10^{11}$\,\Lsun.
This estimate agrees with the conclusion that the bulk of the galaxy luminosity
emerges from such a compact region \citep{gon12,sak13}. In addition,
our continuum models for $\Sigma_{\mathrm{IR}}=2.2\times10^8$\,L$_{\odot}$\,pc$^{-2}$
predict $T_{\mathrm{dust}}=115-160$\,K for the photosphere with
$\tau_{60\,\mu m}\leq0.5$ from the surface (Table~\ref{tab:solid}),
in general agreement with requirements to
account for the high-lying molecular absorption in the far-IR \citep{gon12}.
The HCN\,$\nu_2=1^f\,J=4-3$ to $J=3-2$ flux ratio of $1.9\pm0.3$ \citep{sak10}
is also consistent with strongly saturated HCN vibrational emission.

\underline{Arp 220W}: the western nucleus of Arp~220 has been recently imaged
with extremely high angular resolution at (sub)millimeter wavelengths.
Visibility fitting of the 3\,mm continuum by \cite{sak17} shows that
a 2 Gaussian fitting (or 1 Gaussian and 1 exponential disk) performs much
better than a single Gaussian. While the 3\,mm emission from the
  compact source, with an equivalent $R_{\mathrm{out}}=14$\,pc, is dominated
  by thermal dust emission, the emission from the larger component, with
  $R_{\mathrm{out}}=60$\,pc, is predominantly non-thermal.
  At the higher frequency of 691\,GHz, \cite{wil14} infer
  $R_{\mathrm{out}}\leq80$\,pc, and in the far-IR \cite{gon12} estimate
  $R_{\mathrm{out}}=47-89$\,pc (both corrected for a distance of 85\,Mpc).
  For the reason stated above, the more extended component at 3\,mm is
  probably emitting in the far-IR, so that
  we adopt $R_{\mathrm{out}}=47-60$\,pc yielding
  $\Delta\Omega=(4.1-6.7)\times10^{-2}$\,arc\,sec$^2$.
With the HCN flux
of $\sim28$\,Jy\,km\,s$^{-1}$ \citep{mar16}, 
$F_{\mathrm{HCN\,\nu2\,3-2}}/\Delta\Omega=400-700$\,Jy\,km\,s$^{-1}$\,arc\,sec$^{-2}$.
This is significantly lower than the brightness of the line in NGC~4418,
in agreement with the significantly lower excitation of the far-IR
  absorption lines in Arp~220, and can be explained with
$\Sigma_{\mathrm{IR}}=1.1\times10^8$\,L$_{\odot}$\,pc$^{-2}$ 
(Fig.~\ref{hcngreenhousecolH2}a)\footnote{As noted in \S\ref{overall},
  the modeled $F_{\mathrm{HCN\,\nu2\,3-2}}/\Delta\Omega$ would have to be multiplied
  by $\Delta V/(67\,\mathrm{km\,s^{-1}})$, with the result that the observed
  value could be explained with even lower $\Sigma_{\mathrm{IR}}$ for
  the high $\Delta V$ inferred from CO observations \citep{sco17}, although we
  also note that $\Delta V$ would have to be corrected by optical depth
  broadening and that a lower $X_{\mathrm{HCN}}/\Delta V$ would also
  partially compensate for the increase of $\Delta V$.}.
The implied luminosity
is $(0.76-1.2)\times10^{12}$\,\Lsun, in good agreement with the luminosity
inferred from the analysis of the far-IR absorption lines
\citep[$(0.91-1.1)\times10^{12}$\,\Lsun\ after correcting for the
  adopted distance;][]{gon12}. For
$\Sigma_{\mathrm{IR}}=1.1\times10^8$\,L$_{\odot}$\,pc$^{-2}$,
the predicted $T_{\mathrm{dust}}$ across the $\tau_{60\,\mu m}\leq0.5$ photosphere
is $100-130$\,K, also in agreement with the inferred values from
the far-IR absorption lines. We also favor the AGN model that
  predicts high central continuum brightness $T_{\mathrm{B}}$ in the
  millimeter, in excess of 500\,K for
  $N_{\mathrm{H2}}\sim10^{26}$\,cm$^{-2}$ (Fig.~\ref{hcngreenhousepar}b),
  similar to the value measured by \cite{sak17}; by contrast, the SB model
  for the same value of $\Sigma_{\mathrm{IR}}$ yields maximum brightness of
  $\approx400$\,K. The main drawback of our model is that it predicts
  a low HCN\,$\nu_2=1^f\,J=4-3$ to $J=3-2$ flux ratio of $\lesssim1.7$ owing
  to the extreme column densities, while
the observed value is $2.0\pm0.16$ \citep{mar16}.

\underline{Arp 220E}: the eastern nucleus of Arp~220 has been also imaged with
high angular resolution in the (sub)millimeter. \cite{sak17} found
$\Delta\Omega=0.14$\,arc\,sec$^2$ at 3\,mm
(i.e. $R_{\mathrm{out}}=87$\,pc for the extended component),
and \cite{wil14} determined 
$\Delta\Omega=0.15$\,arc\,sec$^2$ at 434\,$\mu$m. Using the HCN flux
of $4.6-5.6$\,Jy\,km\,s$^{-1}$ by \cite{mar16}, we infer
$F_{\mathrm{HCN\,\nu2\,3-2}}/\Delta\Omega=30-40$\,Jy\,km\,s$^{-1}$\,arc\,sec$^{-2}$.
On the other hand, the HCN\,$\nu_2=1^f\,J=4-3$ to $J=3-2$ flux ratio is
$2.0\pm0.2$ \citep{mar16}, consistent with $N_{\mathrm{H2}}\sim10^{24.8}$\,cm$^{-2}$.
Our model with the minimum
$\Sigma_{\mathrm{IR}}=1.4\times10^7$\,L$_{\odot}$\,pc$^{-2}$ may still
overestimate the line brightness, so that we adopt
$(0.5-1)\times10^7$\,L$_{\odot}$\,pc$^{-2}$
yielding a luminosity of $(1.3-2.4)\times10^{11}$\,\Lsun\ at 85\,Mpc,
consistent with the estimate by \cite{wil14}.

\underline{IC 860}: Recent interferometric observations have revealed
  an extremely compact source at (sub)millimeter wavelengths, with equivalent
  $R_{\mathrm{out}}=(13-14.5)$\,pc \citep{aal19}; however, the
  HCN\,$\nu_2=1^f\,J=3-2$ line is more extended, $R_{\mathrm{out}}\approx20$\,pc,
  suggesting that the infrared emission has at least a similar size.
  We have nevertheless adopted
  the conservative range $R_{\mathrm{out}}=(14.5-20)$\,pc and thus
  $\Delta\Omega=(0.8-1.5)\times10^{-2}$\,arc\,sec$^2$. Using the HCN flux
of $\approx4$\,Jy\,km\,s$^{-1}$ by \cite{aal15b}, we infer
$F_{\mathrm{HCN\,\nu2\,3-2}}/\Delta\Omega=260-500$\,Jy\,km\,s$^{-1}$\,arc\,sec$^{-2}$,
which is mostly consistent with
$\Sigma_{\mathrm{IR}}\sim5.5\times10^7$\,L$_{\odot}$\,pc$^{-2}$
(Fig.~\ref{hcngreenhousecolH2}a).
The infrared luminosity of the nucleus is then $(4-7)\times10^{10}$\,\Lsun,
$30-50$\% of the total infrared luminosity of the galaxy.

\underline{Zw 049.057}: An obscured and compact nucleus was identified
by \cite{fal15} from high-lying far-IR molecular absorption, with
a most plausible radius of $15-25$\,pc at 56\,Mpc
($\Delta\Omega=(1.0-2.7)\times10^{-2}$\,arc\,sec$^2$)
and a column density of $N_{\mathrm{H2}}\sim10^{24.5-25}$\,cm$^{-2}$.
The HCN\,$\nu_2=1^f\,J=3-2$ line was detected by \cite{aal15b} with a flux
of $6.8$\,Jy\,km\,s$^{-1}$, yielding $250-700$\,Jy\,km\,s$^{-1}$\,arc\,sec$^{-2}$.
From Fig.~\ref{hcngreenhousecolH2}, we estimate
$\Sigma_{\mathrm{IR}}=(0.5-2)\times10^8$\,L$_{\odot}$\,pc$^{-2}$ for the
upper and lower limit of $\Delta\Omega$, respectively. This gives a
luminosity of $(1.0-1.4)\times10^{11}$\,\Lsun\ for the compact core, in
agreement with the estimate by \cite{fal15} from far-IR
molecular absorption lines ($(0.7-1.2)\times10^{11}$\,\Lsun).
Our predicted range of $T_{\mathrm{dust}}$ across the far-IR photosphere
brackets the range favored by \cite{fal15} ($90-130$\,K).

\underline{Mrk 231}: A highly excited component was inferred from
high-lying OH absorption in the far-IR \citep{gon14b}, with an
estimate radius of $55-73$\,pc at 192\,Mpc. We thus adopt
$\Delta\Omega=(1.1-1.9)\times10^{-2}$\,arc\,sec$^2$, which with the
observed HCN flux of $1.6$\,Jy\,km\,s$^{-1}$ \citep{aal15a} yields
$F_{\mathrm{HCN\,\nu2\,3-2}}/\Delta\Omega=80-150$\,Jy\,km\,s$^{-1}$\,arc\,sec$^{-2}$.
This component is not expected to have the extremely high column
densities of the previous sources, but
$N_{\mathrm{H2}}\sim(0.7-2)\times10^{24}$\,cm$^{-2}$ \citep{gon14b}.
This is consistent
with a high $\Sigma_{\mathrm{IR}}=(1.1-2.2)\times10^8$\,L$_{\odot}$\,pc$^{-2}$
(Fig.~\ref{hcngreenhousecolH2}), which translates into a luminosity
of $(1.8-2.1)\times10^{12}$\,\Lsun, the expected luminosity of the nucleus.
Our range of photospheric $T_{\mathrm{dust}}$ in Table~\ref{tab:solid} includes
the values favored from far-IR molecular absorption
\citep[$95-120$\,K,][]{gon14},
  and suggests that the model with
  lower $\Sigma_{\mathrm{IR}}\sim10^8$\,L$_{\odot}$\,pc$^{-2}$
  (with $T_{\mathrm{dust}}-\mathrm{photo}=100-135$\,K) better fits the data.
The HCN\,$\nu_2=1^f\,J=4-3$ line has not been observed, and 
our model predicts a $\nu_2=1^f\,J=4-3$ to $J=3-2$ flux ratio significantly
higher than for the other sources, $\sim2.5$.

\section{Discussion}
\label{sec:discussion}

The greenhouse effect presented here, or the effect of increasing
dust temperatures due to trapping of mid- and far-IR radiation 
in environments of extreme column densities, has the advantage of
explaining, through the calibration presented in
Fig.~\ref{hcngreenhousecolH2}, three general observations: 
the bright emission of the HCN vibrational lines in buried sources,
the dust temperatures in the photosphere required to explain
the high-lying molecular absorption in the far-IR, and the high
brightness and compactness of the (sub)millimeter continuum, all with
involved source luminosities that are consistent with values calculated
from independent approaches.

As pointed out in \S\ref{sec:descont}, our models are in principle
  applicable to a single source or to an ensemble of non-radiatively
  interacting sources. However, luminous HCN vibrational emission 
  in well-studied cases is arising from well defined compact galactic nuclei
  rather than from individual giant molecular clouds (GMCs) widespread over
  a kilo-parsec disk, even though ``hot cores'' in our galaxy also show  
  HCN\,$\nu_2=1$ emission. There are probably several reasons that
    can account for this. First, the values of
    $\Sigma_{\mathrm{IR}}\sim10^8$\,L$_{\odot}$\,pc$^{-2}$ we infer in the most
    buried BGNs are higher than the values
    $\lesssim10^6$\,L$_{\odot}$\,pc$^{-2}$
    typically inferred in galactic hot cores
    \citep[e.g.][]{nom04,dot06}, although in some cases
    $\Sigma_{\mathrm{IR}}\gtrsim10^7$\,L$_{\odot}$\,pc$^{-2}$ \citep{ces10}.
    In addition, the timescale of this buried
  phase in individual clouds, $\sim10^5$\,yr \citep[e.g.][]{wil01,dot06}, is 
  much shorter than the typical dynamical time scale of galactic disks. 
  When the buried phase turns on in independent
  clouds, their appearance is spread out over the full dynamical
  timescale of the disk and their contribution to the luminosity of the
  galaxy will be low. In galaxies where high HCN vibrational fluxes are
  detected, the syncronization required by large gas masses participating
  in the buried phase can only take place within a common sphere of
  influence, the galactic nucleus. This accounts for the bimodality observed in
  the high-lying OH\,65\,$\mu$m absorption in galaxies, suggesting 
  ``coherent'' structures \citep{gon15}.
  Finally, in case of multiple luminosity sources, the greenhouse effect
    and HCN vibrational excitation should be generally more enhanced
when the sources are packed and radiatively coupled with each other than
when they are widespread and radiatively decoupled.
This is because packed sources have mutual heating and mutual contribution to
the overall shielding required for an efficient greenhouse effect.
The gas velocity dispersion $\Delta V$ is also higher in this latter
scenario, contributing to further boost the HCN vibrational emission.

The spherical symmetry used in the present study, however, 
assumes isotropic column densities from the center and no clumpiness.
This oversimplified smoothed density structure may overestimate
the dust temperature as compared with real systems. If the gas and dust
are mainly concentrated into clumps, there will be an increasing number of
sightlines with lower column densities along which the radiation tends
to escape \citep[e.g.][]{rot12}, decreasing $T_{\mathrm{dust}}$. Likewise,
a flat structure like a disk will have minimum column densities along
the direction perpendicular to the disk plane, and radiation will tend to
escape in that direction. We have indeed evidence for a clumpy structure
in the galaxies considered in this work, as the analysis of the HCN band
at 14\,$\mu$m yields excitation temperatures of $\sim300$\,K \citep{lah07}
while the photosphere in our models has temperatures significantly lower.
In addition, most individual sources considered here (Arp~220, NGC~4418,
  Zw~049.057, and IC~860) are estimated to have
  $N_{\mathrm{H2}}>10^{24.5}$\,cm$^{-2}$, for which our AGN models predict
  little --but observations show prominent-- $9.7$\,$\mu$m absorption.
  The mid-IR continuum in front of which the silicate absorption is produced
  is arising from regions at $350-400$\,K \citep{gon12}, clearly
  warmer than the $T_{\mathrm{dust}}$ at the far-IR photosphere of our models.
  This continuum may be due to leakage of mid-IR radiation from the very
  nuclear region, or generated by a surrounding star formation component.
  Judging from the SED of NGC~4418, where the output power is most likely
  dominated by the BGN, the leakage of mid-IR radiation is estimated as
  $L_{\mathrm{5-20\,\mu m}}/L_{\mathrm{5-1000\,\mu m}}\approx10$\%, although part of
  the mid-IR emission may arise from surrounding super star clusters as
  observed with VLBI at radio wavelengths \citep{var14}.

The caveat here is the contrast between the column densities
through clumps and along sightlines that only cross interclump
material --the background smooth gas distribution. At least in the
innermost nuclear regions of the galaxies considered in this work,
clouds are expected to (partially) lose their individuality and blend
into a fluctuating-dense medium \citep[e.g.][]{sco97,dow98}.
We also note that
the column densities of $\gtrsim10^{25}$\,cm$^{-2}$ inferred in the nuclear
regions of (U)LIRGs where HCN vibrational emission is detected are
beam-averaged,
and thus high columns apply to a significant range of solid angles --unless
the nucleus is seen close to edge on.
If these compact nuclei are mainly supported by radiation pressure,
  the rotation will slow down and the inner disk will evolve to become more
  spheroidal, like a cocoon with high columns even in the polar
  direction\footnote{This effect, combined with the inferred expansion
    of the nuclear regions in ULIRGs \citep{gon17}, may be crucial for the
    formation of spheroidal bulges.}.
Viewing the problem with a different perspective, and considering the high
HCN abundance that is still required to explain the HCN\,$\nu_2=1$ emission,
one could hypothesize that
strong HCN vibrational emission arises only in galaxies
where the greenhouse effect turns on as a result of the high column
densities averaged over solid angles.
By contrast, in galaxies where OH shows prominent
outflows that are expected to be wide-angle, the column densities averaged
over solid angles will be significantly lower due to sweeping out the nuclear
ISM by the outflow, and the greenhouse effect will
also be much less important with the consequent weakness of the HCN
vibrational emission \citep{fal19}.

When the column densities averaged over solid-angles become so large that
the region becomes opaque to its own mid- and even far-IR emission,
the increase of $T_{\mathrm{dust}}$ is unavoidable. As the temperature increases,
the dust emits at shorter wavelengths for which the optical depths are
even higher, thus enhancing the radiation trapping effect. With
$\tau_{\mathrm{20\mu m}}\sim300$ for $N_{\mathrm{H2}}=10^{25}$\,cm$^{-2}$
(Fig.~\ref{greenhouse}a), 20\,$\mu$m photons have negligible probability
to escape and the radiation field becomes nearly isotropic. The interior
of BGNs are infrared-dominated regions.

Our calibration in Fig.~\ref{hcngreenhousecolH2} implies source
  luminosities that are in agreement, for all sources considered in this
  paper, with independent estimates (\S\ref{datacomp}). 
The calibration involves a high
$X(\mathrm{HCN})\approx10^{-6}$, in rough agreement with the value
inferred in the far-IR photosphere of NGC~4418 from HCN rotational
lines seen in absorption \citep{gon12} and with the values
  inferred from the HCN\,14\,$\mu$m band \citep{lah07}. Nevertheless,
the effect of $X(\mathrm{HCN})$ in our model calculation
  is coupled with the effects of other fiducial parameters, such as
$q$, $\Delta V$, and the $\kappa$-curve (Fig.~\ref{kabs}).
Allowing for the possible errors in these parameters,
we estimate that the fiducial $X(\mathrm{HCN})$ is probably accurate
within a factor $\sim2$ for $T_{\mathrm{dust}}\gtrsim200$\,K.
On the one hand, the general chemical picture depicted by \cite{har10}
  is supported here, in the sense that high temperatures lock an
  important fraction of oxygen into hydrides like H$_2$O and OH,
  as observed in the far-IR, leaving an effective carbon-rich gas-phase
  chemistry that boosts the abundances of cyanopolynes. On the
  other hand, the abundance of HCN in the chemical models drops 
  quickly for $T_{\mathrm{gas}}<400$\,K, while we favor a high HCN
  abundance down to at least $T_{\mathrm{dust}}\sim200$\,K. It is possible
  that generalized shock chemistry keeps a
  substantial fraction of the gas
  in the external regions of the cocoons with $T_{\mathrm{gas}}>T_{\mathrm{dust}}$.

The involved HCN column densities are so high in BGNs that it becomes
hard, at least from our starburst approach, to
distinguish between AGN-dominated and SB-dominated regions from 
the observed HCN $\nu_2=1^f\,J=3-2$ and $4-3$ lines alone.
Nevertheless, high-resolution observations of these lines provide
  very useful constraints on the spatial extent where the greenhouse effect
  turns on, and on the velocity field of the gas in the inner regions that
  could provide evidence for a central point-like concentration of mass.
  If $\Sigma_{\mathrm{IR}}$ can be estimated from these
  observations, as performed in \S\ref{datacomp} and including spatial
  information if the lines are resolved, high-resolution
  observations of the (sub)millimeter continuum would provide the peak
  brightness and then favor an AGN or SB origin of the luminosity. The very
  strong brightness peak at $2.6$\,mm in Arp\,220W \citep{sak17}, combined
  with our inferred $\Sigma_{\mathrm{IR}}\sim10^8$\,L$_{\odot}$\,pc$^{-2}$ and
  the high point-like mass concentration \citep{sco17}, allow
  us to favor an energetically significant AGN in this source.

In general, discriminating between an AGN or SB origin of the luminosity
relies on the upper value that $\Sigma_{\mathrm{IR}}$ could attain in a compact
starburst; we do not rule out that starbursts even more compact and
  intense than considered in this study are possible.
For the highest $N_{\mathrm{H2}}=10^{25}$\,cm$^{-2}$ considered in
this work, $\Sigma_{\mathrm{IR}}\gtrsim2\times10^8$\,L$_{\odot}$\,pc$^{-2}$
makes AGN and SB models basically indistinguisable upon the diagnostics
considered here, assuming that
this value is physically possible for a starburst.
On the other hand, convective energy transfer can make the
  $T_{\mathrm{dust}}$ distribution of an AGN
  closer to that of a SB. We also note that the most deeply buried BGNs
  ($N_{\mathrm{H2}}>10^{25}$\,cm$^{-2}$) have photon-diffusion timescales 
  ($\gtrsim10^4$\,yr) that are similar to the flickering time of quasar-like
  AGN \citep[$10^{4-5}$\,yr; e.g.][]{sch15,ich19}, so that an AGN-powered
  BGN with high IR luminosity can have a
  faded AGN at its center. In such a case, even if the BGN cocoon around
  the AGN has a small gap, we may see little AGN signature in the leaked
  short-wavelength radiation. In BGNs, an AGN may resemble a SB if these
  effects become significant.

\section{Conclusions}
\label{sec:conclusions}

We have modeled in spherical symmetry both the $T_{\mathrm{dust}}$ profile
and the HCN vibrational emission, with special emphasis in models with
high column densities, to be applied to the nuclear region of buried
galaxies. Both AGN and starburst models have been generated.
Our main findings are:
\begin{itemize}
\item[1.] Trapping of the continuum radiation at mid- and far-infrared
  wavelengths extraordinarily increases the dust temperatures in the
  innermost regions of the modeled sources. This greenhouse effect
  enhances by more than one order of magnitude the radiation density in
  the mid-IR responsible for the vibrational excitation of the cyanopolynes.  
\item[2.] The increase of $T_{\mathrm{dust}}$ in the innermost regions is
  capable of generating continuum brightness of several hundred K at millimeter
  wavelengths in sources where the optical depth at these frequencies becomes
  significant.
\item[3.] The models for the continuum also predict dust temperatures close
  to the surface of the source, i.e. at the far-IR photosphere, that are in
  agreement with those required to explain the high-lying far-IR
  molecular absorption
  observed with {\it Herschel}/PACS in the most buried and warmest sources.
\item[4.] We use the results of the continuum models to calculate the
  excitation and emission of HCN from the $J=3-2$ and $4-3$ lines of the
  bending ($\nu_2=1$) vibrational state. The vibrationally excited lines
  are much more useful than the ground-state ($\nu=0$) lines to look deeper
  inside the hot, obscured galactic nuclei.
\item[5.] For the highest column density we have considered,
  $N_{\mathrm{H2}}=10^{25}$\,cm$^{-2}$, the above HCN lines are optically thick
  at least in the innermost regions for luminosity surface densities above
  $\Sigma_{\mathrm{IR}}\sim10^7$\,L$_{\odot}$\,pc$^{-2}$. The radius of this
  HCN $\nu_2=1$ photosphere increases with increasing $\Sigma_{\mathrm{IR}}$.
\item[6.] For sufficiently high column densities, which we estimate to be
  $10^{25}-10^{25.5}$\,cm$^{-2}$, we expect a drop of the brightness of the
  HCN vibrational lines toward the center of the source, due to line
  absorption of the bright (sub)millimeter continuum.
\item[7.] Using the abundance $X(\mathrm{HCN})=10^{-6}$ and an intrinsic
  $\Delta V=67$\,km\,s$^{-1}$, we reproduce the observed brightness of the
  HCN vibrational lines in several galaxies, with inferred luminosities that
  agree with independent estimates.
\item[8.] The HCN $\nu_2=1^f\,J=3-2$ and $J=4-3$ lines have significant
  optical depths in buried sources, and the impact of very high $T_{\mathrm{dust}}$
  in AGN models is moderate on the populations of the involved low-$J$ levels.
  Nevertheless, the combination with high-resolution measurements
  of the (sub)millimeter continuum and dynamical estimates of the central
  mass can provide useful diagnostics to favor an AGN or starburst origin of
the source luminosity.
\end{itemize}

\acknowledgments
We thank Francesco Costagliola for his help in writing scripts to
generate models sequentially, and an anonymous referee for
  useful comments that improved the manuscript. We thank
  Susanne Aalto for enabling us to use the data of IC~860 prior to
  the publication, and to participants at the Sesto 2019 workshop
  for comments on this work, in particular Dr. Jay Gallagher
  for his question on convection in BGNs.
E.GA is a Research Associate at the Harvard-Smithsonian
Center for Astrophysics, and thanks the Spanish 
Ministerio de Econom\'{\i}a y Competitividad for support under project
ESP2017-86582-C4-1-R. 
KS acknowledges the grant-in-aid  MOST 107-2119-M-001-022.
This research has made use of NASA's Astrophysics Data System (ADS)
and of GILDAS software (http://www.iram.fr/IRAMFR/GILDAS).

\software{GILDAS}


{}

\appendix

\section{The dust temperature profile}
\label{app}

\subsection{The basic approach}

\begin{figure}
\begin{center}
\includegraphics[angle=0,scale=.7]{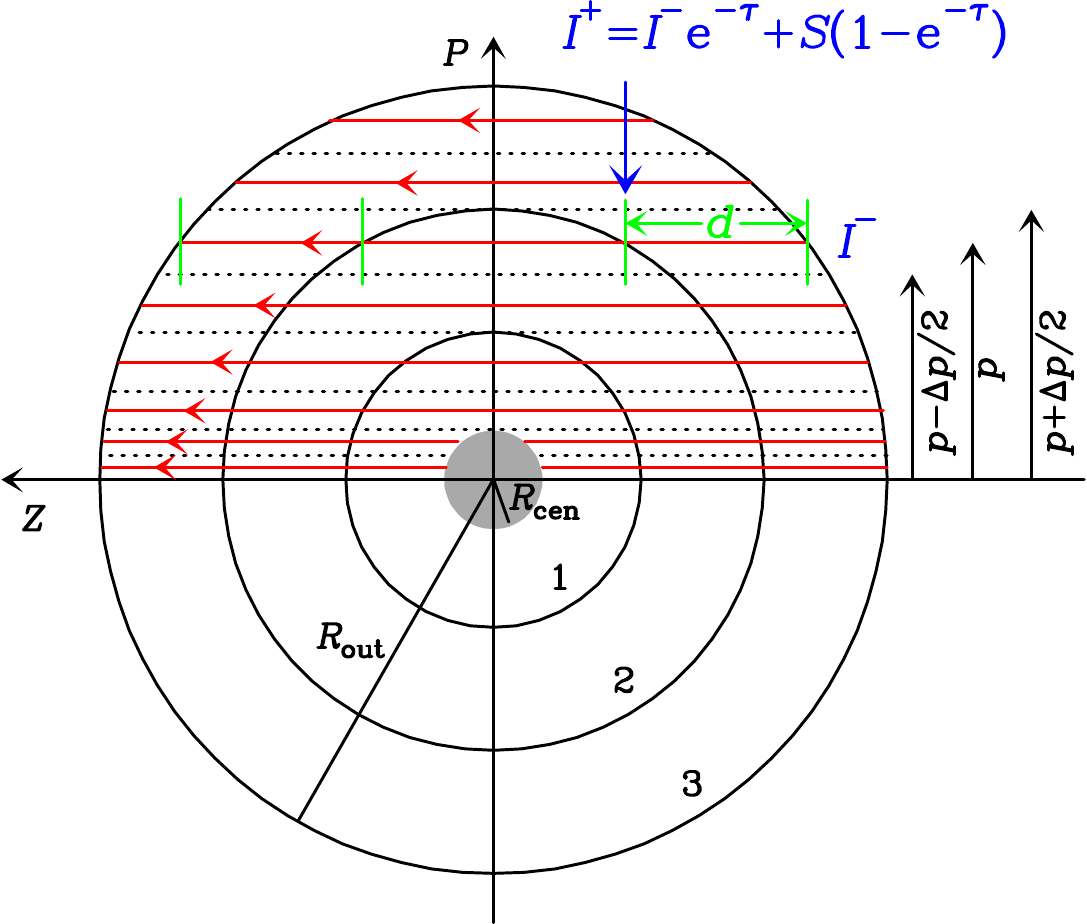}
\end{center}
\caption{Sketch of the modeled source. The radiation field
  is simulated by means of parallel rays (in red), each one representing
  the intensity in an interval $[p-\Delta p/2, p+\Delta p/2]$. After crossing
  a shell, the intensity is updated according to
  eq.~(\ref{inten}). These rays are used to compute the cooling and heating
  of every shell (eqs.~\ref{cool} and \ref{heat}), as well as the emergent
  SED and the radiation pressure on dust. 
} 
\label{pi}
\end{figure}

Figure \ref{pi} shows an squematic approach of the modeled source. The source
is divided into a number $N_{\mathrm{sh}}$ of spherical shells, within which all
physical parameters are uniform. 
The basic assumption of the approach to calculate the dust temperature profile
is that the radiation from the heating source(s) is {\it locally} absorbed by
dust and re-emitted in the infrared. For AGN models, a central compact
blackbody source with temperature $T_{\mathrm{cen}}=1300$ K is assumed to
re-radiate the AGN luminosity $L_{\mathrm{IR}}^{\mathrm{cen}}$, with radius
$R_{\mathrm{cen}}$:
\begin{equation}
L_{\mathrm{IR}}^{\mathrm{cen}}=4\pi\,R_{\mathrm{cen}}^2 \sigma\,T_{\mathrm{cen}}^4
\end{equation}
For SB models, we assume that the energy deposited (and absorbed) per unit
time in shell $m$ is proportional to both the dust mass and the density of
that shell, $\Gamma_m^{\mathrm{SB}}\propto \rho_m M_m$ (normalized as
$L_{\mathrm{IR}}^{\mathrm{SB}}=\sum_m \Gamma_m^{\mathrm{SB}}$), and re-emited at the
equilibrium dust temperature of the shell. The approximation that the 
bulk of the dust is heated by the {\it mid- and far-infrared radiation field}
within the source enables us to neglect scattering.

Owing to the spherical symmetry of the source, the radiation field is
simulated by means of a set of parallel ``rays'' that crosse the source, each
one characterized by an impact parameter $p$ and simulating the intensity
between $p-\Delta p/2$ and $p+\Delta p/2$ in a number of frequency intervals.
When crossing shell $m$, the intensity of the ray in each frequency interval
is updated as
\begin{equation}
  I^+_{p,m,\nu}=I^-_{p,m,\nu}\,\exp\{-\tau_{p,m,\nu}\}+S_{m,\nu}\,
  \left(1-\exp\{-\tau_{p,m,\nu}\}\right),
  \label{inten}
\end{equation}
where $I^-_{p,m,\nu}$ and $I^+_{p,m,\nu}$ are the incident and emergent
intensities, $S_{m,\nu}$ is the source function, and $\tau_{p,m,\nu}$ is
the optical depth of the ray through shell $m$ at frequency $\nu$:
\begin{eqnarray}
  S_{m,\nu} & = & \frac{2 h \nu^3}{c^2}\,
  \frac{1}{\exp\{\frac{h\nu}{kT_m}\}-1} \\
  \tau_{p,m,\nu}  & = & \rho_m\,\kappa_{\nu} \, d_{p,m}.
\end{eqnarray}
In the above equations, $T_m$ is the dust temperature in shell $m$,
$\rho_m$ is the density of dust, $\kappa_{\nu}$ is the mass absorption
coefficient of dust at frequency $\nu$, and $d_{p,m}$ is the length
of the path travelled by the ray in shell $m$. For rays crossing the central
source, the intensity of the ray is updated correspondingly.

Similar to the method used for lines and described in \cite{gon97}, an integral 
approach is used to compute the heating and cooling of dust in any shell.
The cooling of dust grains in shell $m$ is given by
\begin{equation}
  \Lambda_m=8\pi^2\,\int d\nu  \int_0^{R_m} dp \, p \, S_{m,\nu} \,
  \left(1-\exp\{-\tau_{p,m,\nu}\}\right),
  \label{cool}
\end{equation}
where $R_m$ is the outer radius of shell $m$. Similarly, the heating
of dust grains in shell $m$ is
\begin{equation}
  \Gamma_m=8\pi^2\,\int d\nu  \int_0^{R_m} dp \, p \, I^-_{p,m,\nu} \,
  \exp\{-\tau_{p,m,\nu}\}+\Gamma_m^{\mathrm{SB}},
  \label{heat}
\end{equation}
and the equilibrium dust temperatures are found through an iterative method
by equalizing the cooling and heating in all shells:
\begin{equation}
  \Gamma_m-\Lambda_m=0.
  \label{tdusteq}
\end{equation}
The integrals in eqs.~(\ref{cool}) and (\ref{heat}) are calculated by
using the rays that simulate the radiation field. These same rays are also used
to compute the emergent spectral energy distribution (SED)
and the radiation pressure on dust (see below), ensuring the
overall consistency of the method. In all our models, energy is conserved
to better than $1$\%.

\subsection{Convergence}

Equation \ref{tdusteq} is solved iteratively, starting with the $T_{\mathrm{dust}}$
profile of either the optically thin solution or of the solution of
another model.
In each iteration, the full Jacobian matrix is computed as the rays cross
the source, and a Newton-Raphson procedure calculates the correction
$\Delta T_{\mathrm{dust}}$ in all shells. The convergence criterion is that
the relative variation of temperatures, $\Delta T_{\mathrm{dust}}/T_{\mathrm{dust}}$,
is lower than $10^{-4}$ in all shells.

\begin{figure*}
\begin{center}
\includegraphics[angle=0,scale=.5]{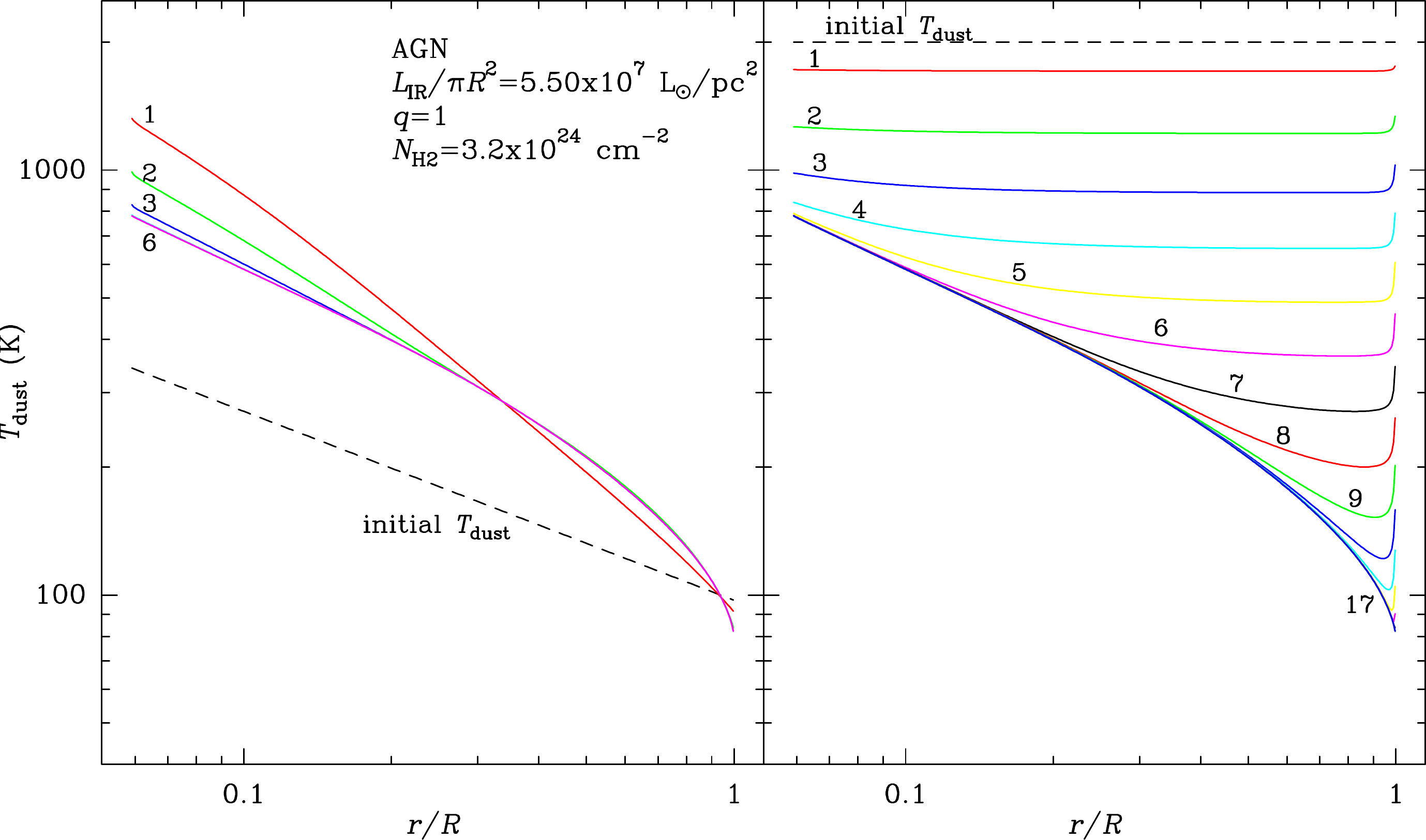}
\end{center}
\caption{Example of convergence of our models. The two panels show
  with coloured curves the computed $T_{\mathrm{dust}}$ after each iteration 
  (labeled with the iteration number), for the same model parameters but
  different initial temperatures.
  In the left panel, the initial $T_{\mathrm{dust}}$ were close to the optically
  thin solution, while $T_{\mathrm{dust}}$ was taken to be the (unphysical) value
  of 2000 K in the right-hand panel to check for possible local minima
  (dashed curves). The final
  $T_{\mathrm{dust}}$ profile, attained after 6 and 17 iterations in the left-hand
  and right-hand panels, is the same in both calculation. Since the first
  iteration of the Newton-Raphson approach usually overestimates the
  $T_{\mathrm{dust}}$--correction, $0.5\times\Delta T_{\mathrm{dust}}$ was used
  in the first iteration.
  } 
\label{tdustini}
\end{figure*}

No local minimum was found in our approach, and the same equilibrium
$T_{\mathrm{dust}}$ profile was obtained regardless of the initial temperatures
(see Fig.~\ref{tdustini}). When starting with the opticallt thin solution,
convergence was achieved in $5-8$
iterations even in the most optically thick models.

Although the model implicitely conserves energy (i.e. $L_{\mathrm{IR}}$ calculated
from the emergent SED is equal to $L_{\mathrm{IR}}^{\mathrm{cen}}$ in case
of an AGN model, or $L_{\mathrm{IR}}^{\mathrm{SB}}$ in case of a SB model),
and eq.~(\ref{tdusteq}) is accomplished in every shell, accurate
$T_{\mathrm{dust}}$ profiles are only obtained if a sufficiently fine grid
is used. The condition of convergence to a unique (exact) solution is
that every shell is optically thin at all wavelengths involved in the
absorption and emission. The optical depth at the peak of the 10\,$\mu$m
silicate feature is lower than 1 for $N_{\mathrm{H2}}<10^{22}$\,cm$^{-2}$, and
this constraint was used to establish the maximum thickness of every shell
in all our models, with a minimum number of shells of 100 to describe properly 
the $T_{\mathrm{dust}}$ profile. In models with the maximum
$N_{\mathrm{H2}}=10^{25}$\,cm$^{-2}$, the number of shells was $\approx1000$.


\subsection{Test}

\begin{figure*}
\begin{center}
\includegraphics[angle=0,scale=.5]{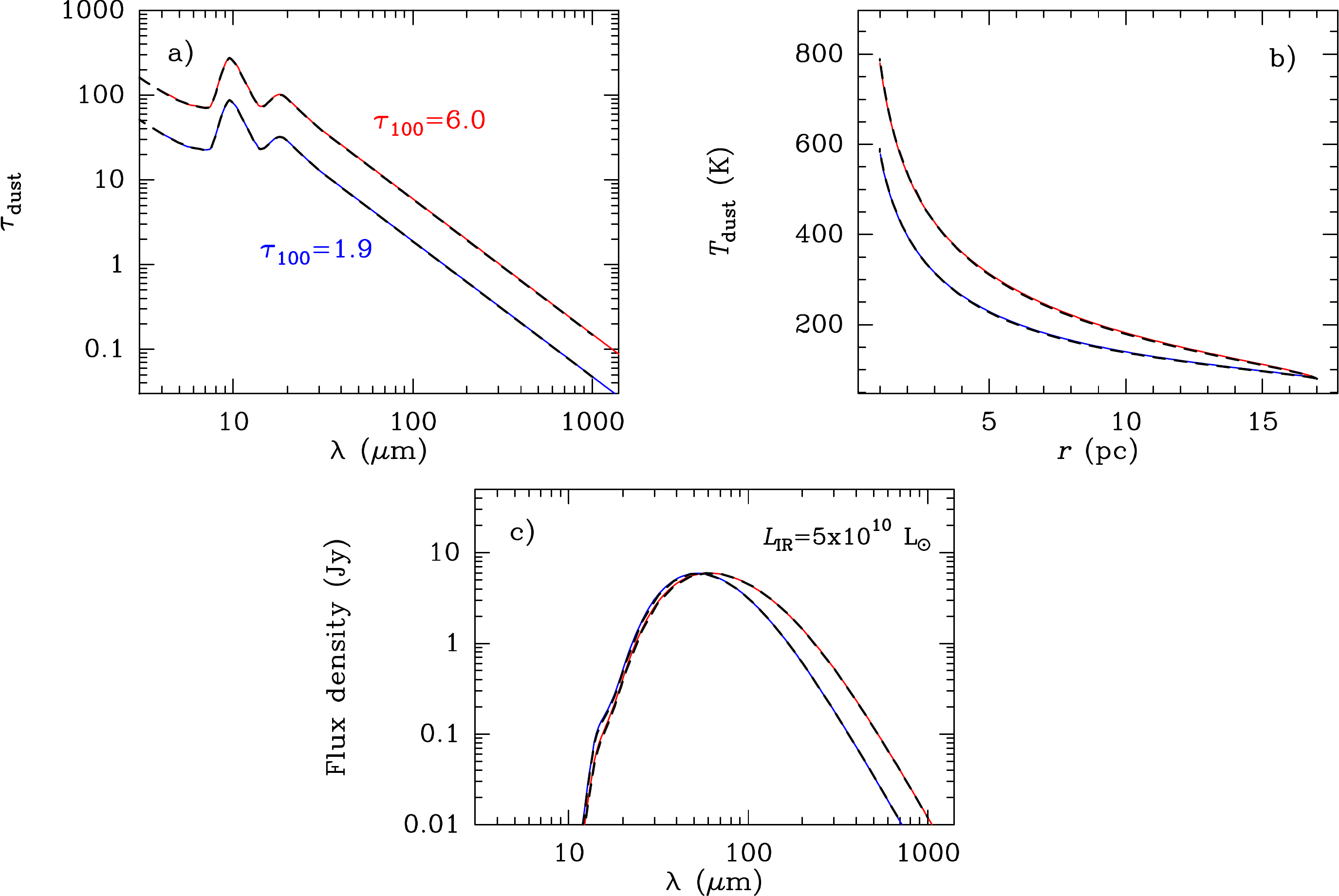}
\end{center}
\caption{Comparison between the results of two of our models (coloured curves)
  and those obtain with the V4 version of the code DUSTY \citep{ive97,ive99}
  (dashed black lines). The models are both optically thick (panel a), and
  the heating source is punctual with a luminosity of
  $5\times10^{10}$ L$_{\odot}$ observed at $59.1$ Mpc. These calculations
  use the red $\kappa_{\nu}$-curve of Fig.~\ref{kabs}.
  As shown in panels b and c, results from both codes are indistinguishable.
  } 
\label{dustdusty}
\end{figure*}

The code was benchmarked in two optically thick models with the version
V4 of the code DUSTY \citep{ive97,ive99}, yielding indistinguishable
$T_{\mathrm{dust}}$ profiles and emergent SEDs (Fig.~\ref{dustdusty}).
In both comparison models, the heating source is assumed to be punctual
(i.e. our AGN models).

\subsection{Radiation pressure}

Radiation pressure on dust is accurately calculated by using the same rays
that simulate the radiation field (Fig.~\ref{pi}). As a ray with impact
parameter $p$ is crossing the source toward the interior
(i.e. for $z<0$ in Fig.~\ref{pi}), the radiation exerts an inward pressure;
the net inward force on shell $m$ is given by
\begin{equation}
  F_m^{\mathrm{inward}}=\frac{8\pi^2}{c}\,\int d\nu  \int_0^{R_m} dp \, p \, 
  \cos\theta \, I^-_{p,m,\nu} \, (1-\exp\{-\tau_{p,m,\nu}\}), 
  \label{finward}
\end{equation}
where $\cos\theta=\sqrt{1-(p/r)^2}$ corrects for the radial component.
Likewise, when the ray is crossing the source toward the outside
(for $z>0$ in Fig.~\ref{pi}), the net outward force on shell $m$ due to
radiation pressure has the same expression:
\begin{equation}
  F_m^{\mathrm{outward}}=\frac{8\pi^2}{c}\,\int d\nu  \int_0^{R_m} dp \, p \, 
  \cos\theta \, I^-_{p,m,\nu} \, (1-\exp\{-\tau_{p,m,\nu}\}),
  \label{foutward}
\end{equation}
The net (outward) force on shell $m$ is the difference between both,
$F_m^{\mathrm{outward}}-F_m^{\mathrm{inward}}$.

\subsection{Overall results and fitting}

\begin{figure*}
\begin{center}
\includegraphics[angle=0,scale=.5]{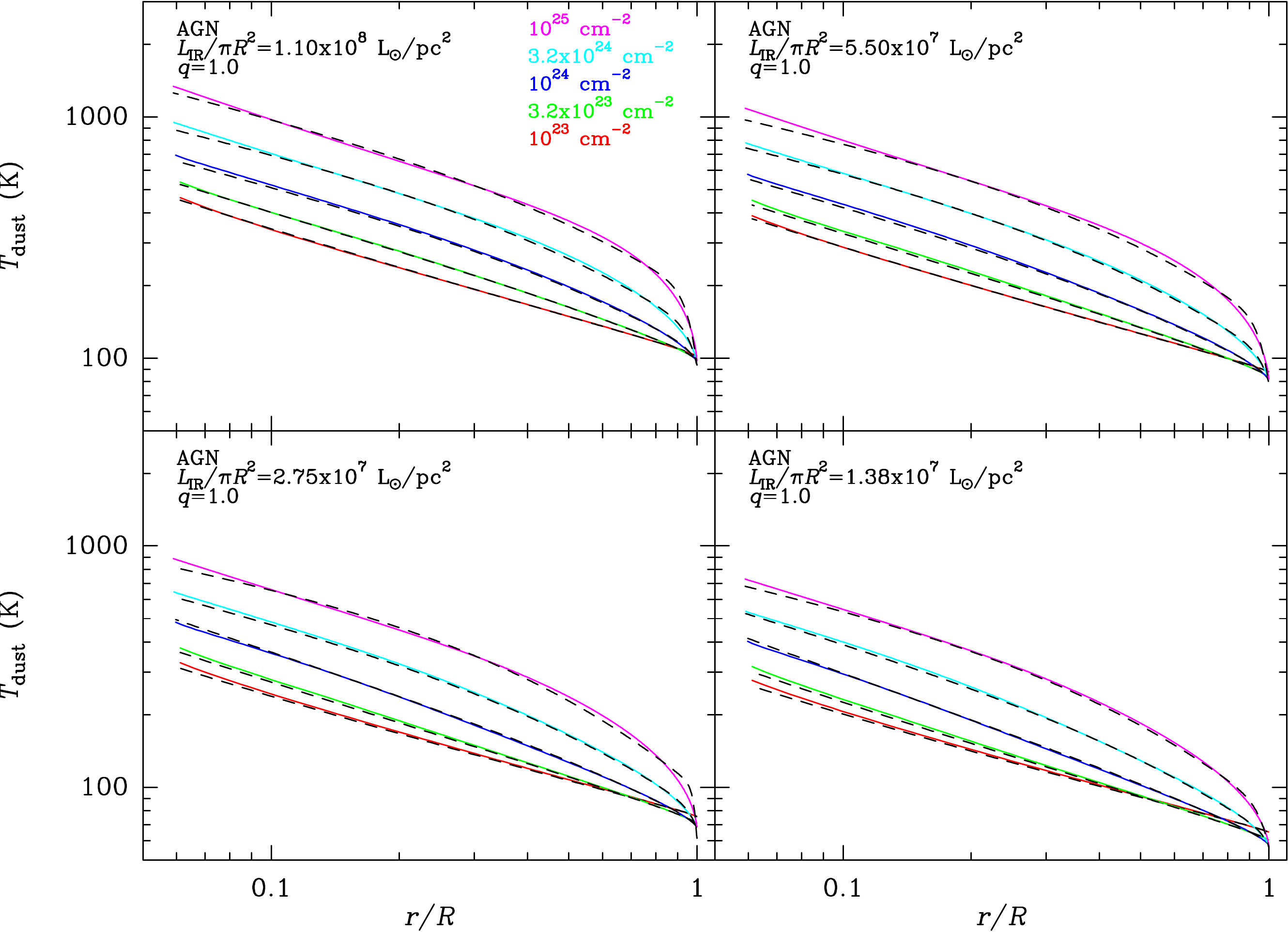}
\end{center}
\caption{The $T_{\mathrm{dust}}$ profiles for AGN models with $q=1.0$
  ($\rho\sim r^{-q}$). Each panel shows results for fixed
  $L_{\mathrm{IR}}/\pi R^2$ and different H$_2$ column densities (as indicated
  in the upper-left panel). The calculations
    use the red $\kappa_{\nu}$-curve of Fig.~\ref{kabs}.
  The dashed black curves indicate the fits to the $T_{\mathrm{dust}}$
  profiles using
  eq.~(\ref{eq:fit}), with parameters listed in Tables~\ref{tab:fitagn} and
  \ref{tab:fitsb}.
} 
\label{tdagnq1p0}
\end{figure*}

\begin{figure*}
\begin{center}
\includegraphics[angle=0,scale=.5]{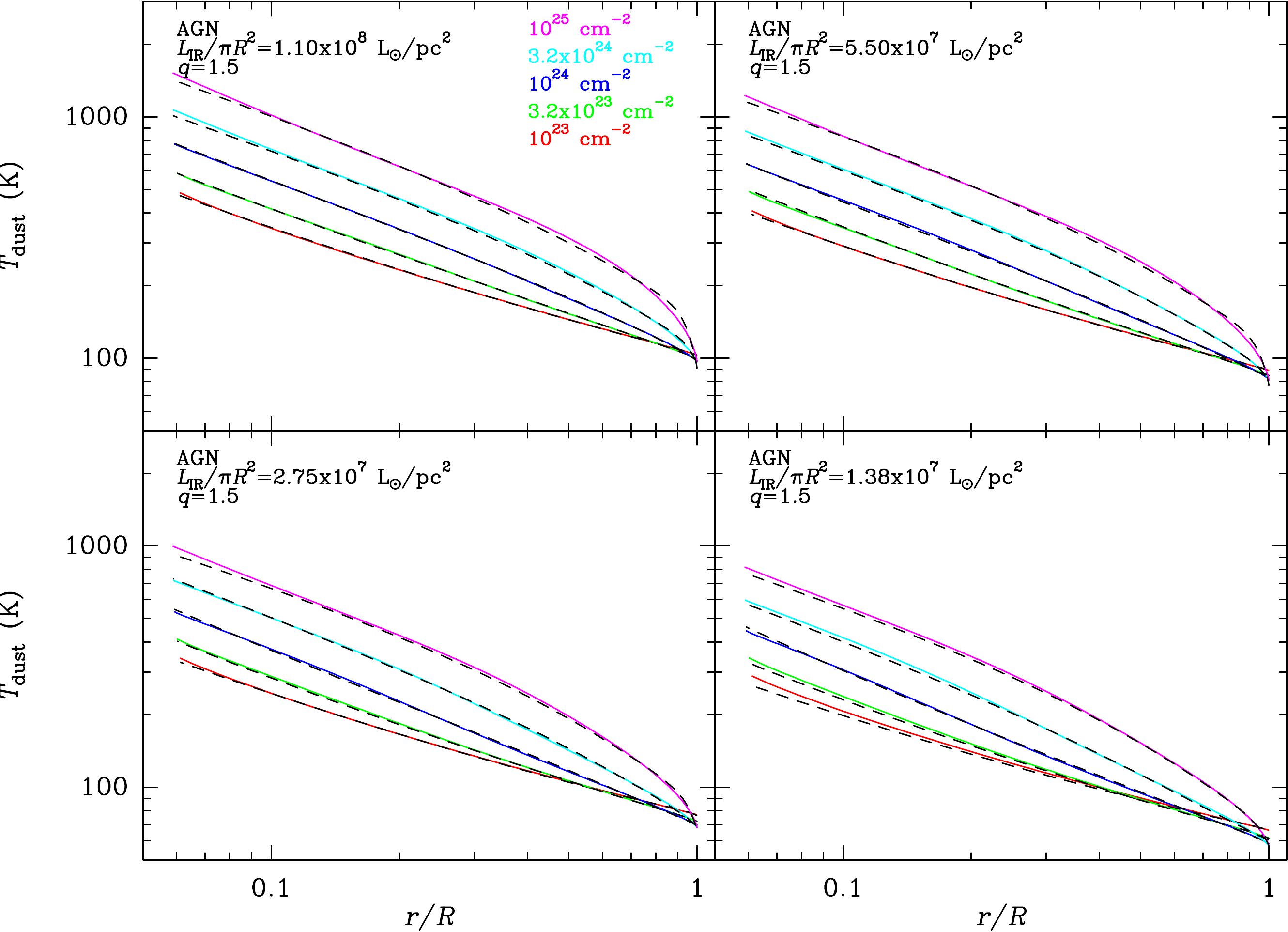}
\end{center}
\caption{Same as Fig.~\ref{tdagnq1p0} but for AGN models with $q=1.5$.
} 
\label{tdagnq1p5}
\end{figure*}

\begin{figure*}
\begin{center}
\includegraphics[angle=0,scale=.5]{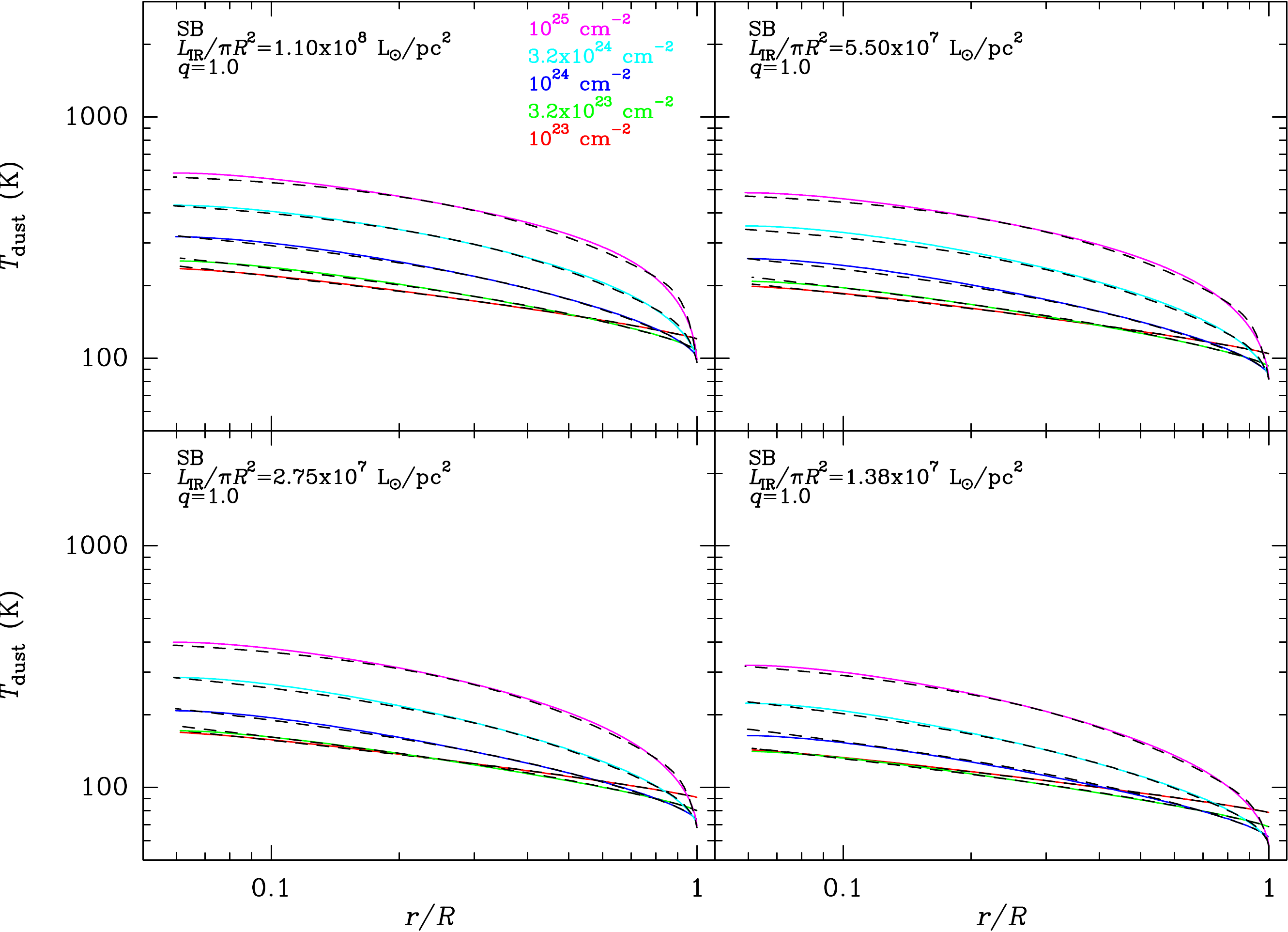}
\end{center}
\caption{Same as Fig.~\ref{tdagnq1p0} but for SB models with $q=1.0$.
} 
\label{tdsbq1p0}
\end{figure*}

\begin{figure*}
\begin{center}
\includegraphics[angle=0,scale=.5]{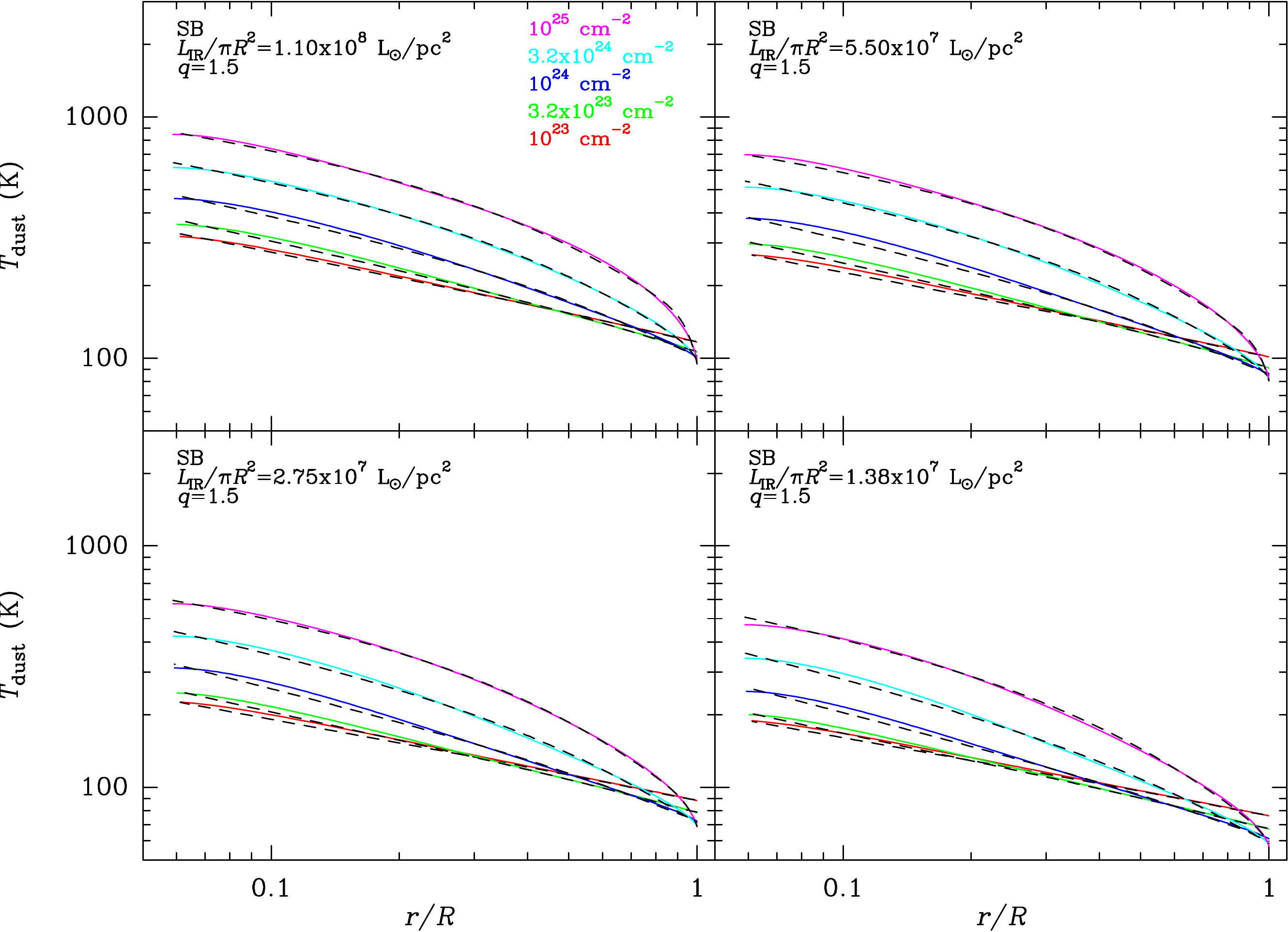}
\end{center}
\caption{Same as Fig.~\ref{tdagnq1p0} but for SB models with $q=1.5$.
} 
\label{tdsbq1p5}
\end{figure*}

The $T_{\mathrm{dust}}$ profiles as a function of the normalized radius
$r_n\equiv r/R$ depend on the spatial distribution of the heating source(s)
(AGN or SB), the surface brightness (characterized as $L_{\mathrm{IR}}/(\pi R^2)$),
the density profile ($\rho\sim r^{-q}$), the column density
(characterized as $N_{\mathrm{H2}}$ by assuming a gas-to-dust ratio
of 100 by mass), and only very slightly on the $\kappa_{\nu}$-curve of
  Fig.~\ref{kabs}. Results can then be easily scaled to any size $R$.
A subset of $T_{\mathrm{dust}}$ profiles is shown in
Figs.~\ref{tdagnq1p0}-\ref{tdsbq1p5}. Each curve has been fitted to a
modified Schechter function:
\begin{equation}
  \log_{10} T_{\mathrm{dust}}= A \, r_n^{\alpha} \, \exp\{-\beta r_n\} \,
  \frac{1}{1+b\,r_n^{\gamma}},
  \label{eq:fit}
\end{equation}
where the last factor $(1+b\,r_n^{\gamma})^{-1}$ is included
to approximately account for the sharp decrease of $T_{\mathrm{dust}}$
close to the surface for high $N_{\mathrm{H2}}$. For each model, the 5 parameters
$A$, $\alpha$, $\beta$, $b$, and $\gamma$ are fitted to minimize $\chi^2$,
and their values are listed in Tables~\ref{tab:fitagn} (AGN models) and
  \ref{tab:fitsb} (SB models). Eq.~\ref{eq:fit} gives
  $T_{\mathrm{dust}}$ accurate to better than 13\% for all models at all radii.

   \begin{table*}\small
      \caption{Fitting values for the $T_{\mathrm{dust}}$ profiles of AGN models}
         \label{tab:fitagn}
\begin{center}
          \begin{tabular}{lcccccccc}   
            \hline
            \noalign{\smallskip}
            Type & $L_{\mathrm{IR}}/\pi R^2$ & $N_{\mathrm{H2}}$ & $q$ &
            $A$ & $\alpha$ & $\beta$ & $b$ & $\gamma$ \\
            & ($10^7$\,L$_{\odot}$/pc$^{-2}$) & ($10^{24}$\,cm$^{-2}$) &
            & & ($10^{-2}$) & ($10^{-2}$) & ($10^{-3}$) & \\
            \noalign{\smallskip}
            \hline
            \noalign{\smallskip}
 AGN & 11.01 & 10.00 & 1.0 &  2.668 & $ -5.648$ & $ 16.764$ & $146.265$ &  20.773  \\
 AGN & 11.01 & 10.00 & 1.5 &  2.536 & $ -8.024$ & $ 15.725$ & $107.428$ &  20.365  \\
 AGN & 11.01 &  3.16 & 1.0 &  2.549 & $ -5.516$ & $ 18.780$ & $ 74.416$ &  20.608  \\
 AGN & 11.01 &  3.16 & 1.5 &  2.404 & $ -8.210$ & $ 16.134$ & $ 35.358$ &  20.576  \\
 AGN & 11.01 &  1.00 & 1.0 &  2.362 & $ -6.579$ & $ 15.215$ & $ 24.373$ &  20.647  \\
 AGN & 11.01 &  1.00 & 1.5 &  2.213 & $ -9.659$ & $ 10.109$ & $  6.375$ &  21.118  \\
 AGN & 11.01 &  0.32 & 1.0 &  2.172 & $ -8.233$ & $  7.812$ & $  8.106$ &  20.747  \\
 AGN & 11.01 &  0.32 & 1.5 &  2.062 & $-10.521$ & $  2.918$ & $  2.187$ &  20.036  \\
 AGN & 11.01 &  0.10 & 1.0 &  2.062 & $ -9.089$ & $  2.173$ & $  4.829$ &  20.271  \\
 AGN & 11.01 &  0.10 & 1.5 &  2.000 & $-10.362$ & $ -0.941$ & $  3.613$ &  21.048  \\ 
 AGN &  5.51 & 10.00 & 1.0 &  2.635 & $ -4.868$ & $ 20.881$ & $132.058$ &  20.962  \\
 AGN &  5.51 & 10.00 & 1.5 &  2.497 & $ -7.637$ & $ 19.768$ & $ 86.143$ &  20.481  \\
 AGN &  5.51 &  3.16 & 1.0 &  2.472 & $ -5.739$ & $ 20.961$ & $ 56.616$ &  20.667  \\
 AGN &  5.51 &  3.16 & 1.5 &  2.321 & $ -8.561$ & $ 17.684$ & $ 20.085$ &  20.665  \\
 AGN &  5.51 &  1.00 & 1.0 &  2.243 & $ -7.443$ & $ 14.488$ & $ 16.834$ &  21.090  \\
 AGN &  5.51 &  1.00 & 1.5 &  2.112 & $-10.263$ & $  9.888$ & $ -4.572$ &   5.112  \\
 AGN &  5.51 &  0.32 & 1.0 &  2.067 & $ -8.844$ & $  6.632$ & $  6.585$ &  21.918  \\
 AGN &  5.51 &  0.32 & 1.5 &  1.964 & $-11.304$ & $  1.615$ & $  2.226$ &  16.081  \\
 AGN &  5.51 &  0.10 & 1.0 &  1.987 & $ -9.366$ & $  1.819$ & $  4.110$ &  22.515  \\
 AGN &  5.51 &  0.10 & 1.5 &  1.929 & $-10.606$ & $ -1.348$ & $  3.617$ &  23.493  \\ 
 AGN &  2.75 & 10.00 & 1.0 &  2.632 & $ -4.143$ & $ 27.796$ & $122.975$ &  59.449  \\
 AGN &  2.75 & 10.00 & 1.5 &  2.438 & $ -7.393$ & $ 23.311$ & $ 60.730$ &  20.678  \\
 AGN &  2.75 &  3.16 & 1.0 &  2.360 & $ -6.356$ & $ 21.630$ & $ 40.503$ &  21.179  \\
 AGN &  2.75 &  3.16 & 1.5 &  2.209 & $ -9.538$ & $ 17.708$ & $  7.453$ &  20.403  \\
 AGN &  2.75 &  1.00 & 1.0 &  2.109 & $ -8.964$ & $ 12.675$ & $ 11.949$ &  19.999  \\
 AGN &  2.75 &  1.00 & 1.5 &  1.977 & $-11.668$ & $  7.044$ & $ -0.686$ &  14.677  \\
 AGN &  2.75 &  0.32 & 1.0 &  1.963 & $ -9.598$ & $  5.346$ & $  5.772$ &  22.366  \\
 AGN &  2.75 &  0.32 & 1.5 &  1.870 & $-11.813$ & $  0.308$ & $  3.141$ &  24.129  \\
 AGN &  2.75 &  0.10 & 1.0 &  1.912 & $ -9.521$ & $  1.432$ & $  3.826$ &  23.814  \\
 AGN &  2.75 &  0.10 & 1.5 &  1.858 & $-10.858$ & $ -1.793$ & $  4.020$ &  22.600  \\ 
 AGN &  1.38 & 10.00 & 1.0 &  2.492 & $ -5.092$ & $ 27.027$ & $ 88.262$ &  20.165  \\
 AGN &  1.38 & 10.00 & 1.5 &  2.346 & $ -7.865$ & $ 25.433$ & $ 36.323$ &  21.058  \\
 AGN &  1.38 &  3.16 & 1.0 &  2.217 & $ -7.653$ & $ 20.534$ & $ 29.480$ &  23.640  \\
 AGN &  1.38 &  3.16 & 1.5 &  2.167 & $ -9.127$ & $ 28.203$ & $-81.325$ &   1.851  \\
 AGN &  1.38 &  1.00 & 1.0 &  1.972 & $-10.267$ & $ 10.214$ & $  9.855$ &  19.857  \\
 AGN &  1.38 &  1.00 & 1.5 &  1.848 & $-13.036$ & $  4.253$ & $  1.653$ &  29.131  \\
 AGN &  1.38 &  0.32 & 1.0 &  1.865 & $-10.261$ & $  4.148$ & $  5.259$ &  22.840  \\
 AGN &  1.38 &  0.32 & 1.5 &  1.780 & $-12.280$ & $ -0.887$ & $  4.103$ &  23.617  \\
 AGN &  1.38 &  0.10 & 1.0 &  1.840 & $ -9.805$ & $  1.112$ & $  3.797$ &  24.068  \\
 AGN &  1.38 &  0.10 & 1.5 &  1.790 & $-10.765$ & $ -2.100$ & $  4.491$ &  24.373  \\
            \noalign{\smallskip}
            \hline
         \end{tabular} 
\end{center}
\end{table*}

   \begin{table*} \small
      \caption{Fitting values for the $T_{\mathrm{dust}}$ profiles of SB models}
         \label{tab:fitsb}
\begin{center}
          \begin{tabular}{lcccccccc}   
            \hline
            \noalign{\smallskip}
            Type & $L_{\mathrm{IR}}/\pi R^2$ & $N_{\mathrm{H2}}$ & $q$ &
            $A$ & $\alpha$ & $\beta$ & $b$ & $\gamma$ \\
            & ($10^7$\,L$_{\odot}$/pc$^{-2}$) & ($10^{24}$\,cm$^{-2}$) &
            & & ($10^{-2}$) & ($10^{-2}$) & ($10^{-3}$) & \\
            \noalign{\smallskip}
            \hline
            \noalign{\smallskip}
 SB  & 11.01 & 10.00 & 1.0 &  2.797 & $  0.121$ & $ 22.374$ & $130.334$ &  21.680  \\
 SB  & 11.01 & 10.00 & 1.5 &  2.720 & $ -3.202$ & $ 24.087$ & $ 82.954$ &  20.418  \\
 SB  & 11.01 &  3.16 & 1.0 &  2.618 & $ -0.645$ & $ 21.688$ & $ 62.731$ &  20.498  \\
 SB  & 11.01 &  3.16 & 1.5 &  2.545 & $ -3.968$ & $ 22.564$ & $ 20.155$ &  20.297  \\
 SB  & 11.01 &  1.00 & 1.0 &  2.395 & $ -1.965$ & $ 15.750$ & $ 22.036$ &  20.438  \\
 SB  & 11.01 &  1.00 & 1.5 &  2.317 & $ -5.408$ & $ 14.695$ & $ -4.138$ &  10.549  \\
 SB  & 11.01 &  0.32 & 1.0 &  2.238 & $ -2.919$ & $  8.951$ & $  6.767$ &  20.822  \\
 SB  & 11.01 &  0.32 & 1.5 &  2.161 & $ -6.360$ & $  6.618$ & $ -2.733$ &  16.831  \\
 SB  & 11.01 &  0.10 & 1.0 &  2.183 & $ -3.206$ & $  4.703$ & $  1.186$ &  20.121  \\
 SB  & 11.01 &  0.10 & 1.5 &  2.131 & $ -6.014$ & $  3.170$ & $ -2.224$ &  17.758  \\ 
 SB  &  5.51 & 10.00 & 1.0 &  2.719 & $  0.106$ & $ 24.519$ & $114.998$ &  19.086  \\
 SB  &  5.51 & 10.00 & 1.5 &  2.662 & $ -2.907$ & $ 27.382$ & $ 62.878$ &  20.524  \\
 SB  &  5.51 &  3.16 & 1.0 &  2.499 & $ -0.954$ & $ 21.858$ & $ 50.524$ &  20.844  \\
 SB  &  5.51 &  3.16 & 1.5 &  2.435 & $ -4.563$ & $ 22.907$ & $  7.946$ &  20.478  \\
 SB  &  5.51 &  1.00 & 1.0 &  2.272 & $ -2.439$ & $ 14.470$ & $ 17.347$ &  20.906  \\
 SB  &  5.51 &  1.00 & 1.5 &  2.191 & $ -6.121$ & $ 12.841$ & $ -4.981$ &  13.836  \\
 SB  &  5.51 &  0.32 & 1.0 &  2.144 & $ -3.252$ & $  8.074$ & $  5.220$ &  20.903  \\
 SB  &  5.51 &  0.32 & 1.5 &  2.070 & $ -6.570$ & $  5.652$ & $ -3.198$ &  13.056  \\
 SB  &  5.51 &  0.10 & 1.0 &  2.109 & $ -3.313$ & $  4.279$ & $  0.685$ &  19.337  \\
 SB  &  5.51 &  0.10 & 1.5 &  2.062 & $ -5.912$ & $  2.906$ & $ -2.423$ &  15.023  \\ 
 SB  &  2.75 & 10.00 & 1.0 &  2.623 & $ -0.097$ & $ 26.483$ & $ 99.361$ &  20.078  \\
 SB  &  2.75 & 10.00 & 1.5 &  2.574 & $ -3.269$ & $ 29.718$ & $ 40.817$ &  20.673  \\
 SB  &  2.75 &  3.16 & 1.0 &  2.362 & $ -1.805$ & $ 20.969$ & $ 40.583$ &  21.209  \\
 SB  &  2.75 &  3.16 & 1.5 &  2.302 & $ -5.386$ & $ 22.021$ & $ -4.402$ &  13.334  \\
 SB  &  2.75 &  1.00 & 1.0 &  2.154 & $ -3.001$ & $ 13.147$ & $ 13.297$ &  20.580  \\
 SB  &  2.75 &  1.00 & 1.5 &  2.068 & $ -7.087$ & $ 10.838$ & $ -5.477$ &  15.250  \\
 SB  &  2.75 &  0.32 & 1.0 &  2.055 & $ -3.449$ & $  7.298$ & $  3.832$ &  21.804  \\
 SB  &  2.75 &  0.32 & 1.5 &  1.985 & $ -6.839$ & $  4.832$ & $ -3.067$ &  13.045  \\
 SB  &  2.75 &  0.10 & 1.0 &  2.037 & $ -3.407$ & $  3.878$ & $  0.396$ &  22.058  \\
 SB  &  2.75 &  0.10 & 1.5 &  1.995 & $ -5.954$ & $  2.675$ & $ -2.425$ &  14.561  \\ 
 SB  &  1.38 & 10.00 & 1.0 &  2.482 & $ -0.833$ & $ 26.809$ & $ 83.318$ &  20.178  \\
 SB  &  1.38 & 10.00 & 1.5 &  2.439 & $ -4.281$ & $ 30.295$ & $ 19.625$ &  19.119  \\
 SB  &  1.38 &  3.16 & 1.0 &  2.215 & $ -2.575$ & $ 19.540$ & $ 32.295$ &  22.092  \\
 SB  &  1.38 &  3.16 & 1.5 &  2.139 & $ -6.685$ & $ 19.307$ & $ -7.106$ &  16.969  \\
 SB  &  1.38 &  1.00 & 1.0 &  2.038 & $ -3.623$ & $ 11.836$ & $  9.638$ &  19.448  \\
 SB  &  1.38 &  1.00 & 1.5 &  1.949 & $ -7.759$ & $  8.939$ & $ -5.580$ &  15.256  \\
 SB  &  1.38 &  0.32 & 1.0 &  1.965 & $ -3.550$ & $  6.545$ & $  2.644$ &  24.918  \\
 SB  &  1.38 &  0.32 & 1.5 &  1.903 & $ -6.949$ & $  4.278$ & $ -3.154$ &  12.858  \\
 SB  &  1.38 &  0.10 & 1.0 &  1.964 & $ -3.517$ & $  3.486$ & $  0.530$ &  81.436  \\
 SB  &  1.38 &  0.10 & 1.5 &  1.927 & $ -5.970$ & $  2.527$ & $ -2.399$ &  14.011  \\
            \noalign{\smallskip}
            \hline
         \end{tabular} 
\end{center}
\end{table*}

\end{document}